\documentclass[aps,floats,pre,showpacs
]{revtex4}

\usepackage{amsfonts,amsmath,mathrsfs} \usepackage{bm} \usepackage{bbm} \usepackage{dcolumn}
\usepackage{epsfig} \usepackage{latexsym} \usepackage{tikz}

\def\tkz#1#2{\mathord{\tikz[remember picture,baseline=(#1.base)]\node[inner xsep=0pt](#1){$#2$};}}
\def\lbar{{\mathchar'26\mkern-10mu \lambda}}

\begin{document}

\title{Hydrodynamic and field-theoretic approaches of light localization in open media}

\author{Chushun Tian}

\affiliation{Institute for Advanced Study, Tsinghua University,
Beijing, 100084, P. R. China}


\date{\today}
\begin{abstract}
{\rm Many complex systems exhibit hydrodynamic (or macroscopic) behavior at large scales
characterized by few variables such as the particle number
density, temperature and pressure obeying a set of hydrodynamic (or macroscopic)
equations. Does the hydrodynamic description exist also for
waves in complex open media?
This is a long-standing fundamental problem in studies on wave localization. Practically,
if it does exist, owing to its simplicity macroscopic
equations can be mastered
far more easily than sophisticated microscopic theories
of wave localization especially for experimentalists.
The purposes of the present paper are two-fold. On the one hand, it is devoted to
a review of substantial recent progress in this subject.
We show that in random open media the wave energy density obeys a highly unconventional
macroscopic diffusion equation at scales much larger than the elastic mean free path.
The diffusion coefficient is inhomogeneous in space;
most strikingly, as a function of
the distance to the interface, it displays novel single parameter scaling
which captures the impact of rare high-transmission states
that dominate long-time transport of localized waves.
We review aspects of this novel macroscopic diffusive phenomenon.
On the other hand, it is devoted to a review of
the supersymmetric field theory of light localization in open media.
In particular, we review its application in establishing a microscopic theory
of the aforementioned unconventional diffusive phenomenon.
}
\end{abstract}


\maketitle

\section{Introduction and motivation}
\label{sec:introduction}

Anderson localization
is one of the most profound concepts in modern condensed matter physics
(see Refs.~\cite{Abrahams10} and \cite{Mirlin08} for recent reviews).
While this phenomenon was originally predicted for
electron systems \cite{Anderson58}, direct observations of
electron wave localization are notably difficult because of
electron-electron interactions.
In the eighties, the universality of Anderson localization as a wave phenomenon was appreciated
by many researchers \cite{John83a,John83,John84,John85,Anderson84}.
The study of classical wave localization
is now a flourishing field \cite{Sheng90,Sheng95,Lagendijk09}. In recent years
the unprecedented
level reached in manipulating dielectric
materials \cite{Wiersma97,Wiersma99,Maret97,Genack00,Fishman07,Wiersma08,Genack06,Maret06,
Maret07,Zhang09,Zhang10}
and elastic media \cite{Weaver98,Hu09} has led to
substantial experimental progress
in studies of Anderson localization in various classical wave systems.
Together with the realization of Anderson localization in ultracold atomic gases
\cite{Billy08,Roati08},
these experiments have raised many fundamental issues, and are triggering a renewal of
localization studies. In addition, the experimental observation of Cao and co-workers
\cite{Cao02} has triggered considerable investigations of
using Anderson localization of light
for random lasing (see Ref.~\cite{Wiersma08a} for a review).


\begin{figure}
 \begin{center}
 \includegraphics[width=8.0cm]{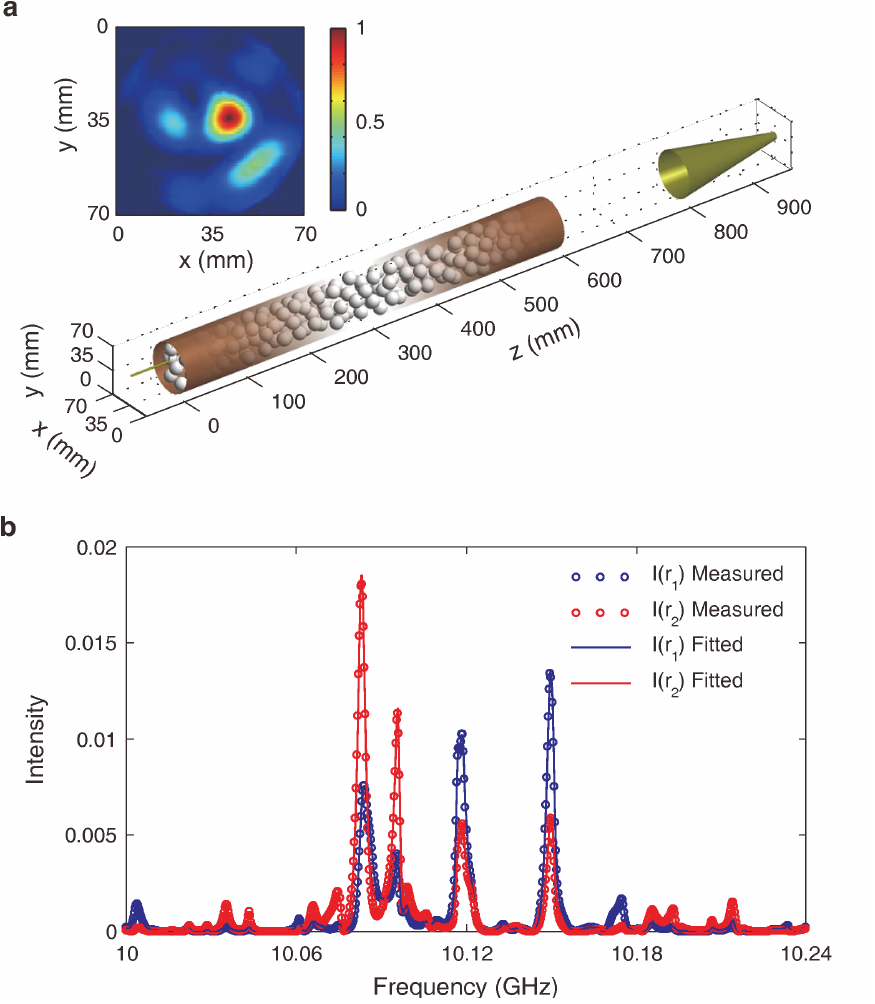}
\end{center}
 \caption{Example of light transport through open media. Upper panel: microwave
 radiation is launched from a horn placed before the sample, a copper tube
 consisting of randomly positioned alumina spheres. Wave intensity
 on the output plane is measured by the detector placed in front of the sample.
 Lower panel: intensity spectra at different positions ${\bf r}_{1,2}$ of the output
 plane. (Adapted by permission from Macmillan Publishers Ltd: Nature \textbf{471},
 345-348, $\copyright$ 2011)}
\label{fig:openmedia}
\end{figure}

Since photons do not mutually interact, it was anticipated
\cite{Anderson84,John84,John87} that
classical electromagnetic waves may serve as an ideal system for
experimental studies of Anderson localization.
Ground-breaking experimental achievement
was made in 2000 when
unambiguous evidence of microwave localization in quasi one-dimensional (Q$1$D) samples
was observed \cite{Genack00}. Since then substantial experimental progress
in electromagnetic wave localization have been achieved. Among them are
dynamics of localized microwave radiation in Q$1$D samples
\cite{Zhang09,Zhang10}, time-resolved transmission of light
through slab media \cite{Maret06,Maret07},
measurements of the spatial distribution of the localized modes \cite{Sebbah06},
and the observation of two-dimensional Anderson localization in photonic lattices
\cite{Fishman07}. It has also been within the reach of microwave experiments
on Q$1$D localized samples the statistics of quasi-normal modes \cite{Genack11}
and transmission eigenvalues \cite{Genack12}. In particular,
the crystallization of transmission eigenvalue distribution
\cite{Dorokhov82,Mello88,Frahm95,Rejaei96,Zirnbauer04,Tian05},
has been observed recently.
While these experimental studies have not yet been carried out in other wave systems,
they further show that classical electromagnetic waves
have great advantages in experimental studies of localization.

\begin{figure}
 \begin{center}
\includegraphics[width=8.0cm]{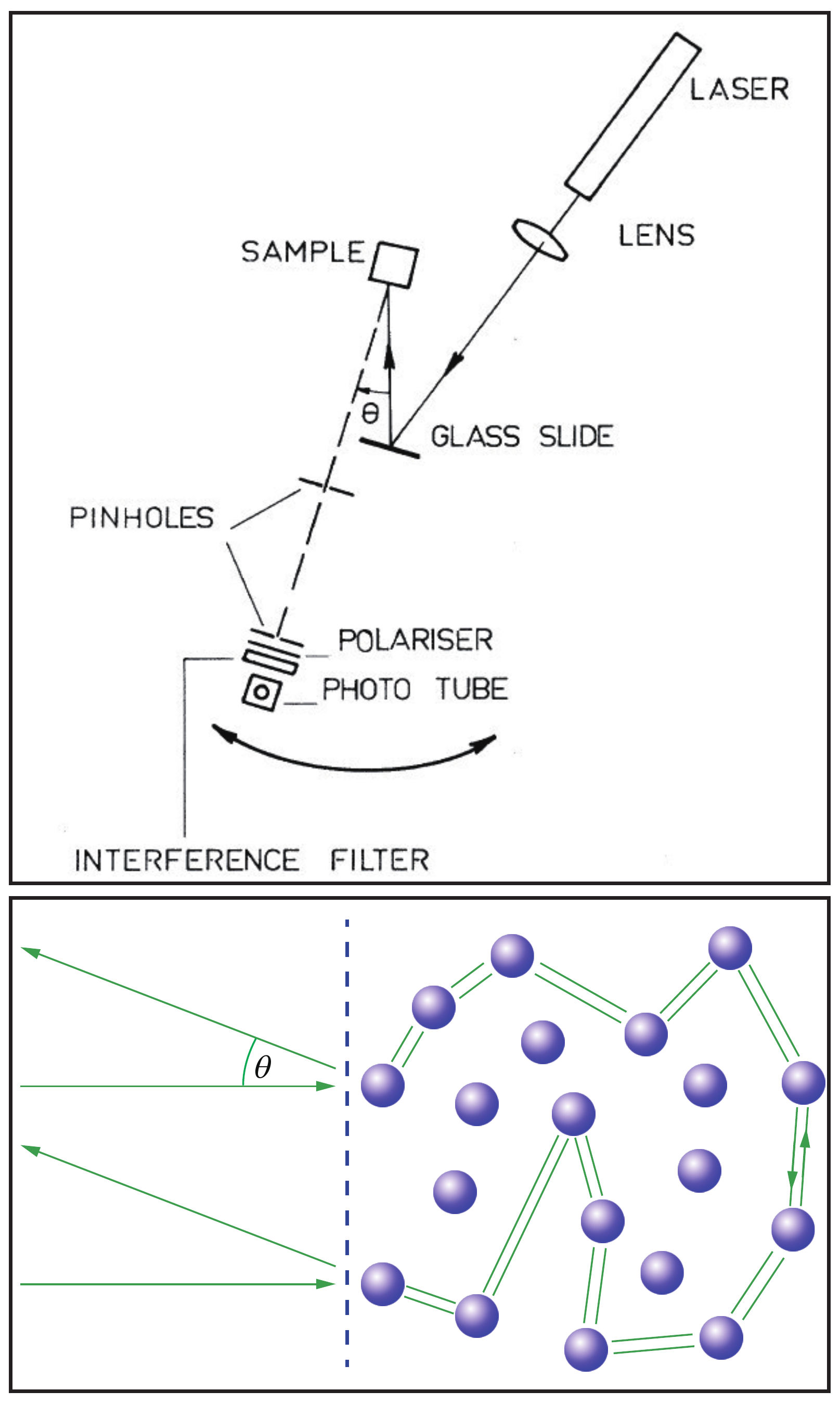}
\end{center}
 \caption{Example of light transport through open media.
 Experimental setup (upper panel) of coherent backscattering (lower panel):
 a beam of light is launched into and backscattered by a semi-infinite random
 medium. (upper panel: from Ref.~\cite{Maret85} with
 reproduction permission from G. Maret $\copyright$ The American Physical Society)}
\label{fig:backscattering}
\end{figure}

Importantly, experimental setups for probing
localization of electromagnetic waves
\cite{Maret06,Genack00,Zhang09,Zhang10,Zhang03,Zhang04,Lagendijk97,Maret99}
and other classical waves \cite{Hu09} typically are very different from those for
quantum matter waves \cite{Billy08,Roati08}. In the former,
wave energies leak out of random media through
interfaces and measurements are performed outside media. Specifically,
in transmission experiments, waves are launched into the system on
one side and detected on the other (Fig.~\ref{fig:openmedia});
in coherent backscattering experiments (see Ref.~\cite{Maret09}
for a review), the light beam is launched into and exit from
the medium at the same
air-medium interface, and the angular variation of the reflected intensity
is measured (Fig.~\ref{fig:backscattering}). Theoretically,
one treats the former system as a finite-sized open medium and the latter one
(often) as a semi-infinite open medium.
Therefore,
these classical wave experiments
address a fundamental issue -- the localization property of open systems -- which, as
we will see throughout this review,
conceptually differs from that of bulk (infinite) systems.

In fact, open systems have been at the core of localization studies for
several decades. In the mid-seventies, Thouless and co-workers noted that
the unique parameter governing the
evolution of eigenstates when the system's size is scaled is
the dimensionless conductance
(which is nowadays called the Thouless conductance), a main transport characteristic
of open (electronic) systems \cite{Thouless74}.
This eventually led to the advent of the milestone discovery -- the single parameter
scaling theory of Anderson localization -- in 1979 \cite{Anderson79}. Shortly later,
Anderson and co-workers further pointed out \cite{Anderson80}
that since the resistance displays a broad distribution
in the (one-dimensional) strongly localized regime, it is
not sufficient to study the scaling behavior of the average conductance.
Instead, one must study how
the entire conductance distribution evolves as the sample size increases.
In fact, long before the Anderson localization theory was developed,
such a distribution was discovered by Gertsenshtein and Vasil'ev
in a study of the exponential decay of radio waves transporting through
waveguides with random inhomogeneities.
Importantly, the large-conductance tail
of the distribution represents
rare localized states peaked near the sample center with
exponentially long lifetime \cite{Lifshits79,Azbel83,Azbel83a,Freilikher03}.
These states have been found experimentally to be
responsible for long-time transport of localized waves
through open systems \cite{Zhang09}.
They have
high transmission values that may be close to unity, in sharp contrast to typical localized states
with exponentially small transmission.
Such peculiar features are intrinsic to open medium.
The high transmission states even promise to have
practical applications: they mimic a `resonator' with
high-quality factors (due to the long lifetime) in optics and thus can be used to
fabricate a random laser \cite{Milner05}. In combination with optical nonlinearity, it can also be used to realize optical bistability \cite{Freilikher10}.

Nowadays experimental and theoretical results on
global transport properties (in the sense of that they provide no information on
wave propagation inside the medium) -- characterized by
conductance, reflection, transmission, etc. -- of both classical and
de Broglie waves in one-dimensional open media
have been well documented (for a review, see, e.g., Ref.~\cite{Mirlin00}).
An essential difference between finite-sized open system and infinite (closed) system
was pointed out by Pnini and Shapiro \cite{Shapiro96}. That is,
the wave field is a sum of traveling
waves for the former and of standing waves for the latter.
Let us mention a few more recent results in order for readers to better appreciate the rich physics
arising from the interplay between localization and openness of the medium.
In Ref.~\cite{Genack06}, it was found that,
surprisingly, even low absorptions
essentially improve the conditions
for the detection of disordered-induced resonances
in reflection as compared with the absorptionless case.
In Ref.~\cite{Fyodorov03}, Fyodorov considered reflection of waves injected
into a disordered medium via a single channel waveguide,
and discovered the spontaneous breakdown of $S$-matrix unitarity.
Interestingly, this may serve as a new signature of Anderson transition in high dimension.
Much less is
known regarding how wave
propagates inside the medium. Nevertheless, they may provide a key to many
new experimental results
such as the spatial distribution of localized modes \cite{Sebbah06}
and the interplay between absorption (gain) and resonant states \cite{Genack06}.

In principle, one may describe the wave field in terms of the superposition of
quasi-normal modes \cite{Zhang09,Genack11,Ching98} namely the eigenmodes of
the Maxwell equation in the presence of open boundary. This exact approach carries
full information on wave propagation but in general must be implemented by
experiments or numerical simulations.
As we are interested in physics at scales much larger than the elastic mean free path,
alternative yet simpler approach might exist. The situation might be similar to
that of many complex systems like fluids. There, many degrees of freedom causing
the complexity of dynamics notwithstanding, at macroscopic scales (much larger than
the mean free path) the system's physics is well described by few variables such as
the particle number density, temperature and pressure and
a set of hydrodynamic (or macroscopic) equations.
In fact, for open diffusive media, such a macroscopic equation is well known which
is the normal diffusion equation of wave energy density reflecting
Brownian motion of classical photons.
For open localized media, the normal diffusion equation breaks down.
The question of fundamental interests and practical importance
-- addressed from experimental viewpoints by many researchers for decades
\cite{Lagendijk00,Skipetrov04,Skipetrov06,Zhang09,Weaver98,Maret09} --
thereby arises: is the macroscopic description
valid for open localized media? Since the eighties
\cite{Berkovits87,Edrei90,Berkovitz90a} there have been substantial efforts in searching for
a generalized macroscopic diffusive model
capable of describing propagation of localized waves in open media.
In the past decade, research activity in this subject
has intensified \cite{Zhang09,Hu09,Lagendijk00,Skipetrov04,Skipetrov06,Tian08,Skipetrov08,Tian10,Skipetrov10}.
In essence, one looks for certain generalization of Fick's law
with the detailed knowledge of multiple wave scattering
entering into the generalized diffusion coefficient.
Such a diffusive model has the advantage of
technical simplicity over many sophisticated first-principles approaches,
and may provide a simple principle for guiding
experimental studies (see Ref.~\cite{Maret09} for a review).

Important progress was achieved by van Tiggelen and co-workers in 2000 \cite{Lagendijk00}.
These authors noticed that
in open media weak localization effects of waves are inhomogeneous in space.
Therefore, they hypothesized the position dependence of
the one-loop weak localization correction to the diffusion coefficient. By further introducing a key assumption
-- the validity of the one-loop self-consistency,
they obtained a phenomenological, nonlinear macroscopic diffusion model
describing static transport of localized waves in open media.
Later, the proposed self-consistent local diffusion
(SCLD) model was extended to the dynamic case \cite{Skipetrov04,Skipetrov06}.
In the past decade the SCLD model has been used to guide
considerable research activities in classical wave localization.
The subjects include: static reflection and transmission of light in
systems near the mobility edge \cite{Lagendijk00}, dynamic reflection and transmission
of Q$1$D weakly localized waves \cite{Skipetrov04},
dynamics of Anderson localization in three-dimensional media \cite{Skipetrov06},
transmission of localized acoustic waves through strongly scattering plates \cite{Hu09},
and transmission and energy storage in random media with gain \cite{Yamilov10}.

This prevailing model was first examined
in Ref.~\cite{Zhang09} in experiments on pulsed microwave transmission in
Q$1$D localized samples. The SCLD model
failed to describe the experimental results. In fact, in these systems,
the long-time transport of localized waves is determined by
rare disorder-induced resonant transmissions.
In addition, the SCLD model fails to describe results
obtained from numerical simulations on static (long-time limit) wave transport
through one-dimensional localized samples \cite{Tian10,Skipetrov10}.
Since the macroscopic diffusive model
describes local (in space) behavior of wave propagation (in the average sense),
it is very `unlikely' that it
could capture simultaneously two prominent
features of resonant transmissions. That is, they are
rare events and objects highly non-local in space.
In fact, it had been commonly suspected that eventually the validity of the macroscopic
diffusion concept in open systems would be washed out by
these rare events.
On the other hand, the predictions of the SCLD model agree with the
experimental results in higher dimensions \cite{Hu09}.
In view of these contradictory observations one inevitably has to re-examine
the fundamental issue:
{\it whether and how do localized waves in open media
exhibit macroscopic diffusion?}

A preliminary but very surprising answer to this question was provided in a first-principles study of
static transport
of localized waves in one-dimensional open media \cite{Tian10}. It turns out that in
these systems, the macroscopic diffusion (or equivalently Fick's law) concept never breaks
down at scales much larger than the elastic mean free path.
Rather, in depending on the distance to
the interface, the diffusion coefficient exhibits novel scaling.
It is such scaling that unifies
the objects of
resonant transmission and
macroscopic diffusion.
The novel scaling behavior is
missed by the phenomenological SCLD model.
In this review, we will discuss the existence of such highly unconventional macroscopic diffusion
in more general situations (e.g., dynamic transport and in high dimensions).
It is believed that waves propagating through
open media may exhibit even richer macroscopic behavior.
As such, the unconventional macroscopic diffusion of waves in
open media may potentially
open a new direction for the study of Anderson localization.
One of the main purposes of this paper is to review
various aspects of this diffusion phenomenon such as its microscopic mechanism,
microscopic theory and perspectives.

Technically, treatments of localization in open media differ dramatically from
those of infinite media. The study of open systems has proved to be
an extremely difficult task especially for high dimensions and in the dynamic case.
Recall that for infinite electron systems,
the diagrammatic technique \cite{Woelfle80a,Woelfle80,Woelfle92,Woelfle10,Larkin79}
has had considerable successes in (approximate) studies of
a variety of localization phenomena, e.g., one-loop weak localization,
strong localization, and criticality of
Anderson transition (see Refs.~\cite{Woelfle92,Woelfle10} for reviews).
In view of these successes, it is natural to extend this technique
to classical wave systems, and this has been done
by many authors \cite{Kroha93,Nieh98,Kirkpatrick85}.
In the past, the non-perturbative diagrammatic theory of Vollhardt and
W{\"o}lfle (VW) \cite{Woelfle80a,Woelfle80}, having the advantage of technical simplicity,
has become a popular approach in studies of classical wave localization.
Furthermore, it has been noticed
\cite{Suslov12,Lagendijk00,Skipetrov04,Skipetrov06,Skipetrov08}
that it is possible to generalize the VW theory
appropriately to describe transport of strongly localized waves in open media.
It should be emphasized that the rational of
the original VW theory \cite{Woelfle80a,Woelfle80}
is built upon the exact Ward identity \cite{Woelfle80,Nieh98,Nieh98a}
and the sophisticate summation over
the dominant infrared divergent diagrams
(see Refs.~\cite{Woelfle92,Suslov07} for a technical review).
This rational is at the root of the one-loop self-consistency and
to the best of my knowledge,  has not yet been established for open media.
These considerations necessitate the invention of an exact theory beyond simple
one-loop perturbation.

There have been few attempts \cite{John83a,John84,John83,John85,Lai12} of generalizing the
replica field theory \cite{Wegner79,Wegner80,Efetov80} and the Keldysh field theory \cite{Kamenev10}
to classical wave systems.
However, these studies \cite{John83a,John84,John83,John85,Lai12} focus on infinite random media.
For open systems some exact solutions for Q$1$D strong localization have been found
by using the replica field theory \cite{Tian05}.
As for the replica field theory, in general, its applicability in the strongly localized regime
needs further investigations due to
the well-known problem of analytic continuation \cite{Efetov97,Zirnbauer84}.
As for the Keldysh field theory, it does not encounter such a difficulty. However,
to the best of my knowledge, it is not clear how to use this theory to obtain
concrete results for strong localization. Finally, the
Dorokhov-Mello-Pereyra-Kumar (DMPK) equation \cite{Beenakker97,Dorokhov82,Mello88} is an exact theory
and a very powerful approach to Q$1$D strong localization in open media.
The special case of the DMPK equation in one dimension was discovered as early as in 1959 \cite{Vasilev59}.
But, it mainly provides information on global transport properties such as
transmission and its fluctuations; it provides no information
on local wave transport properties: it could not describe how
waves propagate from one point to the other
inside media. For example, by using this theory one could not judge the
possibility of
generalizing Fick's law to localized open media.
In addition, it is well-known that the DMPK equation is
valid only in one dimension.

The supersymmetric approach escapes all these key difficulties.
In Ref.~\cite{Chernyak92}, the supersymmetric quantum mechanics was used to study
reflection of classical waves by one-dimensional semi-infinite disordered medium.
The supersymmetric field theory invented by Efetov \cite{Efetov82b,Efetov83,Efetov97}
was used by Mirlin and co-workers to study the deviation from the Rayleigh distribution of classical
light intensity in diffusive Q$1$D samples (see Ref.~\cite{Mirlin00} for a review).
In Ref.~\cite{Tian08}, the supersymmetric field theory was generalized to
high-dimensional open media with internal reflection, and
was employed to investigate
dynamic transport of classical waves in high-dimensional open media.
In particular, the unconventional macroscopic diffusion was discovered
in Ref.~\cite{Tian10} from this first-principles theory \cite{Tian08}.
In fact, the machinery of supersymmetric field theory
has become a standard approach
in studies of disordered electronic systems. Nonetheless,
researchers working on classical wave localization are less familiar with this technique.
On the other hand, there are a number of recently observed phenomena
regarding classical wave transport through open media
which can be thoroughly studied by using this technique.
These include effects of internal reflection on
transmission fluctuations \cite{Genack12} and
dynamics of localized waves \cite{Zhang09}.
Since there are many prominent differences between classical and
electronic waves (e.g., the condition of strong localization \cite{John83a,John84}),
and a review of the application of the supersymmetric field theory in
light localization has been absent,
it is necessary to introduce here this technique -- in the context of
transport of classical electromagnetic waves through open random media -- at a pedagogical level,
which serves another main topic of this review.

The review is written in a self-contained manner,
with the hope that readers wishing to master the supersymmetric technique and
then apply it to classical wave systems
could follow most technical details without
resorting to further technical papers.
Because of this, it could not be a complete introduction
of this advanced theory. In fact, there have been a comprehensive book \cite{Efetov97} and several
excellent reviews
\cite{Efetov83,Mirlin00,Mirlin00a,Mirlin08}
covering various aspects of supersymmetric field theory. In particular,
we will not introduce
non-perturbative treatments \cite{Zirnbauer04,Efetov83a,Zirnbauer91,Hueffmann90,Zirnbauer92,Zirnbauer94},
because these are technically highly demanding.
The present review, implemented with substantial technical details, may be considered
as a `first course in practicing the supersymmetric field theory'.
It aims at helping the researchers working in the field of classical wave localization
to become familiar with formulating large-scale wave dynamics in terms of
supersymmetric functional integral formalism and further to master its
perturbative treatments. With these preparations, readers
may be ready to manipulate its non-perturbative treatments after further resorting to
original technical papers and reviews. We emphasize that although the present review
is written in the context of classical waves, unconventional macroscopic diffusion
to be reviewed below is a universal wave phenomenon in open systems. In particular,
it also exists for de Broglie waves.

For simplicity, we shall focus on scalar waves
throughout this review. We shall not distinguish the concepts of (classical) scalar wave,
electromagnetic wave,
and light. The remainder of this review is organized as
follows. In Sec.~\ref{sec:model} we will review
a number of research activities in studies of transport of localized waves through open media.
These studies may be roughly classified into two categories: one is
based on the macroscopic diffusion picture and the other on the
mode picture. In Sec.~\ref{sec:origin_SUSY} we will discuss in a heuristic manner
the `origin' of supersymmetry in studies of disordered systems. In Sec.~\ref{sec:SUSY}
we will proceed to introduce the supersymmetric field theory suitable for calculating
various physical observables in random open media. Then, in Sec.~\ref{sec:localdiffusion},
we will apply this theory to study a special quantity, the spatial correlation
of wave intensity.
In particular, we will develop a general theory of unconventional macroscopic diffusion
for the general case of a random slab.
In Sec.~\ref{sec:localdiffusion1D}, we will study in details the special
case of one-dimensional unconventional macroscopic diffusion.
In particular, we will present the explicit results for the static local
diffusion coefficient and provide corroborating numerical
evidence. Finally, we conclude in Sec.~\ref{sec:conclusion}. In order to
make this review self-contained,
we include a number of technical details
in Appendices~\ref{sec:supermathematics}-\ref{sec:cancelation}.

\section{Macroscopic diffusion versus mode picture of wave propagation}
\label{sec:model}

In infinite three (or higher)-dimensional systems,
the envelope of eigenfunctions
decays exponentially in space for the distance from
the center far exceeding the localization
length, $\xi$, when disorder is strong. Below or in two dimensions,
even weak disorders lead to localization \cite{Abrikosov79,Berezinskii79,Berezinskii73,Efetov83a,Mott61,Anderson79}.
The (asymptotic) exponential decay of eigenfunctions in space
is a key characteristic of localized waves
in the mode picture. Can localized waves be understood in terms
of the macroscopic diffusion picture?
Recall that in the absence of wave interference effects,
light exhibits pure `particle (classical photons)' behavior, which
in random environments is the well-known Brownian motion
(We shall not consider here the case where the motion of photons
is non-Brownian.) or normal diffusion of photons.
When wave interference is switched on, localization
effects set in. It has been well established that
the latter does not render the macroscopic diffusion concept
invalid \cite{Abrikosov79,Berezinskii79,Efetov83a,Efetov83,Woelfle80a,Woelfle80,
Woelfle92,Woelfle10,Hikami81}. Rather, they (strongly)
renormalize the Boltzmann diffusion constant.  Specifically, given the point like source
$J_t({\bf r}')$ located at ${\bf r}'$, the disorder averaged intensity
profile at time $t$ is given by \cite{Chernyak92} $I({\bf r},t)=\int\frac{d\omega}{2\pi}\frac{d\tilde \omega}{2\pi}
e^{i\tilde \omega t}{\cal Y}({\bf r},{\bf r}';\tilde \omega)
J_{\omega+\frac{\tilde \omega}{2}}({\bf r}')J_{\omega-\frac{\tilde \omega}{2}}^*({\bf r}')$,
where $J_{\omega}({\bf r}')$ is the spectral decomposition of the source.
Here, we have introduced
the spatial correlation function defined as
\begin{eqnarray}
{\cal Y} ({\bf r},{\bf r}';\tilde \omega)  \equiv \left\langle G^A_{(\omega+\frac{\tilde \omega^+}{2})^2} ({\bf r},{\bf
 r}') \, G^R_{(\omega-\frac{\tilde \omega^+}{2})^2} ({\bf r}',{\bf r})\right\rangle
 \label{DCdefinition}
\end{eqnarray}
in the frequency domain.
Here, $G^{R,A}_{(\omega\mp\frac{\tilde \omega^+}{2})^2}$ are the retarded (advanced) Green functions whose explicit
definitions will be given in Sec.~\ref{sec:trick}.
Throughout this review we use $\langle\cdot\rangle$ to denote the disorder average.
Most importantly, the correlation function satisfies \cite{Woelfle92,Efetov97}
\begin{eqnarray}\label{eq:130}
    (-i{\tilde \omega}- D({\tilde \omega})\nabla^2)
    {\cal Y}({\bf r},{\bf r}';{\tilde \omega})=\delta({\bf r}-{\bf r}').
\end{eqnarray}
It is important that here, the diffusion coefficient, $D({\tilde \omega})$, is frequency-dependent.
The latter accounts for the retarded nature of the response to the spatial inhomogeneity
of the wave energy density.

How are the mode and macroscopic diffusion pictures modified
in open media? For the mode picture,
because of the energy leakage through the system's boundary,
the Hermitian property is destroyed.
Consequently, one describes transport in terms of quasi-normal modes
\cite{Genack11,Zhang09,Ching98} each of which has
a finite lifetime. Loosely speaking, wave energy is pumped into these modes,
stored there and emitted
in the course of time.
For the macroscopic diffusion picture,
substantial conceptual issues arise. In this section, we will review
great efforts made in extending the macroscopic diffusion concept to
localized wave transport through open media.

\subsection{Electromagnetic wave propagation: macroscopic diffusion picture}
\label{sec:SCLD}

Macroscopic diffusive approach to light transport may be dated back to
the first half of last century, when
the Boltzmann-type kinetic equation was used to study
radiation transfer in certain astrophysical processes \cite{Chandrasekhar}.
In fact, it is a canonical procedure of deriving a normal diffusion equation --
valid on the scale much larger than the transport mean free path, $l$ -- from the Boltzmann
kinetic equation.
In the normal diffusion equation, the diffusion constant $D(\tilde \omega) = cl/d \equiv D_0$, where $d$ is the dimension and $c$ the wave velocity.
Notice that the latter may be reduced by the scattering resonance \cite{Niuwenhuizen} which
we shall not further discuss, and from now on we set $c$ to unity.

The Boltzmann transport theory essentially
treats light scattering off dielectric fluctuations as Brownian motion of
classical particles, the `photons', and ignore the wave nature of light.
However, the latter gives rise to interference effects
which have far-reaching consequences. A canonical example
is the coherent backscattering of light from semi-infinite diffusive media (Fig.~\ref{fig:backscattering},
lower panel). There, two photon paths following
Brownian motion may counterpropagate and thus constructively
interfere with each other \cite{Altshuler83,Bergmann84},
leading to enhanced backscattering \cite{Maret09,Golubentsev84,Lagendijk85,Maret85,Akkermans86}.
Effects of wave interference are even more pronounced in low dimensions
($d\leq 2$) or in three-dimensional strongly scattering random media,
where light localization eventually occurs \cite{Anderson84,John84,John85}.
To better understand the macroscopic diffusion picture of wave transport in open localized media
we begin with a brief summary of the
(conventional) macroscopics of wave propagation in infinite media.

\subsubsection{Infinite random media}
\label{sec:diffusioninfinite}

In this case, the diffusion of wave energy follows Eq.~(\ref{eq:130}).
The leading order wave interference correction to
$D_0$ -- the well-known weak localization correction due to
constructive interference between two counterpropagating paths --
was first found (for electronic systems) using the diagrammatic method
\cite{Woelfle80a,Woelfle80,Larkin79}, which is
\begin{equation}\label{eq:123}
    \delta D^{(1)}({\tilde \omega})=-\frac{D_0}{\pi\nu}\int\frac{d^d{\bf k}}{(2\pi)^d}\frac{1}{-i\tilde \omega + D_0 {\bf k}^2},
\end{equation}
with $\nu$ is the local density of states of particles.
The result is perturbative and valid only if $\delta D^{(1)}({\tilde \omega})/D_0\ll 1$.
In spite of this limitation, Eq.~(\ref{eq:123}) has many important implications:
the lower critical dimension of
Anderson transition is two, because it suffers infrared divergence
for dimension $d\leq 2$ \cite{Anderson79}. That is,
the system is always localized in low dimensions ($d\leq 2$) while
exhibits a metal-insulator transition in higher dimensions ($d> 2$).

(i) In dimension $d\leq 2$, $\delta D^{(1)}({\tilde \omega})$ is much smaller than
$D_0$ for large frequencies, $\tilde \omega\gg D_0/\xi^2$.
This implies that at short times ($\ll \xi^2/D_0$),
wave transport largely follows normal diffusion with a small suppression
due to interference. For low frequencies, $\tilde \omega\ll D_0/\xi^2$ which
corresponds to long times ($\gg \xi^2/D_0$),
one finds using non-perturbative methods \cite{Abrikosov79,Berezinskii79,Efetov83a,Woelfle80a,Woelfle80,Berezinskii73}
that the dynamic diffusion coefficient, $D(\tilde \omega)$,
crosses over to $ \sim -i\tilde \omega \xi^2$.
Notice that $D({\tilde \omega})\stackrel{\tilde \omega\rightarrow 0}{\longrightarrow} 0$
is a hallmark of strong localization. Namely,
wave energy diffusion stops.
(ii) In dimension $d>2$, result (\ref{eq:123}) does not diverge in the
infrared limit ($\tilde \omega=0$) but suffers an ultraviolet
divergence. To cure the latter, one needs to introduce the ultraviolet
cutoff $\sim l^{-1}$ for the integral over ${\bf k}$.
This introduces a dimensionless coupling constant
$\nu D_0 l^{d-2} \sim D_0/\delta D^{(1)}({\tilde \omega}=0)$.
For large coupling constant, the interference correction (\ref{eq:123}) is negligible;
for a coupling constant order of unity, the interference correction and
$D_0$ are comparable. The latter is the onset
of the Anderson transition, and $g_0\equiv \nu D_0 l^{d-2}={\cal O}(1)$ is equivalent
to the well-known Ioffe-Regel criterion \cite{Ioffe69}. Above the critical point,
the system is in the metallic phase and
$D(\tilde \omega)$ approaches some non-zero constant in the limit $\tilde \omega\rightarrow 0$;
below the critical point, the system is in the localized phase and
$D(\tilde \omega)$ vanishes again as $\sim -i\tilde \omega\xi^2$.
(iii) At the critical point (in dimension $d>2$), the low-frequency behavior of the
diffusion coefficient is given by \cite{Wegner76}
$
    D(\tilde \omega)\sim (-i\tilde \omega)^{\frac{d-2}{d}}
\stackrel{\tilde \omega\rightarrow 0}{\longrightarrow}0 $.
These results have also been predicted for completely different systems
-- the quantum kicked rotor \cite{Casati79,Fishman10} --
by using first-principles theories \cite{Altland10,Altland11}, and
have been observed experimentally \cite{Raizen95,Deland08,Deland10}.

Let us make two remarks for the macroscopic diffusion equation (\ref{eq:130}).
First, the dynamic diffusion coefficient $D(\tilde \omega)$
is homogeneous in space. Secondly, this shows that wave interference does not
destroy Fick's law: the energy flux is proportional to
the (negative) energy density gradient, given by
$-D(\tilde \omega)\nabla I({\bf r};\tilde \omega)$ in the frequency domain.
According to this law, the response to inhomogeneous wave energy density is
local in space and non-local in time (retarded effect).


\subsubsection{Open random media}
\label{sec:SCLDintroduction}

The study of light propagation in open random media has a long history
\cite{Chandrasekhar}, and since the eighties researches on this subject
have been strongly motivated by the
search for light localization.
As mentioned above, in the diffusive regime,
wave interference corrections to the Boltzmann diffusion constant are
negligible and photons follow normal
diffusion or Brownian motion. The effect of the air-medium interface is
to introduce certain boundary conditions
that implement the normal diffusion equation (see Ref.~\cite{Niuwenhuizen} for a review).
The boundary condition is essential to calculations of
coherent backscattering lineshape (for diffusive samples) \cite{Golubentsev84,Akkermans86}.
In this case, the boundary condition effectively enlarges the
diffusive sample, and the effective interface is located outside the medium
at a distance of $\zeta$ -- the so-called extrapolation length -- to the genuine interface.
The relation between the extrapolation length and the internal reflection coefficient
was studied by various authors \cite{Lagendijk89,Zhu91,Genack93}.

{\it Can wave propagation in open localized media be described by certain
macroscopic diffusion equation?}
This fundamental problem has attracted the attention of many researchers.
The first attacks were undertaken more than two decades ago (e.g., Refs.~\cite{Berkovits87,Berkovitz90a,Edrei90}),
motivated by experiments on coherent backscattering of light from
strongly disordered media.
The earlier attempts resort to phenomenological
generalization of the scale-dependent diffusion coefficient
developed for infinite media to open media \cite{Lee86,Imry82}.
While these theories had been debated, an important observation was made in Ref.~\cite{Lagendijk00}.
That is, because photons near
the air-medium interface easily escape from the medium, the returning probability
density near the interface must be smaller than that deep in
the medium. Consequently, wave interference (or localization) effects must be
inhomogeneous in space. The strength of interference effects increases as
waves penetrate into the medium.
Indeed, both the supersymmetric field theory \cite{Tian08} and the diagrammatic
theory \cite{Skipetrov08} developed very recently for open media
justify this important observation. Specifically,
it is shown
that near the interface the diffusion coefficient
may be largely unrenormalized even though strong localization
develops deep in the medium (cf. Sec.~\ref{sec:localdiffusion1D}).
Interestingly, a similar conclusion was reached
in the study of superconductor-normal metal hybrid structures even earlier \cite{Altland98}.

To take the inhomogeneity of wave interference effects into account,
the phenomenological SCLD model was introduced
in a series of papers \cite{Lagendijk00,Skipetrov04,Skipetrov06}
which include two key assumptions.
The first is to hypothesize
a local (position-dependent) diffusion coefficient, $D({\bf r};\tilde \omega)$,
that locally generates a macroscopic energy flux via Fick's law.
Specifically, instead of Eq.~(\ref{eq:130}), the correlation function
${\cal Y}({\bf r},{\bf r}';\tilde\omega)$ satisfies
\begin{equation}\label{eq:96}
    \left[-i\tilde\omega - \nabla \cdot D({\bf r},\tilde \omega)\nabla\right]
    {\cal Y}({\bf r},{\bf r}';\tilde\omega) = \delta({\bf r}-{\bf r}').
\end{equation}
The crucial difference from Eq.~(\ref{eq:130}) is
the position-dependence of the diffusion coefficient.
The second is the one-loop consistency which is a phenomenological generalization
of the VW theory \cite{Woelfle80a,Woelfle80}.
That is, the local diffusion coefficient is given by
\begin{equation}\label{eq:11}
    \frac{1}{D({\bf r};\tilde \omega)}=\frac{1}{D_0} \left[1+\frac{d}{4\nu}{\cal Y} ({\bf r},{\bf
r};\tilde \omega)\right].
\end{equation}
It was argued \cite{Lagendijk00,Skipetrov04,Skipetrov06}
that the SCLD model is valid for both weakly and strongly disordered systems.
As mentioned in the introductory part,
the SCLD model has guided a number of research activities on
localization
in open media. In particular, the prediction of the SCLD model
and measurements in experiments on localized elastic waves
in three dimensions agree well \cite{Hu09}.

\begin{figure}
 \begin{center}
 \includegraphics[width=8.0cm]{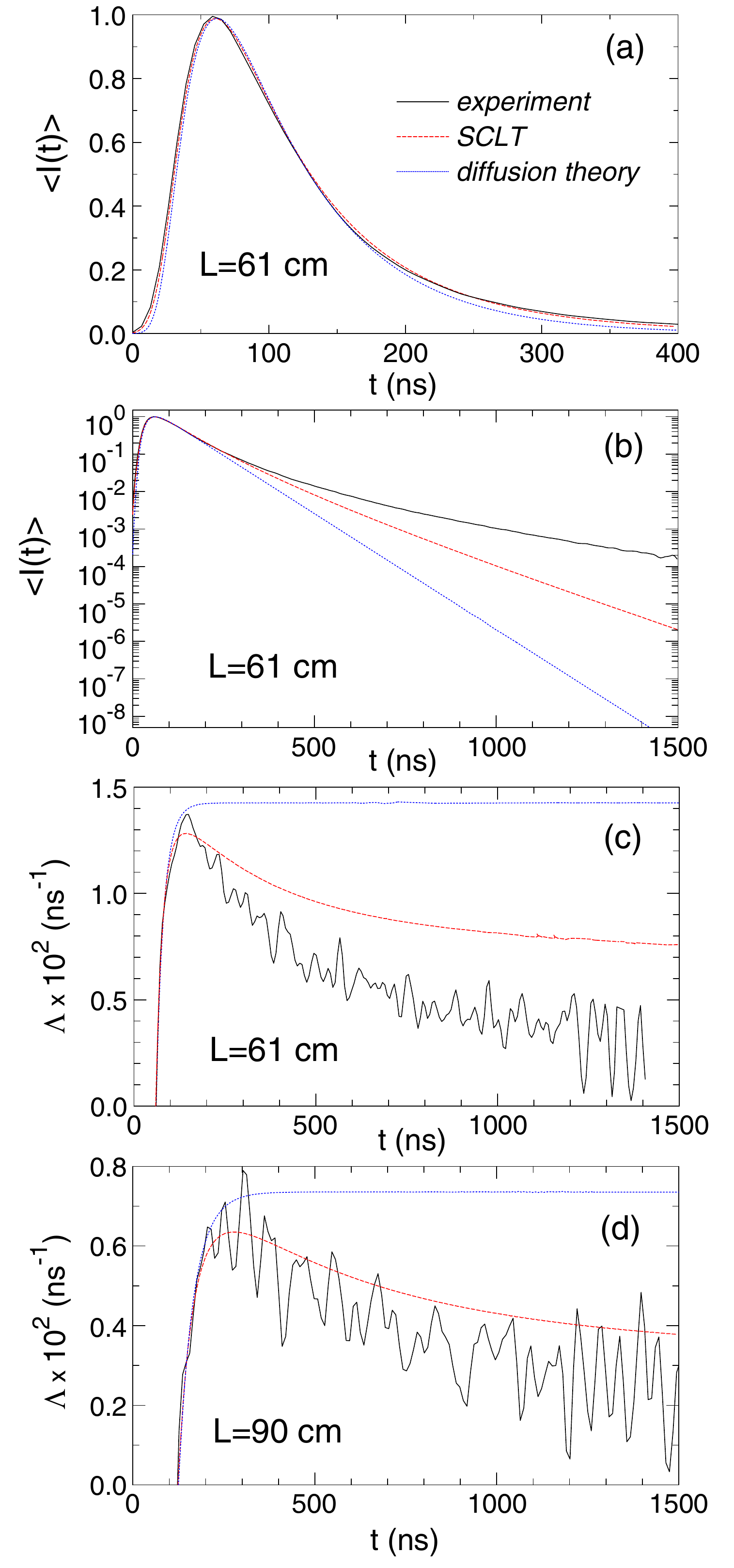}
\end{center}
 \caption{A Gaussian pulse
 is launched into Q$1$D samples
 of different sample lengths ($61$ and $90$ cm), and the averaged intensity $\langle I(t)\rangle$
and decay rate $\langle \Lambda(t)\rangle$ on the output plane are measured.
The experimental results (solid, in black)
are compared with the predictions of the SCLD (namely `SCLT' in
panel (a)) model (dashed, in red) and
of classical diffusion theory (dotted, in blue). Although the prediction of the SCLD model
and the measurements agree well at short times ((a)), dramatic deviations show up at long times ((b)-(d)).
The curves are normalized to the peak value in (a) and (b). (from Ref.~\cite{Zhang09}
with reproduction permission from Z. Q. Zhang $\copyright$ The American Physical Society)}
\label{fig:dynamics}
\end{figure}

\begin{figure}[h]
 \centering
 \includegraphics[width=8.0cm]{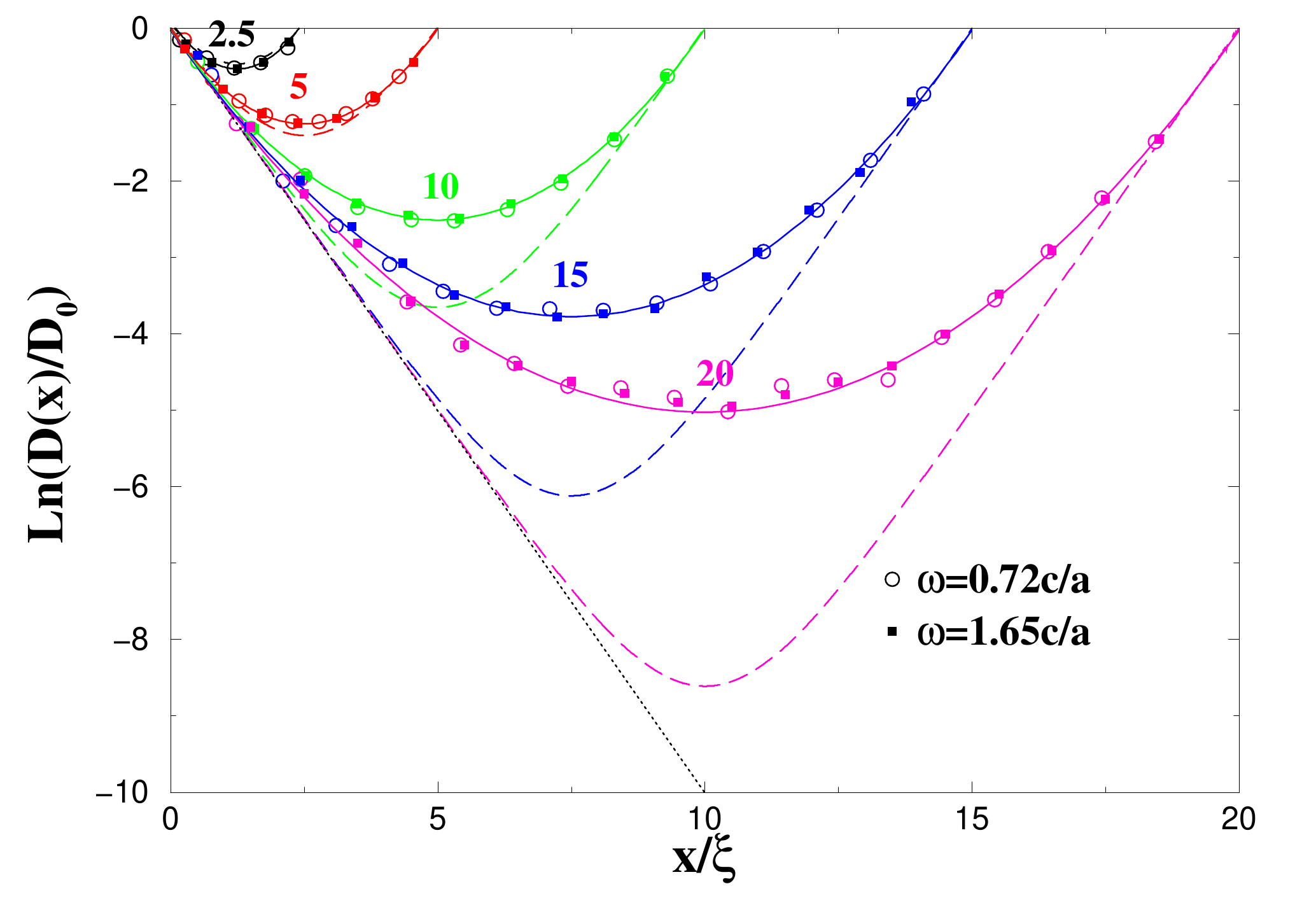}
 \caption{Results of $D(x)/D_0$ obtained from numerical simulations (squares and circles),
 the analytic prediction (\ref{DSL}) (solid lines) and the SCLD model (dashed lines)
 are compared. Numerical simulations are performed for two (angular) wave frequencies,
 $\omega=1.65 c/a$ (square) and $\omega=0.72 c/a$
 (circle), and for five different sample lengths, $L/\xi=2.5,5,10,15 $ and $20$.
(from Ref.~\cite{Tian10} with reproduction permission from C. S. Tian,
S. K. Cheung, and Z. Q. Zhang $\copyright$ The American Physical Society)}
 \label{fig:1Dlocal}
\end{figure}

In (quasi) one-dimension strong disagreement between the predictions of the SCLD model
and measurements in experiments and numerical simulations
has been seen \cite{Zhang09,Tian10}.
In the dynamic case, $\tilde \omega\neq 0$,
both experiments and numerical simulations
on the dynamics of microwave pulse propagation through Q$1$D samples
have been carried out \cite{Zhang09}, and the results of the time-resolved
transmission are compared
with the prediction of Eqs.~(\ref{eq:96}) and (\ref{eq:11}).
As shown in Fig.~\ref{fig:dynamics}, although the SCLD model can account very well for
the time-resolved
transmission at intermediate times, it fails completely at longer times.
In the steady state ($\tilde \omega=0$),
numerical simulations of the wave intensity profile across the sample have
been carried out and the local diffusion coefficient was computed
\cite{Tian10}. The numerical results are compared with the predictions of the
SCLD model, and dramatic deviations are found
(see Fig.~\ref{fig:1Dlocal}).

It has been further shown \cite{Zhang09,Tian10} that in (quasi) one-dimensional systems
transport of localized waves is dominated by rare disorder-induced resonant
transmissions which lead to far-reaching consequences (see also Sec.~\ref{sec:localdiffusion1D}).
In fact, by using the first-principles microscopic theory it has been
shown \cite{Tian10} that for $x$ deep inside the samples, i.e.,
$\xi \ll {\rm min} (x,\, L-x)$,
\begin{equation}
D(x) \sim e^{-\frac{x(L-x)}{L\xi}}.
\label{DSL}
\end{equation}
This result has been fully confirmed by numerical simulations
(see Fig.~\ref{fig:1Dlocal}).
Notice that this expression is symmetric with respect
to the sample mid-point, i.e., $D(x)=D(L-x)$.
Most importantly, near the sample center, the enhancement from the prediction
of the SCLD model is exponentially large $\sim e^{\frac{({\rm min}(x,\,L-x))^2}{L\xi}}$.
As to be shown in Sec.~\ref{sec:localdiffusion1D}, this dramatic enhancement
finds its origin at the novel scaling displayed by $D(x)$ \cite{Tian10}.
That is, $D(x)$ depends on $x$ via the scaling factor $x(L-x)/(L\xi)$.

\subsubsection{Is the concept of local diffusion universal?}
\label{sec:universal}

In some optical systems (e.g., Faraday-active medium
\cite{Golubentsev84,John88}) the one-loop weak localization
may be strongly suppressed and eventually time-reversal symmetry may be broken.
In these cases, the SCLD model is no longer applicable since it crucially relies on the
one-loop self-consistency or the time-reversal symmetry.
On the other hand, there have been rigorous studies showing that
systems with/without the time-reversal symmetry (more precisely,
corresponding to the Gaussian orthogonal/unitary ensemble (GOE/GUE)
in the random matrix theory \cite{Beenakker97})
have largely the same strong
localization behavior, and the only difference is the numerical factor of the localization length
\cite{Efetov83a,Zirnbauer91,Zirnbauer92,Zirnbauer94}.
An important question therefore arises:
Is local diffusion an intrinsic macroscopic phenomenon of (localized) open media?
This question was first studied in Ref.~\cite{Tian10}. By using the first-principles
theory to be reviewed below the authors showed that Eq.~(\ref{DSL}) is valid for both GOE and GUE systems,
and the symmetry only affects the numerical coefficient of the localization length, $\xi$.

\subsection{Electromagnetic wave propagation: mode picture}
\label{sec:transport_mode}

Alternatively, transport of waves through
open media may be understood in terms of quasi-normal modes.
Each quasi-normal mode (labeled by $n$) is characterized by the (complex) wave amplitude,
$\varphi_n ({\bf r})$, the central frequency $\omega_n$,
and linewidth $\Gamma_n>0$. The last represents the rate of
wave energy leakage through the interfaces. Upon pumping wave energies into the system,
a number of quasi-normal modes are excited, and the wave energies are stored
in these modes. In terms of the quasi-normal mode picture,
the essential differences between diffusive and localized samples are as
follows. For the former system the quasi-normal modes are extended in space, and
wave energies are readily transported through the sample.
Correspondingly, the lifetimes of quasi-normal modes are short, resulting in
a characteristic linewidth
greatly exceeding the mean spacing
between neighboring eigenfrequencies. For the latter system, there are quasi-normal modes
\cite{Lifshits79,Azbel83,Azbel83a} whose coupling to outside environments is exponentially small
with linewidths much smaller than the mean
eigenfrequency spacing. At long times, wave energies are stored mainly
in these long-lived modes which therefore play decisive roles in transport of
localized waves.

\subsubsection{Dynamic single parameter scaling model}
\label{sec:DSPS}

Dynamics of localized waves undergoes an essential change in the
modal distribution of energy with time: the
transmission is mainly due to short-lived overlapping
modes at early times and by long-lived
localized states at longer times.
The long-time dynamics of localized waves is dominated by
the spectrally isolated and long-lived
localized modes, and
is well captured by a phenomenological model, the
so-called dynamic single parameter scaling (DSPS) model \cite{Zhang09,Zhang10},
developed for (quasi) one-dimensional systems.

Consider a one-dimensional sample of length $L\gg \xi$ with transparent interfaces.
For a resonantly excited localized mode
peaked at a distance $x$ from, say, the left interface ($x=0$),
the steady-state intensity at
the right interface ($x=L$) relative to
the incident wave is \cite{Azbel83,Azbel83a},
\begin{equation}\label{eq:14}
    T_x=e^{-\gamma (L-2x)},\quad 0<x\leq L/2.
\end{equation}
Notice that $T_x$ is symmetric with respect to the sample center,
i.e., $T_x=T_{L-x}$. According
to the one-parameter scaling hypothesis \cite{Anderson80}, the Lyapunov exponent
(namely the inverse localization length) $\gamma$ follows the Gaussian distribution,
\begin{equation}\label{eq:15}
    P(\gamma)=\sqrt{\frac{L\xi}{\pi}}\exp\left[-\frac{L\xi}{4}(\gamma-\xi^{-1})^2\right].
\end{equation}
For (spectrally isolated and long-lived) localized states, the decay rate is the
ratio of the sum of the outgoing fluxes at two interfaces to the
integrated wave energy inside the sample,
\begin{equation}\label{eq:12}
    \Gamma(\gamma,x)\sim
    \gamma \frac{1+e^{-\gamma(L-2x)}}{2e^{\gamma x}-e^{-\gamma(L-2x)}-1}.
\end{equation}
Here, an overall numerical factor has been ignored since it is
irrelevant for further discussions.
Assuming that
the modes are uniformly distributed inside the sample, we find the transmitted intensity to be
\begin{equation}\label{eq:13}
    I(t) \equiv I(x=L,t) \sim L^{-1} \int_{4/L}^\infty d\gamma \int_0^{
L/2} dx\, T_x \Gamma^2 e^{-\Gamma t} P(\gamma).
\end{equation}
The theoretical prediction of the DSPS model (Eqs.~(\ref{eq:15}) and (\ref{eq:12})) for
the decay rate $-d\ln I(t)/dt$ is in good agreement with
both the experimental measurements and the numerical simulations \cite{Zhang09}.

\subsubsection{Complete modal analysis}
\label{sec:mode}

Complete modal analysis has been further performed experimentally by Wang and Genack \cite{Genack11}.
The underlying general principle is as follows. (In fact,
a similar principle has been adopted in
studies of transport through quantum chaotic systems, see Ref.~\cite{Altland10}
for a review.) For a linear medium, the wave field,
denoted as $E({\bf r},t)$,
is the superposition of quasi-normal modes,
\begin{equation}\label{eq:17}
    E({\bf r},t)=\sum_n c_n \varphi_{n}({\bf r})\,e^{-i(\omega_n -i\Gamma_n) t},
\end{equation}
where the coefficients, $c_n$'s, are fixed by the initial
condition. The wave intensity profile
is given by
\begin{equation}\label{eq:19}
    |E({\bf r},t)|^2= \sum_n |c_n \varphi_{n}({\bf r})|^2\,e^{-\Gamma_n t} + \sum_{n\neq n'}
c_n c_{n'}^*\varphi_n ({\bf r})\varphi_{n'} ({\bf r})^* e^{-i(\omega_n-\omega_{n'})t}
e^{-(\Gamma_n+\Gamma_{n'})t/2}.
\end{equation}
The first term is an incoherent contribution reflecting wave transport through
individual modes. In contrast, the second term introduces
interference between different modes. Since the system is finite, the central frequencies
are discrete with a characteristic spacing $\Delta \omega$. At long times,
$t\gg 1/\Delta \omega$,
the second term becomes negligible and Eq.~(\ref{eq:19}) is simplified to
\begin{equation}\label{eq:20}
    |E({\bf r},t)|^2 \stackrel{t\gtrsim 1/\Delta \omega}{\longrightarrow} \sum_n |c_n \varphi_{n}({\bf r})|^2\,e^{-\Gamma_n t}.
\end{equation}
This shows that transport at long times is dominated by the long-lived modes
with small $\Gamma_n$.

It is not an analytic tractable task to obtain complete knowledge
on the spectrum and the wave amplitude.
Wang and Genack realized \cite{Genack11} that
quantitative analysis may be substantially simplified
in combination with experimental measurements. Experimentally,
a pulse is incident from one interface and propagates to the other.
The field pattern at the output surface
is recorded, which depends only on the coordinates in the transverse plane.
According to Eq.~(\ref{eq:17}), it can be decomposed in terms of the
so-called volume field speckle pattern, i.e., $
c_n \varphi_{n}({\bf r})$ with the longitudinal coordinate
fixed at $L$.
Then, both the spectrum
and the corresponding mode speckle pattern
can be determined
experimentally, see Fig.~\ref{fig:openmedia}, lower panel for typical
intensity spectra which is the squared modular of mode speckle pattern.
They in principle afford a full account of dynamic and static transmission.
Experiments confirmed a broad range of decay rates
and that at long times, the transmission is indeed dominated
by incoherent contributions \cite{Genack11}.

\section{`Origin' of supersymmetry}
\label{sec:origin_SUSY}

{\it How does the supersymmetry enter into the theory of Anderson localization?}
In this section we will discuss in a heuristic manner
how the supersymmetric trick is prompted in studies of
disordered systems particularly in Anderson localization systems \cite{Efetov82,Efetov82a}.
Furthermore, we will present some crude technical hints showing how the use of this trick
eventually leads us to a supermatrix field theory.
The introduction of the supermatrix field lies at the core of
nonlinear supermatrix $\sigma$ model theory of localization, as
will become clearer in the remainder of this review.
To understand the content of this
section the readers need to have some
basic knowledge on the Grassmann algebra
and supermathematics, a preliminary introduction to which
is given in Appendix~\ref{sec:supermathematics}.

\subsection{The supersymmetric trick}
\label{sec:trick}

Consider a random medium embedded in the air background.
Microscopically, the wave field, $E$, is described by the Helmholtz
equation \cite{John84,John85,Landau}
\begin{equation}
\left\{\nabla^2 + \omega^2(1
+\epsilon({\bf
r}))\right\}E({\bf r}) = 0,
\label{Helmholtz}
\end{equation}
where $\omega$ is the (circular) frequency.
This equation also describes the propagation of elastic waves \cite{John83a,John83,Sheng90,Kirkpatrick85}, and is a
good approximation \cite{Niuwenhuizen} to electromagnetic waves provided that the
vector character (which is essential, for example, to multiple scattering
in a Faraday-active medium in the presence of a
magnetic field \cite{John89,Tiggelen96}) is unimportant.
$\epsilon({\bf r})$ is the fluctuating dielectric field (with zero mean).
Interestingly, classical scalar wave equation (\ref{Helmholtz}) bears a firm analogy to
the Schr{\"o}dinger equation:
${\hat H}\equiv -\nabla^2 -\omega^2 \epsilon ({\bf r})$ plays the role of the `Hamiltonian' and $\omega^2$
the `particle energy'. Notice that here the `potential',
$\omega^2 \epsilon ({\bf r})$, is `energy'-dependent.
(The interesting property of the energy-dependence of the potential immediately
leads to important consequences of light localization which are profoundly different
from electron localization. That is, in the zero frequency limit, the system has an extended
state where the wave field is uniform in space. Inheriting from this, in one and two
dimension the localization length diverges in the limit $\omega\rightarrow 0$ while
in higher dimension the system is extended for sufficiently low $\omega$,
see Sec.~\ref{sec:correlator_infinite} for further discussions.)

Let us recall the canonical method of calculating the Green function --
the path integral formalism \cite{Altland}. Specifically, for
wave propagation in an infinite random medium
described by Eq.~(\ref{Helmholtz}),
similar to quantum mechanics \cite{Altland,Abrikosov65},
we may introduce the
retarded (advanced) Green function $G^{R,A}_{\omega^2}$ \cite{John84},
\begin{equation}
(\omega_\pm^2- {\hat H} )G^{R,A}_{\omega^2}({\bf r},{\bf
 r}')=\delta({\bf r} -{\bf r}'),
\label{Greenfunction}
\end{equation}
where $\omega_\pm = \omega \pm i\delta $ and $\delta$ is a positive infinitesimal.
The fluctuating dielectric $\epsilon({\bf r})$
follows Gaussian distribution with
\begin{equation}
\left\langle \epsilon({\bf r})\epsilon({\bf
r}')\right\rangle = \Delta ({\bf r}-{\bf r}'),
\label{eq:175}
\end{equation}
where $\Delta ({\bf r}-{\bf r}')=\Delta (|{\bf r}-{\bf r}'|)$ is the correlation function.
Then, one may cast the Green functions into the functional integral over some vector field,
$\phi_s ({\bf r}),\,\phi_s^* ({\bf r})$, i.e.,
$$
G^{R,A}_{\omega^2}({\bf r},{\bf
 r}') = \mp i\frac{\int \phi_s({\bf r}) \phi_s^*({\bf r}')
e^{-S[\phi_s]}d[\phi_s^*]d[\phi_s]}{\int
e^{-S[\phi_s]}d[\phi_s^*]d[\phi_s]}
= \mp i\frac{\int d[\phi_s^*]d[\phi_s] \phi_s({\bf r}) \phi_s^*({\bf r}')
e^{-S[\phi_s]}}{\int d[\phi_s^*]d[\phi_s]
e^{-S[\phi_s]}},
$$
where $S[\phi_s]\equiv \mp i\int d{\bf r}
\phi_s^*(\omega_\pm^2-{\hat H})\phi_s $ is a bilinear action.
(Throughout this review we use the notation `$d[\cdot]$' or `$D[\cdot]$' to denote
the functional measure.)
Here, the vector field $\phi_s^*({\bf r}),\phi_s({\bf r})$ are either
ordinary complex (with the subscript $s={\rm B}$)
or anticommuting (with the subscript $s={\rm F}$) variables. The latter are called Grassmmanians.
It should be stressed that for Grassmannians, the complex conjugate
is purely formal: an anticommuting variable and its
complex conjugate should be understood as independent variables.
Notice that throughout this review
the independent anticommuting degrees of freedom are even
which allows us to freely move the measure under the integral:
it can be placed either before or after the integrand.

It is very important that the disorder,
i.e., $\omega^2 \epsilon({\bf r})$, enters into both the denominator
(namely the normalization factor)
and the numerator of the functional integral. This makes the subsequent
disorder average a formidable task.
Our aim therefore is to develop a first-principles theory
such that disorder is eliminated from the normalization factor.
This is accomplished by the supersymmetric trick \cite{Efetov82,Efetov82a}.
Specifically, we promote $\phi_s({\bf r})$
to a two-component supervector (or graded) field,
$\phi({\bf r})$, and $\phi^*_s({\bf r})$ to $\phi^\dagger({\bf r})$,
the Hermitian conjugate of $\phi({\bf r})$, i.e.,
(For simplicity, in this section we ignore the time-reversal symmetry of the Helmholtz equation (\ref{Helmholtz}).)
\begin{eqnarray}
\label{eq:82}
  \phi({\bf r})\equiv \left(
               \begin{array}{c}
                 \phi_{\rm F}({\bf r}) \\
                 \phi_{\rm B}({\bf r}) \\
               \end{array}
             \right),\qquad \phi^\dagger({\bf r})\equiv (\phi_{\rm F}^*({\bf r}),\phi_{\rm B}^*({\bf r})).
\end{eqnarray}
As the vector field $\phi_{\rm F}({\bf r})$ ($\phi_{\rm B}({\bf r})$) describes
fermionic (bosonic) particles, $\phi({\bf r})$ represents a particle which is a
mixture of fermion and boson: the term `super' thereby follows.

By using identity (\ref{eq:120}) in Appendix~\ref{sec:supermathematics},
we may rewrite the Green functions as
\begin{eqnarray}
    \pm i G^{R,A}_{\omega^2}({\bf r},{\bf
 r}') &=& \frac{\delta^2{Z}[J^\dagger,J]}{\delta J_{\rm F}({\bf r}')\delta J^\dagger_{\rm F}({\bf r})}
 \bigg|_{J^\dagger,J=0},\label{eq:83}\\
 {Z}[J^\dagger,J] &\equiv& \int d[\phi^\dagger]d[\phi]
 \exp\left\{- \int d{\bf r}\left(\mp i\phi^\dagger({\bf r})(\omega_\pm^2-{\hat H})\phi({\bf r})
 +J^\dagger({\bf r})\phi({\bf r}) + \phi^\dagger({\bf r})J({\bf r})\right)\right\}.
 \nonumber
\end{eqnarray}
In the absence of the external source,
$J^\dagger = J = 0$, the action is invariant under the unitary transformation in the supervector
space: this is the so-called supersymmetry or $\Bbb{Z}_2$-grading.
The most striking property of Eq.~(\ref{eq:83})
is that $Z[J^\dagger=J=0]$ is unity.
Indeed, we have here a supermatrix, $M$ (cf. Eq.~(\ref{eq:146})), as the Gaussian kernel, with
the matrix elements $M_{\rm FF}=M_{\rm BB}=\mp i(\omega_\pm^2-{\hat H})$ and
$M_{\rm FB}=M_{\rm BF}=0$. From Eq.~(\ref{eq:119}),
it follows that in the absence of sources, $J^\dagger=J=0$, the
partition function $Z={\rm sdet} M=
{\rm det}(\omega_\pm^2-{\hat H})/{\rm det}(\omega_\pm^2-{\hat H})=1$,
where the denominator (numerator) results from the integral over the commuting
(anticommuting) variables, $\phi^*_{\rm B}, \phi_{\rm B}$ ($\phi^*_{\rm F}, \phi_{\rm F}$).
Here, `sdet' stands for the superdeterminant (see Eq.~(\ref{eq:58}) in Appendix~\ref{sec:supermathematics}
for the definition) and `det' for the ordinary determinant.
The striking property of $Z[J^\dagger=J=0]=1$ is due to
equal number of anticommuting and commuting
components. It is this property that (i) leads to a compact expression
after disorder averaging (for disorders now enter only into the exponent
of the numerator), (ii) keeps the full effect of the disorder averaging,
and (iii) renders the theory free of the analytic continuation problem
which is encountered in the replica field theory \cite{Tian05,Efetov97,Zirnbauer84}.

In fact, $Z[J^\dagger=J=0]=1$ is a special case of the general theorem
discovered in Refs.~\cite{Parisi79,Efetov83,McKane80,Wegner83}.
According to this theorem, for
a function $f(x)$ satisfying $f(+\infty)<\infty$, we have
\begin{equation}\label{eq:188}
    \int d\phi^\dagger d\phi f(\sqrt{(\phi, \phi)})=f(0),
\end{equation}
with $\sqrt{(\phi, \phi)}$ being the `length' of the supervector $\phi$
(cf. Eq.~(\ref{eq:113})).
Remarkably, it states that (if the integrand is rotationally invariant in the supervector
space,) the integral is given by the value of the integrand at the
boundary of integration domain.
To apply this theorem to present studies we expand
$\phi({\bf r})$ in terms of the eigenmodes of $\hat H$,
i.e., $\phi({\bf r})=\sum_n\phi_n\varphi_n ({\bf r})$, where
$\varphi_n ({\bf r})$ satisfies $\hat H \varphi_n ({\bf r})=\omega^2_n \varphi_n ({\bf r})$.
Substituting the expansion into $Z$ we find
\begin{eqnarray}
    \label{eq:189}
 {Z}[J^\dagger=J=0] = \prod_n \int d\phi^\dagger_nd\phi_n
 e^{\pm i(\omega_\pm^2-\omega_n^2)\phi^\dagger_n\phi_n}=1,
\end{eqnarray}
where the convergence of the integrand is guaranteed by the positive infinitesimal $\delta$.


\subsection{Origin of supermatrix field theory}
\label{sec:origin}

Above we explained the necessity of introducing functional integral over the
supervector field if we would wish to get rid of disorders from the normalization
factor. But this is not the end of the story. As will become clearer, it is the very beginning
instead! Below we will adopt the heuristic approach of Bunder {\it et. al.} \cite{Zirnbauer05}
to explain -- in an intuitive manner -- why eventually we will
deal with an effective theory of supermatrix field, rather
than the above-mentioned supervector field. To this end
the source term is unimportant and we therefore set
$J^\dagger,J$ to zero.

Let us perform the disorder averaging. As a result,
\begin{equation}
Z \rightarrow \langle Z\rangle =\int d[\phi^\dagger]d[\phi]
\exp\left\{-\int d{\bf r}d{\bf r}' \left[i\phi^\dagger({\bf r}')
\delta ({\bf r}'-{\bf r}) \left(\nabla^2+\omega^2
\right)\phi({\bf r})+ \frac{\omega^4 }{2}\Delta({\bf r}-{\bf r}')\left(\phi^\dagger({\bf r})
\phi({\bf r})\right)\left(\phi^\dagger({\bf r}')
\phi({\bf r}')\right)\right]
\right\}.
 \label{eq:191}
\end{equation}
The mathematical structure of the exponent is now undergoing a dramatic change:
it is no longer quadratic in the supervector field. Rather, a quartic term appears.
The latter may be viewed as the effective interaction among
 `elementary particles' represented by $\phi$.

On the other hand, experiences have shown that
interesting physics in disordered systems arises from
multiple scattering. This suggests that we should not treat the `interaction' perturbtively.
Instead, we have to keep track of its full effects.
To this end, we recall the well-known identity:
$\int dx \delta (x-x_0)=1$. Suppose that its analog,
\begin{equation}\label{eq:190}
    \int D\tilde Q({\bf r},{\bf r}') \delta
    (\tilde Q ({\bf r},{\bf r}')-\phi({\bf r})\otimes \phi^\dagger({\bf r}'))=1
\end{equation}
(for all fixed ${\bf r}, {\bf r}'$), exists in supermathematics.
Then, the Dirac function in this `identity'
enforces $\tilde Q ({\bf r},{\bf r}')$ to have the same structure
as the dyadic product $\phi({\bf r})\otimes \phi^\dagger({\bf r}')$. Therefore,
it must be a supermatrix, i.e.,
\begin{eqnarray}
\tilde Q ({\bf r},{\bf r}') \equiv \left(
                                     \begin{array}{cc}
                                       \tilde Q_{\rm FF} ({\bf r},{\bf r}') & \tilde Q_{\rm FB} ({\bf r},{\bf r}') \\
                                       \tilde Q_{\rm BF} ({\bf r},{\bf r}') & \tilde Q_{\rm BB} ({\bf r},{\bf r}') \\
                                     \end{array}
                                   \right).
\label{eq:193}
\end{eqnarray}
Here $\tilde Q_{\rm FF},\tilde Q_{\rm BB}$
($\tilde Q_{\rm FB},\tilde Q_{\rm BF}$) are (anti)commuting variables.
(More precisely, the former (latter) belongs to the subset ${\cal A}^+$
(${\cal A}^-$) of the Grassmann algebra and has even (odd) parity,
cf. Appendix~\ref{sec:supermathematics}.)
The measure is defined as
$D\tilde Q \equiv
\pi^{-1} d\tilde Q_{\rm FB}
d\tilde Q_{\rm BF}
d\tilde Q_{\rm FF}
d\tilde Q_{\rm BB}
$, where we have ignored the arguments ${\bf r},{\bf r}'$ to make the formula compact.

Now, let us use the identity (\ref{eq:162}) to rewrite Eq.~(\ref{eq:191}) as
\begin{equation}
\langle Z \rangle=\int d[\phi^\dagger]d[\phi]
\exp\left\{\int d{\bf r}d{\bf r}' {\rm str} \left[i\delta ({\bf r}'-{\bf r}) \left(\nabla^2+\omega^2
\right) \phi({\bf r})\otimes \phi^\dagger({\bf r}')
+ \frac{\omega^4 }{2}\Delta({\bf r}-{\bf r}')
(\phi({\bf r})\otimes \phi^\dagger({\bf r}'))
(\phi({\bf r}')\otimes \phi^\dagger({\bf r}))\right]
\right\},
 \label{eq:194}
\end{equation}
where `str' is the supertrace (see Eq.~(\ref{eq:94}) for the definition).
Inserting the `identity' (\ref{eq:190}) into it, we obtain
\begin{eqnarray}
\langle Z \rangle &=& \int d[\phi^\dagger]d[\phi]\left\{\int D[\tilde Q]\delta
    (\tilde Q ({\bf r},{\bf r}')-\phi({\bf r})\otimes \phi^\dagger({\bf r}'))\right\}\nonumber\\
    && \times \exp\left\{\int d{\bf r}d{\bf r}' {\rm str} \left[i\delta ({\bf r}'-{\bf r}) \left(\nabla^2+\omega^2
\right) \phi({\bf r})\otimes \phi^\dagger({\bf r}')
+ \frac{\omega^4 }{2}\Delta({\bf r}-{\bf r}')
(\phi({\bf r})\otimes \phi^\dagger({\bf r}'))
(\phi({\bf r}')\otimes \phi^\dagger({\bf r}))\right]
\right\}.
 \label{eq:195}
\end{eqnarray}
Suppose that it is legitimate to exchange the integral order.
Integrating out the $\phi$-field first gives
\begin{eqnarray}
\langle Z \rangle = \int D[\tilde Q] J[\tilde Q]
\exp\left\{\int d{\bf r}d{\bf r}' {\rm str} \left[i\delta ({\bf r}'-{\bf r}) \left(\nabla^2+\omega^2
\right) \tilde Q({\bf r},{\bf r}')
+ \frac{\omega^4 }{2}\Delta({\bf r}-{\bf r}')
\tilde Q({\bf r},{\bf r}')\tilde Q({\bf r}',{\bf r})\right]
\right\},
 \label{eq:196}
\end{eqnarray}
where we have used the `identity': $\int D\tilde Q \delta (\tilde Q-\phi\otimes \phi^\dagger)
f(\phi\otimes \phi^\dagger)\equiv \int D\tilde Q \delta (\tilde Q-\phi\otimes \phi^\dagger)
f(\tilde Q)$ and
\begin{equation}\label{eq:192}
    J[\tilde Q ({\bf r},{\bf r}')] \equiv \int d\phi^\dagger ({\bf r}') d\phi({\bf r}) \delta
    (\tilde Q ({\bf r},{\bf r}')-\phi({\bf r})\otimes \phi^\dagger({\bf r}')).
\end{equation}
The latter is the analog of the well-known identity:
$\int dx \delta (f(x))=\sum_i |f'(x_i)|^{-1}$ with $f(x_i)=0$.

Thus, we have achieved an important result: disorder-averaged correlation
functions can be traded to a functional integral over a supermatrix field.
This is the core of the so-called superbosonization \cite{Zirnbauer05}.
Nevertheless, the above derivations are intuitive and largely formal.
In particular, we have paid no attention to the
mathematical foundation of manipulations with the Dirac function of supermatrices.
In fact, the superbosonization can be established on the level of
mathematical rigor without involving the Dirac function of Grassmannians
\cite{Zirnbauer05}. However, this requires advanced knowledge on
supermathematics and we shall not proceed further here. In the next section, we will follow
the more conventional route of using the super-Hubbard--Stratonovich (HS) transformation
to introduce the supermatrix field, which was first done by Efetov \cite{Efetov83}.

\section{Supersymmetric field theory of light localization in open media}
\label{sec:SUSY}

In the remainder of this review, we will study wave
propagation in large scales
by using the supersymmetric field theoretic approach.
Due to the interplay between wave interference and wave energy leakage
through the air-medium interface,
localization physics of classical waves in an open medium is even richer
in comparison to that in an infinite medium.
At the technical level, the corresponding field-theoretic description
dramatically differs from that for infinite media.
In this section, we will review the
supersymmetric field theory of light localization in
random open media developed in Ref.~\cite{Tian08}.
We will first
reproduce the supermatrix $\sigma$ model of Efetov in the context
of classical waves in an infinite medium;
Then, we will derive the air-medium coupling action that crucially constrains the supermatrix
field on the interface.

In reality, localization properties are probed by quantities such as
the correlation function ${\cal Y}({\bf r},{\bf r}';\tilde \omega)$,
the wave intensity distribution \cite{Mirlin00}, the transmission distribution
\cite{Dorokhov82,Mello88,Frahm95,Rejaei96,Zirnbauer04,Tian05}, etc.. The microscopic expressions of
these observables involve the product of
the retarded and advanced Green functions. Thus,
we have to double the supervector so as to account for
the distinct analytic structures of these two Green functions:
this defines the advanced/retarded (`ar') space,
with index $m=1,2$.
Because Helmholtz equation (\ref{Helmholtz})
is invariant with respect to the time-reversal operation (the complex conjugation),
we may consider the wave field and its complex conjugate as independent variables.
To accommodate this degree of freedom we need to further double the supervector,
which defines the time-reversal (`tr') space, with index
$t=1,2$. Finally, the doubling (\ref{eq:82}) defines the fermionic/bosonic (`fb') space,
with index $\alpha={\rm F},\, {\rm B}$. We have thereby introduced
an $8$-component supervector field $\psi({\bf r})=\{\psi_{m\alpha t}({\bf r})\}$.

\subsection{Nonlinear supermatrix $\sigma$ model for infinite media}
\label{sec:NSM}

Let us start from the case of infinite media.
It is necessary to reproduce
the nonlinear supermatrix $\sigma$ model for light localization
and discuss substantially the underlying technical ideas. The reasons are as follows.
(i) Although the derivations for classical scalar waves
are largely parallel to those of (spinless) de Broglie waves \cite{Efetov97}, below many
detailed treatments are different from Ref.~\cite{Efetov97}. (ii) In doing so,
we wish to show that the mathematical rigor of
the nonlinear $\sigma$ model must be understood correctly.
That is, it is not a `rigorous' mapping of the microscopic
equation (\ref{Helmholtz}); rather, it is an effective (low-energy) theory
derived from the latter under some parametric conditions.
(iii) Although the low-energy field theory turns out to be the same for
classical and de Broglie waves, the conditions justifying
such a theory are not. This leads to far-reaching
consequences. In fact, as we will see below, localization physics of
light differs from that of de Broglie waves in many important
aspects. Therefore, it is necessary to have
at our disposal a complete list of these conditions.
(iv) Reflections of the derivations may provide significant insights into
developing a first-principles localization theory incorporating vector wave character
and (or) linear gain effects.
Both are fundamental issues in studies of classical electromagnetic wave localization.
(v) Although theoretical works using
the supersymmetric field theory to study light localization have appeared,
as mentioned in the introductory part, to the best of my knowledge,
detailed derivations of this theory in the context of classical waves have so far been absent.
(vi) We hope that a technical review on
the nonlinear supermatrix $\sigma$ model of light localization
in infinite media may help the readers unfamiliar with Efetov's theory
to better appreciate how this technique
unifies mathematical rigor and physical transparency, and 
powerful it is in studying light localization.

Similar to Eq.~(\ref{eq:83}), one may express the product of the retarded and advanced Green
function in terms of certain derivative of the partition function, ${\cal Z}[J^\dagger,J]$,
with respect to the source, $J^\dagger({\bf r}),J({\bf r})$. The partition function is now a functional integral
over the supervector field, $\psi$, and its Hermitian conjugate, $\psi^\dagger$,
\begin{equation}
{\cal Z}[J^\dagger,J]\equiv z \int d[\psi^\dagger]d[\psi]
\left\langle\exp\left\{-\int d{\bf r} \left[i\psi^\dagger K \left(\nabla^2+\omega^2+\omega^2
\epsilon({\bf r})-\omega\tilde \omega^+\Lambda \right)\psi+J^\dagger K \psi
+\psi^\dagger K J\right]
\right\}\right\rangle,
 \label{Lagrange}
\end{equation}
where $\Lambda \equiv \sigma_3^{\rm ar}
\otimes {\mathbbm{1}}^{\rm fb}\otimes {\mathbbm{1}}^{\rm tr} $, with
$\sigma_i^X,\, i=0,1,2,3$ and $\sigma_0^X\equiv {\mathbbm{1}}^X$ the Pauli
matrices defined on the space $X=$`ar', `fb', and `tr',
$K$ is a metric tensor,
and to make the formula compact we have dropped out the
arguments of the fields. The numerical coefficient $z$
depends on the metric tensor $K$. Similar to the example discussed in Sec.~\ref{sec:trick},
this normalization factor is independent of disorder \cite{Zirnbauer85}, and its
explicit value is not given here since it does not affect the subsequent analysis.
We will postpone giving the explicit form of the metric tensor
till Sec.~\ref{sec:fluctuation}. At this moment we merely mention that the choice of
$K$ ensures the convergence of the above functional integral.
In deriving Eq.~(\ref{Lagrange})
we have omitted all
the $\tilde \omega^2$ terms, since we are interested in large-scale physics
where the characteristic time ($\sim \tilde \omega^{-1}$)
is much larger than the inverse (angular) frequency $\omega^{-1}$, i.e.,
\begin{equation}\label{eq:144}
    \tilde \omega \ll \omega.
\end{equation}

To proceed further, we consider a simpler case for
dielectric fluctuations, i.e., $\Delta ({\bf r}-{\bf r}')
=\Delta \, \delta({\bf r}-{\bf r}')$
with $\Delta$ being the disorder strength.
Performing the disorder averaging, we obtain
\begin{equation}
{\cal Z}[J^\dagger,J]=z\int d[\psi^\dagger]d[\psi]
\exp\left\{-\int d{\bf r} \left[i\psi^\dagger K \left(\nabla^2+\omega^2
-\omega\tilde \omega^+\Lambda \right)\psi+ \frac{\Delta \omega^4 }{2}\left(\psi^\dagger K \psi\right)^2
+J^\dagger K\psi
+\psi^\dagger K J\right]
\right\}.
 \label{Lagrangeaverage}
\end{equation}
Importantly, in the absence of the frequency and source terms, i.e.,
$\tilde \omega^+ = J^\dagger = J=0$, the action namely the exponent
is invariant under the gauge
transformation $U$: $\psi \rightarrow U\psi$, provided that $U$ preserves the metric tensor,
i.e.,
\begin{equation}\label{eq:124}
    U^\dagger K U=K.
\end{equation}
Furthermore, by the construction of the supervector field it
satisfies the `reality condition',
\begin{eqnarray}\label{eq:23}
    \psi^\dagger=(C\psi)^{\rm T},\quad C=\left(
                                    \begin{array}{cc}
                                       -i\sigma_2^{\rm tr} & 0 \\
                                      0 & \sigma_1^{\rm tr} \\
                                    \end{array}
                                  \right)^{\rm fb}\otimes {\mathbbm{1}}^{\rm ar},
\end{eqnarray}
where the superscript, `T', stands for the transpose.
Its invariance under the gauge transformation requires
\begin{equation}\label{eq:125}
    U^\dagger = CU^{\rm T}C^{\rm T}.
\end{equation}
All the supermatrices $U$ satisfying Eqs.~(\ref{eq:124}) and (\ref{eq:125})
constitute the symmetry group of the system and is denoted as $G$.

\subsubsection{Effective interactions and low-momentum transfer channels}
\label{sec:interaction}

Similar to Eq.~(\ref{eq:191}), the action in Eq.~(\ref{Lagrangeaverage})
describes the dynamics of an `elementary particle' which is a mixture of
fermion and boson.
Remarkably, the quartic term -- arising from the disorder averaging --
introduces the effective interaction between elementary particles.
Written in terms of the components of the supervector, it is
\begin{eqnarray}
  \frac{\Delta \omega^4 }{2}\,\int d{\bf r}\left(\psi^\dagger K\psi\right)^2
  &=&
  \frac{\Delta \omega^4 }{2}\,\int d{\bf r} \psi^\dagger_i K_{ij} \left(\psi^\dagger_{i'} K_{i'j'} \psi_{j'}\right) \psi_j
  \nonumber\\
  &=&
  \frac{\Delta \omega^4 }{2}\,\int d{\bf r} \psi^\dagger_i K_{ij} \left(\psi_j \otimes \psi^\dagger_{i'} K_{i'j'}\right) \psi_{j'}
  \nonumber\\
  &=&
  \frac{\Delta \omega^4 }{2}\,\int d{\bf r} \psi_i (-kK)^{\rm T}_{ij} \left(\psi^\dagger_j \otimes \psi^\dagger_{i'} K_{i'j'}\right) \psi_{j'},
\label{eq:147}
\end{eqnarray}
with $k=\sigma_3^{\rm fb}
$ which accounts for the anticommutating relation of Grassmannians.
Here, the subscripts $i,j,\cdots$ are
the abbreviations of the index $(m\alpha t)$ of the supervector components,
and the Einstein summation convention applies to the index. Recall that if
the elementary particle is coupled to an external field, $(V_{\rm ext})_{ij}$,
then the action acquires an additional term $\int d{\bf r} \psi^\dagger_i K_{ij} (V_{\rm ext})_{jk} \psi_k$
(see, for example, the $\omega\tilde \omega^+\Lambda$ term in the action).
Comparing this structure with the interaction (\ref{eq:147}), we see that the particle
self-generates some effective fields dictating the particle's motion.
Specifically, the first equality suggests an effective scalar field
$\sim \psi^\dagger_{i} K_{ij} \psi_{j}$ (because this is a number), while
the other two (which are related to each other via the reality condition (\ref{eq:23}))
of a supermatrix structure $\sim \psi \otimes \psi^\dagger K$
(because this is a dyadic product).

\begin{figure}
 \begin{center}
 \includegraphics[width=8.0cm]{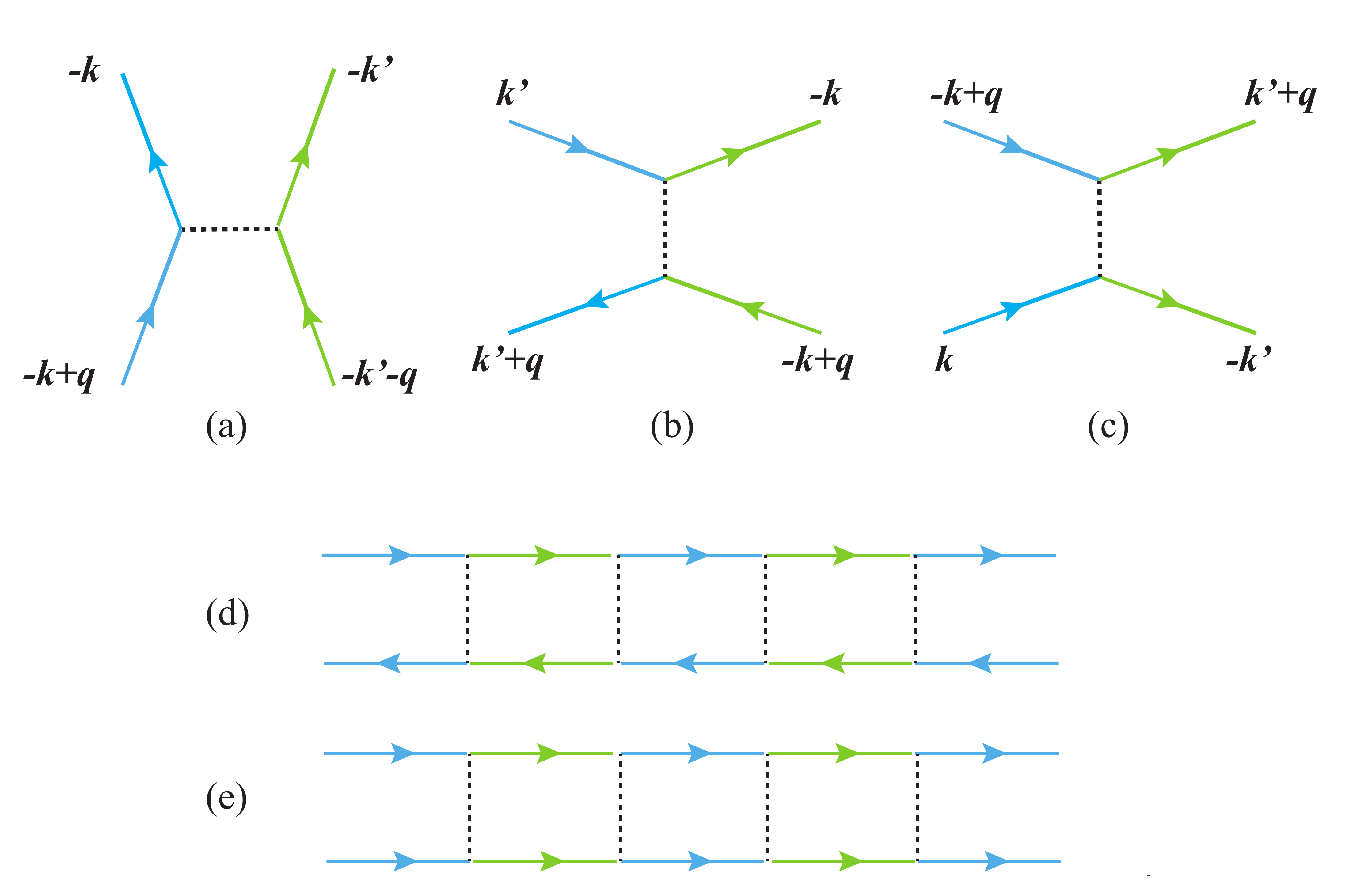}
\end{center}
 \caption{The (effective) particle-particle interaction may result in a small momentum (${\bf q}$) transfer via the direct
 (a), the diffuson or exchange (b), or the cooperon (c) channel. Successive small momentum transfer
 via the second (third) channel leads to a ladder diagram (d) ((e)), which is
 the prototype of the diffuson (cooperon) in the diagrammatic perturbation theory.
 The dashed line representing the interaction enforces two spatial coordinates involved
 to be identical, reflecting the $\delta$-correlation nature of disorders.}
\label{fig:channel}
\end{figure}

In the Fourier representation,
these self-generated fields are decomposed into slow and fast modes:
the former fluctuates over a scale much larger than the mean free path $l$,
with $q=|{\bf q}|$ (${\bf q}$ being the Fourier wave number) much smaller than $l^{-1}$;
while the latter over a scale $\lesssim l$ with $q\gtrsim l^{-1}$.
(For ballistic samples, where the mean free path and the system size are comparable,
the fast-slow mode decomposition becomes very subtle, and it turns out that a low-energy
theory is totally different from that to be derived below.
We shall not discuss this issue further, and refer
readers to the original papers
\cite{Khmelnitskii94,Andreev96,Andreev97} and the monograph \cite{Khmelnitskii97}.)
The slow mode possesses mathematical structure inheriting from the self-generated fields.
Each slow-mode structure defines a specific low-momentum (i.e., $ql\ll 1$) transfer channel.
To illustrate this we pass to the Fourier representation and rewrite Eq.~(\ref{eq:147}) as
\begin{eqnarray}
  \frac{\Delta \omega^4 }{2}\,\int d{\bf r}\left(\psi^\dagger K\psi\right)^2
  &=&
  \frac{\Delta \omega^4 }{2}\,\int \frac{d{\bf k}}{(2\pi)^d}\frac{d{\bf k}'}{(2\pi)^d}
  \frac{d{\bf q}}{(2\pi)^d}\, \psi^\dagger_{{\bf k}} K \left(\psi^\dagger_{{\bf k}'}
  K \psi_{-{\bf k}'-{\bf q}}\right) \psi_{-{\bf k}+{\bf q}}\nonumber\\
  &=&
  \frac{\Delta \omega^4 }{2}\,\int \frac{d{\bf k}}{(2\pi)^d}\frac{d{\bf k}'}{(2\pi)^d}
  \frac{d{\bf q}}{(2\pi)^d}\, \psi^\dagger_{{\bf k}} K \left(\psi_{{\bf k}'} \otimes
  \psi^\dagger_{-{\bf k}'-{\bf q}} K\right) \psi_{-{\bf k}+{\bf q}} \nonumber\\
  &=&\frac{\Delta \omega^4 }{2}\,\int \frac{d{\bf k}}{(2\pi)^d}\frac{d{\bf k}'}{(2\pi)^d}
  \frac{d{\bf q}}{(2\pi)^d}\, \psi^{\rm T}_{{\bf k}} K^{\rm T} \left(-k(\psi^\dagger_{{\bf k}'})^{\rm T}
  \otimes \psi^\dagger_{-{\bf k}'-{\bf q}} K\right) \psi_{-{\bf k}+{\bf q}}.
    \label{eq:148}
\end{eqnarray}
Here, $\psi_{\bf k}$ ($\psi^\dagger_{\bf k}$) is the Fourier transformation of $\psi$
($\psi^\dagger$), and stands for annihilating (creating)
an elementary particle of momentum ${\bf k}$ (-${\bf k}$). Notice that
the interaction conserves the total momentum.
Each equality above is respectively represented by an interaction vertex
shown in Fig.~\ref{fig:channel} (a)-(c). The entity of two green (blue) lines
defines a channel through which a momentum ${\bf q}$ is transferred.
(With the momentum ${\bf k}$ or ${\bf k}'$ integrated out,) it varies over a scale
$\sim 1/q\gg l$. Therefore, the entity specifies a slow-mode structure,
and the interaction introduces the slow-mode coupling.

Importantly, since the transferred momentum
is small enough, i.e., $ql\ll 1$, these three channels do not overlap.
Following Ref.~\cite{Altland} we call the first, (a), the direct channel.
Two particles undergoing scattering due to this
interaction acquire
a small momentum change ${\bf q}$ (or $-{\bf q}$).
The second, (b), may be called the diffuson (or exchange following the term 
of Ref.~\cite{Altland}) channel
and the third, (c), the cooperon channel. In fact, if the
momentum ${\bf q}$ transfer through the second (third) channel occurs
successively, then a diagram represented
by (d) ((e)) results, which is
the prototype of the well-known diffuson (cooperon) in the diagrammatic
perturbation theory \cite{Woelfle92,Altland}. Technically, the latter may be achieved 
by inserting Eq.~(\ref{eq:148}) with corresponding low-momentum transfer channel into 
Eq.~(\ref{Lagrangeaverage}), expanding the quartic term and summing up the infinite series.

The fast modes do not affect large-scale physics, and the interaction is dominated
by the slow modes,
\begin{eqnarray}
\frac{\Delta \omega^4 }{2}\,\int \left(\psi^\dagger K\psi\right)^2
d{\bf r}
\approx \frac{\Delta \omega^4 }{2} \int' \frac{d{\bf k}}{(2\pi)^d}\frac{d{\bf k}'}{(2\pi)^d}
  \frac{d{\bf q}}{(2\pi)^d}\,\psi^\dagger_{{\bf k}} K
  \left(\psi^\dagger_{{\bf k}'}
  K \psi_{-{\bf k}'-{\bf q}}+2\psi_{{\bf k}'} \otimes
  \psi^\dagger_{-{\bf k}'-{\bf q}} K\right)\psi_{-{\bf k}+{\bf q}},
\label{eq:43}
\end{eqnarray}
with $\int'\equiv \int_{ql\ll 1}$, where the factor two
results from the reality condition (\ref{eq:23}) namely the reciprocal
relation between the diffuson and cooperon channel. Define
\begin{equation}\label{eq:154}
    J_{{\bf q}} \equiv \int \frac{d{\bf k}}{(2\pi)^d}\psi^\dagger_{{\bf k}}
  K \psi_{-{\bf k}+{\bf q}},\qquad
  M_{{\bf q}} \equiv \int \frac{d{\bf k}}{(2\pi)^d} \psi_{{\bf k}} \otimes
  \psi^\dagger_{-{\bf k}+{\bf q}} K,
\end{equation}
we write Eq.~(\ref{eq:43}) as
\begin{eqnarray}
\frac{\Delta \omega^4 }{2}\,\int \left(\psi^\dagger K\psi\right)^2
d{\bf r}
\approx \frac{\Delta \omega^4 }{2} \int'
\frac{d{\bf q}}{(2\pi)^d}\,J_{{\bf q}}J_{-{\bf q}} -\Delta \omega^4
\int'
\frac{d{\bf q}}{(2\pi)^d}\,{\rm str} (M_{{\bf q}}M_{-{\bf q}}),
\label{eq:155}
\end{eqnarray}
where in obtaining the second term we have used the identity (\ref{eq:162}).

\subsubsection{Introduction of the supermatrix $Q$-field}
\label{sec:HS_transformation}

The two exponents, $I_1\equiv \exp\left(-\frac{\Delta \omega^4 }{2} \int'
\frac{d{\bf q}}{(2\pi)^d}\,J_{{\bf q}}J_{-{\bf q}}\right)$ and
$I_2\equiv \exp\left\{\Delta \omega^4 \int'
\frac{d{\bf q}}{(2\pi)^d}\,{\rm str}(M_{{\bf q}}M_{-{\bf q}})\right\}$,
can be decoupled via the HS decoupling which
transforms a quartic interaction into a quadratic one.
The super-HS decoupling below for the quartic interaction of the
supervector field was introduced by Efetov
(see Refs.~\cite{Efetov83,Efetov97} for a technical review and the original derivations).
Here, we wish to adopt the  treatments of Ref.~\cite{Zirnbauer84}.

By inserting the identity:
\begin{equation}\label{eq:161}
    1 \equiv {\cal N}/{\cal N},\quad {\cal N} \equiv \int D[{\cal E}_{{\bf q}}]\exp\left\{-
\frac{1}{2\Delta}\int'
  \frac{d{\bf q}}{(2\pi)^d} {\cal E}_{{\bf q}}{\cal E}_{-{\bf q}}\right\}
\end{equation}
into $I_1$, we obtain
\begin{eqnarray}\label{eq:157}
I_1 = {\cal N}^{-1}
\int D[{\cal E}_{{\bf q}}]\exp\left\{-
\frac{1}{2\Delta}\int'
  \frac{d{\bf q}}{(2\pi)^d} {\cal E}_{{\bf q}}{\cal E}_{-{\bf q}}\right\} \exp\left(-\frac{\Delta \omega^4 }{2} \int'
\frac{d{\bf q}}{(2\pi)^d}\,J_{{\bf q}}J_{-{\bf q}}\right),
\end{eqnarray}
where ${\cal E}_{-{\bf q}}={\cal E}_{\bf q}^*$.
Then, for the integral we make the change of variable:
${\cal E}_{{\bf q}} \rightarrow {\cal E}_{{\bf q}} + i\Delta \omega^2 J_{{\bf q}}$
which does not affect the convergence. As a result,
\begin{equation}\label{eq:158}
I_1 = {\cal N}^{-1}
\int D[{\cal E}_{{\bf q}}]\exp\left\{
-i\omega^2
\int' \frac{d{\bf q}}{(2\pi)^d} {\cal E}_{-{\bf q}}
J_{{\bf q}} -
\frac{1}{2\Delta}\int'
  \frac{d{\bf q}}{(2\pi)^d} {\cal E}_{{\bf q}}{\cal E}_{-{\bf q}}\right\}.
\end{equation}
Therefore, we achieve the HS decoupling in the direct channel:
on the right-hand side the action is linear in $J_{\bf q}$.
Conjugate to the `external' field structure
$\sim \psi^\dagger K\psi$ which
is a commuting variable (because Grassmannians appear in pairs),
the decoupling field ${\cal E}_{\bf q}$ is a complex scalar field.
This is the ordinary HS decoupling \cite{Altland}.

For the exponent $I_2$, by inserting the unity:
\begin{equation}\label{eq:163}
    1\equiv \int D[Q_{\bf q}]
\exp\left\{-
\Delta^{-1}\int'
\frac{d{\bf q}}{(2\pi)^d}\,{\rm str}(Q_{{\bf q}}Q_{-{\bf q}})\right\},
\end{equation}
where the supermatrix $Q_{\bf q}$ has the same symmetry structure as
the dyadic product $\psi\otimes
\psi^\dagger K$, and the convergence is assumed, we obtain
\begin{equation}\label{eq:159}
I_2 = \int D[Q_{\bf q}]\exp\left\{-
\Delta^{-1}\int'
\frac{d{\bf q}}{(2\pi)^d}\,{\rm str}(Q_{{\bf q}}Q_{-{\bf q}})+
\Delta \omega^4 \int'
\frac{d{\bf q}}{(2\pi)^d}\,{\rm str}(M_{{\bf q}}M_{-{\bf q}})\right\}.
\end{equation}
Making the variable transformation:
$Q_{{\bf q}}\rightarrow Q_{{\bf q}}-
\Delta \omega^2 M_{{\bf q}}$,
we find
\begin{equation}\label{eq:160}
    I_2 = \int D[Q_{\bf q}]\exp\left\{2 \omega^2 \int'
\frac{d{\bf q}}{(2\pi)^d}\,{\rm str}(M_{{\bf q}}Q_{-{\bf q}}) -\Delta^{-1}\int'
\frac{d{\bf q}}{(2\pi)^d}\,{\rm str}(Q_{{\bf q}}Q_{-{\bf q}})\right\}.
\end{equation}
This is the super-HS decoupling in the diffuson-cooperon channel:
on the right-hand side the action is linear in $M_{\bf q}$.
In contrast to Eq.~(\ref{eq:158}),
the super-HS decoupling in the diffuson-cooperon channel has a normalization factor of unity.
This is a consequence of equal
commuting (anticommuting) degrees of freedom which
can be easily found to be $16$ by explicitly writing down the matrix elements of
the dyadic product $\psi\otimes \psi^\dagger$.

Combining the HS decouplings (\ref{eq:158}) and (\ref{eq:160}), we have
\begin{eqnarray}
&& e^{-\frac{\Delta \omega^4 }{2}\,\int \left(\psi^\dagger K\psi\right)^2
d{\bf r}}\nonumber\\
&=&
{\cal N}^{-1}
\int D[{\cal E}_{{\bf q}}]\exp\left\{
-
i\omega^2 \int' \frac{d{\bf q}}{(2\pi)^d} \int \frac{d{\bf k}}{(2\pi)^d} {\cal E}_{-{\bf q}}
\psi^\dagger_{{\bf k}}
  K \psi_{-{\bf k}+{\bf q}} -
\frac{1}{2\Delta}\int'
  \frac{d{\bf q}}{(2\pi)^d} {\cal E}_{{\bf q}}{\cal E}_{-{\bf q}}\right\}
\nonumber\\
& \times & \int D[Q_{{\bf q}}]\exp\left\{
-2 \omega^2 \int'\frac{d{\bf q}}{(2\pi)^d} \int \frac{d{\bf k}}{(2\pi)^d}
 \psi^\dagger_{{\bf k}}K
Q_{-{\bf q}} \psi_{-{\bf k}+{\bf q}} -
\Delta^{-1} \int'
  \frac{d{\bf q}}{(2\pi)^d} {\rm str} Q_{{\bf q}}Q_{-{\bf q}}\right\}.
\label{eq:44}
\end{eqnarray}
In the derivation above we have used the identity (\ref{eq:162}).
Most importantly, by the HS decoupling we transfer the interaction -- a quartic term -- into
an action quadratic in $\psi$.
The price is the introduction of two new fields, the complex scalar field ${\cal E}_{{\bf q}}$
and the supermatrix field $Q_{{\bf q}}$.

Let us substitute Eq.~(\ref{eq:44}) into Eq.~(\ref{Lagrangeaverage}).
For the first line of Eq.~(\ref{eq:44}), the first term in the exponent of the
numerator is to locally renormalize the average refractive index, i.e.,
$\omega^2 \rightarrow \omega^2 (1+{\cal E}({\bf r}))$,
which is negligible. Therefore, the normalization factor ${\cal N}$ and
the functional integral over ${\cal E}_{{\bf q}}$
cancel. That is, the decoupling in the direct channel
gives a trivial factor of unity and plays no roles.
As a result, upon passing to
the real space representation, we obtain
\begin{eqnarray}
\exp\left\{-\frac{\Delta \omega^4}{2}\, \int
\left(\psi^\dagger K\psi\right)^2\, d{\bf r} \right\}  = \int D[Q]\exp\left\{-\int
\left(2\omega^2\psi^\dagger KQ\psi+\Delta^{-1} {\rm str}Q^2
\right)d{\bf r}\right\}.
\label{HStransformation}
\end{eqnarray}
Notice that the $Q$-field is composed of slow modes and therefore varies
over a scale much larger than the mean free path.
Thanks to Eq.~(\ref{eq:124})
the exponent of the right-hand side is invariant
under the global (in the sense that the transformation is uniform in space)
gauge transformation: $\psi \rightarrow U\psi,\,
Q\rightarrow U^{-1} Q U,\, U\in G$.

Then, we insert Eq.~(\ref{HStransformation}) into Eq.~(\ref{Lagrangeaverage}). For
the action is quadratic in $\psi$, we may integrate out this field exactly, obtaining
(That to exchange the order of integration is legitimate is guaranteed by appropriate
choice of the metric tensor $K$, see Sec.~\ref{sec:fluctuation} for further discussions.)
\begin{equation}
{\cal Z}
= \int
D[Q] e^{-F[Q]-F_J[Q]
}.
\label{average}
\end{equation}
Here, the action
\begin{eqnarray}
F[Q] =  \int d{\bf r}\, {\rm str}\left\{
\Delta^{-1} Q^2- \frac{1}{2}\ln\left[-\nabla^2-\omega^2+ \omega\tilde\omega^+
\Lambda+2i\omega^2
Q\right] \right\}.
\label{F}
\end{eqnarray}
$F_J[Q]$ is the source-dependent action whose details
rely on specific physical observable used to characterize
the system's localization behavior and do not affect localization physics.
Since we are interested in the general structure of
the low-energy field theory in this section, we
do not pay attention to its explicit form.
Importantly, although the functional measure formally includes both $\psi^\dagger$ and $\psi$,
the reality condition (\ref{eq:23}) reduces the independent degrees of freedom
by one half, which leads to the pre-logarithmic factor $\frac{1}{2}$ in Eq.~(\ref{F}).
Equations (\ref{average}) and (\ref{F}) cast physical observables into an expression in terms of
the functional integral over the supermatrix field, $Q({\bf r})$.
In the absence of the $\omega\tilde\omega^+
\Lambda$ term and the external source namely $F_J[Q]$, the action in Eq.~(\ref{average}) is invariant under
the global gauge transformation: $Q\rightarrow U^{-1}Q U$ where
$U\in G$ is homogeneous in space.

\subsubsection{Nonlinear supermatrix $\sigma$ model
}
\label{sec:fluctuation}

We then consider the functional integral over the $Q$-field.
Again we ignore the source term, $F_J[Q]$, for the same reasons as above.
The program is first to find the saddle points (denoted as $Q_0$) of
the action $F[Q]|_{\tilde \omega^+=0}$
and then to integrate out (Gaussian) fluctuations
around them.
Let us substitute $Q=Q_0+\delta Q$ into the action $F[Q]|_{\tilde \omega^+=0}$ and expand it
in terms of the deviation, $\delta Q$. Demanding the linear term to vanish gives
\begin{eqnarray}
Q_0({\bf r}) =
\frac{1}{2}\omega^2 \Delta \mathcal{G}_0 ({\bf r},{\bf r};Q_0), \quad
\mathcal{G}_0({\bf r},{\bf r}';Q)
\equiv i\langle{\bf r}|(-\nabla^2-\omega^2+
2i \omega^2 Q)^{-1}|{\bf r}'\rangle.
\label{eq:52}
\end{eqnarray}
The imaginary part of the matrix Green function $\mathcal{G}_0$
implies that it decays exponentially when
the distance $|{\bf r}-{\bf r}'|$ is greater than a characteristic length
which, as we will see below, is
the elastic mean free path.
In Appendix~\ref{sec:saddlepoint}, we show that
for low frequencies, i.e.,
\begin{equation}\label{eq:145}
    \Delta \omega \nu(\omega)\ll 1 \Leftrightarrow \omega \ll \Delta^{-1/d}
\end{equation}
with $\nu(\omega)\sim \omega^{d-1}$ being the density of states of free photons,
diagonalizing Eq.~(\ref{eq:52}) leads to two
saddle points, $\tilde q\Lambda$ and $\tilde q(-k\Lambda)$, uniform in space (the `mean field' saddle points),
where $\tilde q\equiv \frac{\pi}{4}\Delta \omega \nu(\omega)$.
It is important to notice that these are not the complete solutions to Eq.~(\ref{eq:52}).
Indeed, this equation is invariant under the rotation,
$Q_0\rightarrow T^{-1}Q_0 T,\,T \in G
$, and the two saddle points, $\tilde q \Lambda$ and
$\tilde q (-k\Lambda)$,
are connected via this rotation. Therefore, the solutions
to Eq.~(\ref{eq:52}) constitute a manifold,
\begin{equation}\label{eq:54}
    Q=\tilde q\, T^{-1}\Lambda T.
\end{equation}

As shown in Ref.~\cite{Zirnbauer84}, enforced by the convergence of the functional integral over
$\psi,\psi^\dagger,Q$ and the analytic structure of the
saddle point, the boson-boson (fermion-fermion) block in the fb-space is invariant under
the non-compact (compact) group of transformations. This fixes the metric tensor
(in its diagonal form),
\begin{eqnarray}
\label{eq:121}
  K=\left(
      \begin{array}{cc}
        {\mathbbm{1}}^{\rm fb} & 0 \\
        0 & \sigma_3^{\rm fb} \\
      \end{array}
    \right)^{{\rm ar}}\otimes {\mathbbm{1}}^{\rm tr}.
\end{eqnarray}
Combined with Eqs.~(\ref{eq:124}) and (\ref{eq:125}), this defines
the symmetry group $G=UOSP(2,2|4)$ for the orthogonal ensemble \cite{Zirnbauer85},
which is a pseudounitary supergroup.
The notation to the left (right) of `$|$' stands for the metric
in the bosonic (fermionic) sector:
`$2,2$' refers to the hyperbolic metric, $(+,+,-,-)$
and `$4$' to the Euclidean metric, $(+,+,+,+)$.
Importantly, there is a subgroup $H\subset G$ which is a direct product of two unitary groups, $UOSP(2|2)$,
defined in the advanced-advanced
and retarded-retarded block, respectively. Elements of $H$ generate
rotation rendering $\Lambda$ invariant.
In this sense, elements of $H$ may be considered to be identical. More precisely,
in Eq.~(\ref{eq:54}) $T$ takes the value from the coset space, i.e.,
\begin{equation}\label{eq:128}
    T\in G/H=UOSP(2,2|4)/UOSP(2|2)\otimes UOSP(2|2).
\end{equation}
As we will discuss in details in Sec.~\ref{sec:goldstone},
breaking of this global continuous symmetry leads to
gapless collective modes, the so-called Goldstone modes.
It is these modes,
known as diffusons and cooperons in the perturbative diagrammatic technique \cite{Woelfle80a,Woelfle80,Larkin79},
and their interactions
that carry the full information on localization physics. This will be established
in the following sections. We now
proceed to analyze the action of these modes.

Since we are interested in low-energy physics,
the coupling between the $Q$-field spatial fluctuations and
finite frequency ($\tilde \omega^+\neq 0$) effects is negligible,
(The latter characterizes dynamic effects of fluctuations.)
and they contribute separately to the effective action.
To find the contribution of the former
we set $\tilde \omega^+=0$. Recall that the first order term in the $\delta Q$-expansion
vanishes. By keeping the expansion up to
the second order we find the fluctuation action,
\begin{eqnarray}
\int\frac{d{\bf q}}{(2\pi)^d}
{\rm str}\left\{\omega^4\int\frac{d{\bf k}}{(2\pi)^d}{\cal G}_0\left({\bf k}+\frac{{\bf q}}{2},Q_0\right)\delta Q_{\bf q}
{\cal G}_0\left({\bf k}-\frac{{\bf q}}{2},Q_0\right)\delta Q_{-{\bf q}}+
\Delta^{-1}\delta Q_{\bf q}\delta Q_{-{\bf q}}\right\},
\label{eq:61}
\end{eqnarray}
where we have passed to the Fourier representation: $\delta Q({\bf r})
\rightarrow \delta Q_{{\bf q}}$ and ${\cal G}_0\left({\bf r},{\bf r}',Q_0\right)
\rightarrow {\cal G}_0\left({\bf k},Q_0\right)$. (Notice that according to Eq.~(\ref{eq:52}),
${\cal G}_0\left({\bf r},{\bf r}',Q_0\right)$
is translationally invariant, i.e., ${\cal G}_0\left({\bf r},{\bf r}',Q_0\right)
= {\cal G}_0\left({\bf r}-{\bf r}',Q_0\right)$.)
Suppose first that $Q_0=\Lambda$. We see that fluctuations around it can be decomposed
into two components: one commutes with it while the other anticommutes.
Since the (anti)commutation relation is invariant under the global rotation, we may rotate
$\Lambda$ to arbitrary $Q_0$ in the saddle point manifolds.
At the same time, we obtain two fluctuation components,
the longitudinal ($\delta Q_{\bf q}^l$) and transverse ($\delta Q_{\bf q}^t$) components:
the former (latter) commutes (anticommutes) with $Q_0$.
Physically, fluctuations along the transverse directions
move $Q_0$ to somewhere the saddle point manifold
(cf. Eqs.~(\ref{eq:54}) and (\ref{eq:128})), while fluctuations along
the longitudinal directions bring $Q_0$ out of the saddle point manifold
(see Appendix~\ref{sec:fluctuationaction} for further explanations).

Let us substitute the decomposition into Eq.~(\ref{eq:61}).
Integrating out the longitudinal components, $\delta Q_{{\bf q}}^l$,
gives a factor of unity.
We are left with a functional integral over the transverse component,
$\delta Q_{{\bf q}}^t$. Upon passing to real space, this gives
the fluctuation action (see Appendix~\ref{sec:fluctuationaction} for details),
\begin{equation}\label{eq:67}
    \delta F_1\equiv \frac{\pi\nu D_0}{8}\int d{\bf r}{\rm str}\{\nabla (T({\bf r})^{-1}\Lambda T({\bf r}))\}^2.
\end{equation}
Here, the transport mean free path determining the Boltzmann diffusion
constant $D_0=l(\omega)/d$ is of Rayleigh-type \cite{John84,Edrei90}, i.e.,
\begin{equation}
l(\omega)=\frac{2}{\pi\Delta \omega^2 \nu(\omega)} \sim \frac{1}{\Delta\omega^{d+1}}.
 \label{mfp}
\end{equation}
Therefore, we find that the inequality (\ref{eq:145}) is equivalent to the weak disorder limit, i.e.,
\begin{equation}\label{eq:182}
    \omega l(\omega) \gg 1.
\end{equation}
It implies that the transport mean free path $l(\omega)$ is much
larger than the wavelength $\lbar$.
Since we are interested in the large-scale physics, $\tilde \omega \lesssim D_0/l^2 \sim l^{-1}$,
the inequality (\ref{eq:144}) is guaranteed by (\ref{eq:145}).
It is important to notice that here the weak
disorder condition is established for low instead of high frequencies ($\omega$),
in sharp contrast to electronic systems \cite{Efetov97}.

Next, we consider the contribution due to nonvanishing $\tilde \omega$. Keeping the
$\tilde \omega$-expansion of the action (\ref{F}) up to the
first order, we find
\begin{eqnarray}
\delta F_2&\equiv& -\frac{\omega\tilde\omega^+}{2}
\int d{\bf r} {\rm str} \left\{\Lambda \langle {\bf r}|\left(
-\nabla^2-\omega^2+2i \omega^2 T^{-1}(\tilde q\Lambda) T\right)^{-1}|{\bf r}\rangle\right\}\nonumber\\
&=& \frac{i\omega\tilde\omega^+}{2}
\int d{\bf r} {\rm str} \left\{\Lambda {\cal G}_0 \left({\bf r},{\bf r};T^{-1}(\tilde q\Lambda) T\right)\right\},
\label{eq:71}
\end{eqnarray}
where $T$ is space-dependent.
Ignoring spatial fluctuations of $T$ and
using the saddle point equation (\ref{eq:52}), we simplify it to
\begin{eqnarray}
\delta F_2=i (\omega \Delta)^{-1}\tilde \omega^+
\int d{\bf r} {\rm str} \{\Lambda T({\bf r})^{-1}(\tilde q\Lambda) T({\bf r})\}.
\label{eq:72}
\end{eqnarray}
Notice that $F[Q_0]|_{\tilde \omega^+=0}=0$. Finally, we reduce Eq.~(\ref{average}) to
\begin{equation}
{\cal Z}
= \int
D[Q] e^{-F[Q]-F_J[Q]
}, \qquad
    Q({\bf r})\equiv T({\bf r})^{-1}\Lambda T({\bf r}),\qquad T({\bf r}) \in G/H,
\label{eq:183}
\end{equation}
where the contributions to the action $F[Q]$ (To avoid using too many
symbols we use the same notation as Eq.~(\ref{F}).) are $\delta F_{1,2}$, i.e.,
\begin{equation}
F[Q]\equiv\delta F_1+\delta F_2 =
\frac{\pi\nu(\omega)}{8} \, \int d{\bf r} {\rm str}\, \{D_0(\nabla Q)^2
+2i\tilde\omega^+\Lambda Q\}.
\label{action}
\end{equation}
This is the nonlinear supermatrix $\sigma$ model action.

Now let us briefly discuss the case
where the time-reversal symmetry is broken (namely the unitary ensemble).
In this case, the action (\ref{action}) stays the same. The difference is the
symmetry of the supermatrix $Q$. Indeed, the doubling introduced by time-reversal symmetry
(in tr-space) is absent and the $T$-field is thereby a $4\times 4$ supermatrix defined
in the ar- and fb-spaces.
Consequently, the reality condition (\ref{eq:125}) does not play any role.
(Correspondingly, the pre-logarithmic factor in the action (\ref{F}) disappears.)
The system's symmetry is reduced, and Eq.~(\ref{eq:124}) leads to the coset space
$G/H=U(1,1|2)/U(1|1)\otimes U(1|1)$.

\subsubsection{Ultraviolet divergence and long-ranged disorders
}
\label{sec:geomteric_optics}

Different from elastic waves, photons have arbitrarily small wavelength.
Correspondingly, in high dimensions ($d>2$) the right-hand side of the saddle point equation (\ref{eq:52})
suffers ultraviolet divergence (cf. Eq.~(\ref{eq:56}) and discussions on it in
Appendix~\ref{sec:saddlepoint}). The divergence origins at
that for large $\omega$
the $\delta$-correlated disorder
is no longer a good approximation.
Therefore, one may consider more realistic dielectric fluctuations namely
long-ranged correlation of disorders described by Eq.~(\ref{eq:175}) \cite{John83}.
Then, a fundamental issue of extreme importance arises:
does this kind of disorders affect the universality of Anderson transition of light?
This was first investigated by John and Stephen \cite{John83} by
using the replica field theory. The key is to understand whether
the low-energy field theory of localization is robust against finite-ranged correlation of disorders.
Below, we will derive the low-energy supersymmetric field theory
for this kind of disorders. (The corresponding replica field theory
was derived in Ref.~\cite{John83}.) The program is
very similar to that of deriving Eq.~(\ref{action}). Therefore, we
will address the key differences and only present the final results.

First of all, all the differences stem from the fact that upon performing
the disorder averaging, the effective interaction is non-local
which is $\exp\left\{-\frac{\omega^4}{2}\int
\Delta ({\bf r}-{\bf r}') \left(\psi^\dagger({\bf r}) K\psi({\bf r})\right)
\left(\psi^\dagger({\bf r}') K\psi({\bf r}')\right) d{\bf r}d{\bf r}' \right\}$
(to be compared with the quartic term in Eq.~(\ref{Lagrangeaverage})).
As such, the supermatrix field in the super-HS decoupling
must be non-local also, since it has the same structure as the dyadic
product, i.e., $Q({\bf r},{\bf r}') \sim
\psi({\bf r}) \otimes \psi^\dagger({\bf r}')K$. Consequently, the super-HS decoupling is
modified to
\begin{eqnarray}
&&\exp\left\{-\frac{\omega^4}{2}\int
\Delta ({\bf r}-{\bf r}') \left(\psi^\dagger({\bf r}) K\psi({\bf r})\right)
\left(\psi^\dagger({\bf r}') K\psi({\bf r}')\right) d{\bf r}d{\bf r}' \right\}  \nonumber\\
&=& \int D[
Q]\exp\left\{-\int
\left(2\omega^2\psi^\dagger({\bf r}) K
Q({\bf r},{\bf r}')\psi({\bf r}')
+\Delta^{-1}({\bf r}-{\bf r}'){\rm str}
Q({\bf r},{\bf r}')
Q({\bf r},{\bf r}')
\right)d{\bf r}d{\bf r}'\right\}.
\label{eq:176}
\end{eqnarray}

Similar to the derivation of Eq.~(\ref{F}), with the
$\psi$ field integrated out, we obtain a functional integral over the supermatrix field $
Q({\bf r},{\bf r}')$,
and its the saddle point equation is
\begin{eqnarray}
Q_0({\bf r},{\bf r}') = \frac{1}{2}\omega^2 \Delta({\bf r}-{\bf r}')
{\cal G}_0({\bf r},{\bf r}'; 
Q_0),
\label{eq:177}
\end{eqnarray}
where ${\cal G}_0$ has the same definition as that introduced in Eq.~(\ref{eq:52}).
Here, $Q_0({\bf r},{\bf r}')$ must be understood
as the matrix elements of an operator non-diagonal in real space
and correspondingly, Eq.~(\ref{eq:177}) as an operator equation.
To proceed further we introduce the Wigner transformation defined as
${\cal Q}({\bf R}, {\bf k})\equiv \int d({\bf r}-{\bf r}')
e^{-i{\bf k}\cdot ({\bf r}-{\bf r}')}
Q({\bf r},{\bf r}')$, with the center-of-mass coordinate ${\bf R}=({\bf r}+{\bf r}')/2$. Loosely speaking,
${\bf R}$ and ${\bf k}$ respectively describe the position and momentum of photons.
Wigner transforming Eq.~(\ref{eq:177}), we reduce it to
\begin{equation}\label{eq:178}
    \Delta^{-1}(i\nabla_{\bf k}) {\cal Q}_0({\bf k}) = \frac{1}{2}\omega^2
    {\cal G}_0({\bf k}; {\cal Q}_0({\bf k})),
\end{equation}
where ${\cal G}_0({\bf k}; {\cal Q}_0({\bf k}))$ is the
Wigner transformation of $
{\cal G}_0
$. This leads to a diagonal mean-field saddle point,
$\tilde q(k) \Lambda$, and
modifies the saddle point manifold to
\begin{equation}\label{eq:184}
    {\cal Q}({\bf k})=\tilde q(k) T^{-1}\Lambda T,\quad T\in G/H.
\end{equation}
Here,
$T$ depends neither on ${\bf k}$ nor on the center-of-mass coordinate ${\bf R}$,
and $\tilde q(k)$ is a real positive function of $k$.

As before the fluctuation action results from the transverse fluctuations,
$\delta Q^t({\bf R},{\bf k})
\equiv [{\cal Q}_0({\bf k}), \delta T({\bf R},{\bf k})]
$, around the saddle point ${\cal Q}_0({\bf k})$,
where $\delta T({\bf R},{\bf k})$ parametrizes fluctuations
in the coset space $G/H$.
The fluctuation action is
\begin{equation}\label{eq:179}
    \int\frac{d{\bf q}}{(2\pi)^d} \int\frac{d{\bf k}}{(2\pi)^d}
    {\rm str} \left\{\delta Q^t_{\bf q}({\bf k}
    ){\hat {\cal H}}^t_{{\bf q}}\delta Q^t_{-{\bf q}}({\bf k}
    )\right\},
\end{equation}
with $\delta Q^t_{\bf q}({\bf k})$
being the Fourier transformation of
$\delta Q^t({\bf R},{\bf k})$ with respect to ${\bf R}$.
Here ${\hat {\cal H}}^t_{{\bf q}}$ is a `Hamiltonian' depending on the `external
parameter' ${\bf q}$. In the ${\bf k}$-representation, it reads
\begin{equation}\label{eq:180}
    {\hat {\cal H}}^t_{{\bf q}}({\bf k}) \equiv \Delta^{-1}(i\nabla_{\bf k})+
    \omega^4 {\cal G}_0\left({\bf k}+\frac{{\bf q}}{2};{\cal Q}_0({\bf k})\right)
    {\cal G}_0\left({\bf k}-\frac{{\bf q}}{2};-{\cal Q}_0({\bf k})\right),
\end{equation}
where the first (second) term may be viewed as generalized kinetic energy (potential).

Different from the case of $\delta$-correlated disorders
(cf. Appendix~\ref{sec:fluctuationaction}), the integrand in Eq.~(\ref{eq:179}) does not vanish
at ${\bf q}\rightarrow 0$ in general. Instead,
$\delta Q^t({\bf R},{\bf k})$ are composed of low- and high-lying modes.
To single out the former
we have to expand $\delta Q^t_{\bf q}({\bf k})$ in terms of
the eigenfunctions of the Hamiltonian ${\hat {\cal H}}^t_{{\bf q}}({\bf k})$
denoted as $v_l ({\bf k}; {\bf q})$ which satisfies
\begin{equation}\label{eq:199}
    \hat {\cal H}^t_{{\bf q}}({\bf k}) v_l ({\bf k};{\bf q})
    =\lambda_l({\bf q})v_l({\bf k};{\bf q})
\end{equation}
with $\lambda_l({\bf q})$ being the eigenvalues.
Here the subscript $l=0$ stands for the `ground state' and $l=1,2,\cdots$ for the
`excited states'. Following John and Stephen \cite{John83}
one finds that for general expression of $\Delta(x)$,
at ${\bf q}=0$ Eq.~(\ref{eq:178}) leads to a unique ground state
$v_0 ({\bf k}; 0) \propto \tilde q(k)$ with zero angular momentum.
The ground state eigenvalue vanishes as
$\lambda_0({\bf q}) =c_2 q^2$ in the limit: ${\bf q}\rightarrow 0$,
and all the excitation modes $\lambda_{l\geq 1} ({\bf q})$ are gapped.
Here, the coefficient $c_2$ may be found by applying
the Raileigh-Schr{\"o}dinger perturbation theory to Eq.~(\ref{eq:199}).
Therefore, the low-lying transverse components must have the general form,
$\delta Q^t({\bf R},{\bf k})=[{\cal Q}_0({\bf k}), \delta T({\bf R})]$.
Substituting it into Eq.~(\ref{eq:179}) and performing the hydrodynamic expansion,
we obtain
\begin{equation}\label{eq:187}
    \delta F_1 = \frac{\pi\nu(\omega) D_0(\omega)}{8}\int d{\bf R}
    {\rm str}\{\nabla (T({\bf R})^{-1}\Lambda T({\bf R}))\}^2,
\end{equation}
where the density of states is given by
\begin{equation}\label{eq:186}
    \nu(\omega)=-\frac{2\omega}{\pi}{\rm Im}
    \int\frac{d{\bf k}}{(2\pi)^d} \frac{1}{k^2-\omega^2 + 2i\omega^2 \tilde q(k)}
\end{equation}
and the (bare) conductivity $\nu(\omega) D_0(\omega) \propto c_2$.
Equations (\ref{eq:178}) and (\ref{eq:186}) lead to
the self-consistent Born approximation (or the `coherent
potential approximation' called by John and co-workers \cite{John83,John83a})
of the density of states.

The procedure of deriving the frequency contribution, $\delta F_2$, is similar to that of deriving
the second term of Eq.~(\ref{action}). Now, Eq.~(\ref{eq:71}) becomes
\begin{eqnarray}
\delta F_2 &=& \frac{i\omega\tilde\omega^+}{2}
\int d{\bf R}\int\frac{d{\bf k}}{(2\pi)^d} {\rm str}
\left\{\Lambda {\cal G}_0 \left({\bf k};T^{-1}({\bf R})
(\tilde q({\bf k})\Lambda) T({\bf R})\right)\right\} \nonumber\\
&=&\frac{i\omega\tilde\omega^+}{2}
\int d{\bf R}\int\frac{d{\bf k}}{(2\pi)^d} {\rm str}
\left\{\Lambda T^{-1}({\bf R}) {\cal G}_0 \left({\bf k};
\tilde q({\bf k}) \Lambda \right)T({\bf R})\right\}\nonumber\\
&=& \frac{i\pi\nu(\omega)\tilde\omega^+}{4}
\int d{\bf R} {\rm str}
\left\{\Lambda T^{-1}({\bf R}) \Lambda T({\bf R})\right\},
\label{eq:185}
\end{eqnarray}
where in deriving the last line we have used Eq.~(\ref{eq:186}).
From Eqs.~(\ref{eq:187}) and (\ref{eq:185}) we conclude that the nonlinear
supermatrix $\sigma$ model action
(\ref{action}) is valid even for finite-ranged disorders, provided that
the Hamiltonian ${\hat {\cal H}}^t_{{\bf q}}$ has a unique isolated ground state.
Notice that the explicit expression of $D_0(\omega)$ is generally changed.

\subsubsection{Possible generalizations}
\label{sec:remarks}

Several exact solutions
for strong localization in weakly disordered wires
have been obtained by using the nonlinear $\sigma$ model
(\ref{action}), notably, the
correlation function defined in Eq.~(\ref{DCdefinition}) in Q$1$D infinite wires \cite{Efetov83a},
the first two moment of conductance
\cite{Zirnbauer92,Zirnbauer94} and transmission distribution
\cite{Frahm95,Rejaei96,Zirnbauer94} in finite wires coupled to ideal leads.
These results have revealed a deep connection
between the noncompactness inherited from the hyperbolic metric in
the bosonic sector and strong localization. A further example is the anomalously
localized states in diffusive open samples \cite{Muzykantskii95}.
As mentioned above, the hyperbolic nature of the metric stems
from (i) the convergence of functional integral and (ii) the analytic structure of saddle points.
To build a localization theory incorporating
linear gain effects \cite{Lai12}, one must inevitably treat both (i) and (ii),
since now the sign of the convergence generating factor, $\delta$, is
reversed, i.e., $\tilde \omega^+\rightarrow
\tilde \omega^-$. As such, discussions parallel to those of
Ref.~\cite{Zirnbauer84} must be made cautiously, and modified theory
may help to study many interesting phenomena such as the gain-absorption
duality \cite{Zhang95,Beenakker96}.

It seems possible to generalize the present formalism
to work out a first-principles theory
for electromagnetic wave (vector wave) localization, a long-standing issue in
studies of light localization.
Reflecting the scheme outlined in this section, one may naturally
enlarge the supervector field $\psi$ to
accommodate the vector field structure. However, this step -- right at the beginning
of the entire field theory formulation -- is by no means trivial. In fact,
the additional field components are not trivially independent, and
a crucial step would be to embed the constraint
reflecting the transverse field character of electromagnetic waves
into the functional integral formalism. Then, one may
derive a low-energy field theory by
following the above scheme.

\subsection{Field theory for open media}
\label{coupling}

We have so far focused on infinite (bulk) media,
where the low-energy theory (\ref{action}) is translationally invariant.
The presence of the air-medium interfaces breaks this symmetry. In this part, we
will show that the $Q$-field is constrained on the interface.
The low-energy field theory
thereby obtained describes
many exotic localization phenomena, which will be exemplified in Sec.~\ref{sec:localdiffusion} and
\ref{sec:localdiffusion1D}. In fact, open meseoscopic electron systems
(e.g., quantum disordered wires
or small quantum dots coupled to ideal leads,
superconducting-normetal hybridized structures, etc.)
have been studied extensively (see Refs.~\cite{Efetov97,Beenakker97,Altland98} for reviews).
The effective field theory describing these systems
has been worked out in Refs.~\cite{Efetov97,Iida90,Zirnbauer94,Zirnbauer95}.
All these works study physical observables
such as the conductance or transmission. Whether localized open systems
may exhibit macroscopic diffusion (a kind of hydrodynamic behavior)
was not studied in these works.
In addition, the systems considered there are in quasi one (e.g., quantum wires)
or zero (e.g., quantum dots) dimensions.

In classical wave systems, the experimental setup is
generally more complicated. For example, it may be in higher
dimensions \cite{Maret06,Maret06b,Hu09}. In addition, internal reflection may exist
on the air-medium interface due to the refractive
index mismatch \cite{Genack12}.
The subjects to be investigated are more diverse. Besides of
the physical observables mentioned above,
particular attention has been paid to many other issues, e.g.,
time-resolved transmissions, coherent backscattering lineshape,
and spatial resolution of localized modes.
In Ref.~\cite{Tian08}, the method developed for mesoscopic electronic devices
\cite{Efetov97,Iida90,Zirnbauer94,Zirnbauer95}
was first generalized to open classical wave systems
to address a number of issues of classical wave propagation in random
open media.

\subsubsection{The interface action}
\label{couplingaction}

The propagation of classical waves in the random medium
(the space supporting which is denoted as ${\cal V}_+$)
is described by the effective Green functions,
${\cal G}^{R,A}_{\omega^2}({\bf r}\,, {\bf r}'),\, {\bf r}, {\bf r}' \in {\cal V}_+$.
These two Green functions differ from $G^{R,A}_{\omega^2}({\bf r}\,, {\bf r}')$
only in the domain of ${\bf r}, {\bf r}'$. According to the definition
(\ref{Greenfunction}), for $G^{R,A}_{\omega^2}({\bf r}\,, {\bf r}')$
the variables ${\bf r}, {\bf r}'$ can
be either in the air (The corresponding space
is denoted as ${\cal V}_-$.) or the random medium ${\cal V_+}$. We also
introduce an auxiliary Green
functions, $g^{R,A}_{\omega^2}({\bf r}\,, {\bf r}')$, defined in the air satisfying
\begin{eqnarray}
(\omega_\pm^2- {\hat H}) g^{R,A}_{\omega^2}({\bf r},{\bf
 r}')=\delta({\bf r} -{\bf r}') \,,
\quad
g^{R,A}_{\omega^2}({\bf r},{\bf
 r}')|_{{\bf r} \, {\rm or}\, {\bf
 r}'\in C}=0,\quad {\bf r},
{\bf r}'\in {\cal V}_-,
\label{auxiliaryGreenfunction}
\end{eqnarray}
where $C$ is the air-medium interface.
To proceed further, we need to introduce a theorem due to Zirnbauer
\cite{Zirnbauer95} and refined by Efetov \cite{Efetov97},
originally established for describing the coupling between leads and
mesoscopic electronic devices.
In the present context of classical wave systems, the theorem may be stated as follows: \\

\noindent {\it (Zirnbauer-Efetov) The effective Green function
${\cal G}^{R,A}_{\omega^2}({\bf r}\,, {\bf r}')$ (${\bf r}\,, {\bf r}'\in {\cal V}_+$)
solves
\begin{eqnarray}
\left(
 \omega_\pm^2- {\hat H} \pm i {\hat B} \delta_C \right)\, {\cal G}^{R,A}_{\omega^2}({\bf r},{\bf
 r}') = \delta ({\bf r}-{\bf r}')\,, \quad
\nabla_{{\bf n}({\bf r})}
{\cal G}^{R,A}_{\omega^2}({\bf r},{\bf r}')\big|_{{\bf r}\in C} &=&
0 \,, \quad {\bf r}\in C\,. \label{theorem}
\end{eqnarray}}\\

\noindent Here, ${\bf n}({\bf r})$ is the normal unit vector on
the interface, pointing to ${\cal V}_+$, and the operator ${\hat B}$ is defined as
\begin{eqnarray}
({\hat B} f) ({\bf r}) \equiv \int_C d{\bf r}'\, {\rm Im}\, [B({\bf r},{\bf r}')] f({\bf r}')\,, \quad
B({\bf r},{\bf r}') = \nabla_{{\bf n}({\bf r})}\cdot\nabla_{{\bf n}({\bf r}')} g^R_{\omega^2}({\bf r},{\bf
r}') \,,
\label{effGF6}
\end{eqnarray}
with $f$ a test function on $\Bbb{R}^d$. Finally,
the operator $\delta_C$ is defined as $\int
d{\bf r}\delta_C \, f({\bf r})\equiv \int_{{\bf r}_\perp \in C} d{\bf r}_\perp
f({\bf r}_\perp)$\,.

The theorem states that wave propagation
in random media follows a modified microscopic equation.
It differs from Eq.~(\ref{Greenfunction}) in that the effective `Hamiltonian'
acquires a correction, the
$i\hat B\delta_C$ term which is purely imaginary and located on the interface.
The correction accounts for the wave energy leakage through the interface and therefore
for the dissipative nature of quasi-normal modes.
Correspondingly, by repeating the procedures of Sec.~\ref{sec:HS_transformation},
(We refer readers to Appendix \ref{sec:interfaceaction}
for details.) we find that the action (\ref{action}) acquires a correction,
the interface action $F_{\rm int}[Q]$ given by
\begin{equation}\label{eq:131}
    F_{\rm int}[Q]=-\frac{1}{4}\int d{\bf r}\delta_C\int_{
|{\bf k}_\perp|\leq \omega}\frac{d^{d-1}{\bf k}_\perp}{(2\pi)^{d-1}}\,
    {\rm str}\ln \left(1+\frac{1-R_{{\bf k}_\perp}({\bf r})}{1+R_{{\bf k}_\perp}({\bf r})}
    \Lambda Q({\bf r})\right)
\end{equation}
where the internal reflection coefficient $R_{{\bf k}_\perp}({\bf r})$
generally depends on the transverse momentum, ${\bf k}_\perp$. For simplicity,
we ignore the ${\bf k}_\perp$-dependence of the internal reflection coefficient
from now on, i.e., $R_{{\bf k}_\perp} \equiv R_0$.
In other words, we shall not study the effect arising from that
internal reflection depends on the angle which the ${\bf k}$-vector makes with the interface.
Because both $R_0({\bf r})$ and $Q({\bf r})$ vary over a scale larger than
$l$, in the limit $\omega l\gg 1$ one may further expand the logarithm and
simplify the interface action to \cite{Tian08}
\begin{eqnarray}
F_{\rm int}[Q]  = -\frac{N_{d-1}}{4} \int d{\bf r}\delta_C\,
\frac{1-R_0({\bf r})}{1+R_0({\bf r})}\, {\rm str} [\Lambda Q({\bf r})],
\quad
N_{d-1} =\frac{\omega^{d-1}}{(2\sqrt{\pi})^{d-1} \frac{d-1}{2}\Gamma
\left(\frac{d-1}{2}\right)},
\label{interFresult2}
\end{eqnarray}
where $(1-R_0({\bf r}))/(1+R_0({\bf r}))$ characterizes the strength of the air-medium `coupling'.
(This interface action (\ref{interFresult2}) differs slightly from
that derived in Ref.~\cite{Tian08} because internal reflection is small there.)

\subsubsection{Boundary condition satisfied by the $Q$-field}
\label{boundarycondition}

To proceed further we use the so-called boundary Ward identity
\cite{Altland98}. It states that
the functional integral representation of the local observable,
$\int D[Q] (\cdots) e^{-F[Q]-F_{\rm
int}[Q]}$,
with $(\cdots)$ depending on specific observables
whose details are unimportant for further discussions,
must be invariant under the infinitesimal boundary rotation:
$Q  \rightarrow  e^{-\delta R}Qe^{\delta R}\approx Q-[\delta R,Q]$.
Therefore, it is required that the resulting variation of the interface action
identically vanishes,
\begin{eqnarray}
\delta F_{\rm int} = \int d{\bf r}\delta_C {\rm str}\left\{\delta R\left(\frac{\pi \nu
D_0}{2} Q\nabla_{{\bf n}({\bf r})} Q + \frac{N_{d-1}}{4} \frac{1-R_0}{1+R_0}
[Q,\Lambda]\right)\right\} \equiv 0.
\label{actionchange}
\end{eqnarray}
For $\delta R$ is arbitrary, it gives \cite{note_Ward}
\begin{equation}
\left(2\zeta\, Q\nabla_{{\bf n}({\bf r})} Q + [Q,\Lambda]\right)
\big|_{{\bf r}\in C}=0, \label{BC}
\end{equation}
where
$\zeta$ is given by
\begin{equation}
\zeta=\frac{\tilde l}{2}\frac{1+R_0}{1-R_0},\quad \tilde l\equiv \frac{\pi \nu D_0}{N_{d-1}} \sim l.
 \label{extrapolationlength}
\end{equation}
Equation (\ref{extrapolationlength}) reproduces
the well-known result for the extrapolation length
obtained by completely different methods \cite{Zhu91,Lagendijk89,Genack93}.
However, it should be stressed that the previous methods are valid only for diffusive
samples, while the present theory is valid for both diffusive and localized samples.
In the absence of internal reflection, i.e., $R_0=0$,
Eq.~(\ref{extrapolationlength}) gives the value of
$\frac{2}{3}\, l$ in three dimensions,
in agreement with the result of
Refs.~\cite{Lagendijk00,Niuwenhuizen}. This value is closed to the exact value $\approx 0.71 l$,
obtained by solving the radiative transfer equation \cite{Chandrasekhar,Davison,Hulst}.
(This small deviation is due to the simplification
of $R_{{\bf k}_\perp} \equiv R_0$ made above. The exact value can
be reproduced by the present
field theory by retrieving the ${\bf k}_\perp$-dependence of the internal
reflection coefficient.) In a study of one-dimensional
$SS'S$-superconductor structures \cite{Kupriyanov88},
Kuprianov and Lukichev derived a boundary condition for the Usadel equation
\cite{Usadel70,Larkin77} similar to Eq.~(\ref{BC}), in which the matrix Green function acts like
the supermatrix $Q$-field.

Equation (\ref{BC}) is the boundary condition satisfied by the $Q$-field. Together with the
nonlinear $\sigma$ model action (\ref{action}) it lays down
a foundation for quantitative analysis of wave localization in open media.
The boundary constraint plays an essential role in a field-theoretic
description of localization
in the presence of the interplay
between wave interference and wave energy leakage.
It is valid no matter whether the sample
is diffusive or localized. Noticing that $Q\nabla_{{\bf n}({\bf r})} Q$
mimics the boundary current, we find that Eq.~(\ref{BC})
resembles the so-called `radiative boundary condition'.
The latter has been used by many authors to implement the normal diffusion equation
to study light propagation in diffusive samples \cite{Lagendijk89,Zhu91},
and was justified in the context of one-dimensional random walks \cite{Oppenheim72}.
Indeed, one can show that it is a special case of the more general constraint (\ref{BC}).


\subsection{Spontaneous symmetry breaking and Goldstone modes}
\label{sec:goldstone}

\begin{figure}
 \begin{center}
 \includegraphics[width=8.0cm]{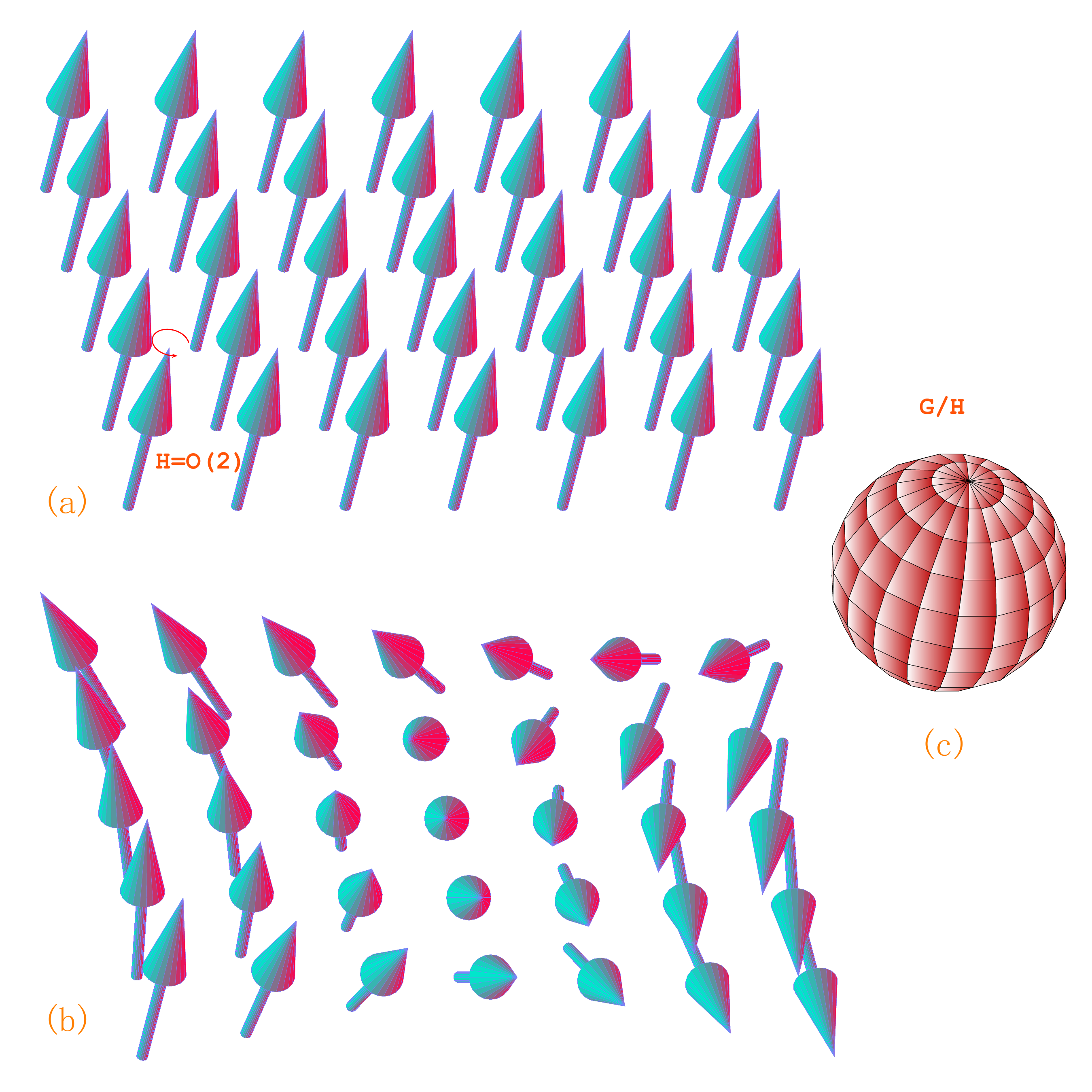}
\end{center}
 \caption{Spontaneous symmetry breaking in a Heisenberg ferromagnet.
 The ground state breaks the global
 rotational symmetry $G=O(3)$ (a). Collective fluctuations around
 this state, the Goldstone modes, are gapless (b). They are associated with the
 breaking of the symmetry $G/H=O(3)/O(2)\cong S^2$ (c).}
 \label{fig:diagram1}
\end{figure}

The supermatrix model (\ref{F}) gives rise to spontaneous symmetry breaking,
and the reduced action (\ref{action}) characterizes
the energy arising from proliferation of the low-lying Goldstone modes
associated with broken continuous symmetries.
To see this let us
first recall a more familiar model namely
the Heisenberg model of ferromagnetism \cite{Heisenberg28,Chaikin}
described by the following Hamiltonian (in unit of temperature),
\begin{equation}\label{eq:25}
    \mathscr{H}[{\bf S}]=-J\sum_{\langle ij\rangle}{\bf S}_i\cdot {\bf S}_j-
    {\bf h}\cdot\sum_i {\bf S}_i,\quad {\bf S}_i^2=1.
\end{equation}
Here without loss of generality the magnitude of the spin ${\bf S}$ is constrained
to unity, the coupling constant
$J>0$, and ${\bf h}$ is the external magnetic field.
In the absence of ${\bf h}$ the system is invariant
under the global three-dimensional rotation giving the symmetry group $G=O(3)$.
Below the critical temperature, this symmetry is broken and all the spins point to
the same direction, say the north pole \cite{Chaikin}. Hence the adjective `spontaneous' follows.
Although in general individual ground state (at ${\bf h}=0$) breaks the symmetry,
all the ground states constitute a degenerate manifold and restores the full symmetry $G$.
Now, an infinitesimal magnetic field ${\bf h}\rightarrow 0$
plays a remarkable role: it selects a particular ground state from this manifold.
In the selected state all the spins are parallel to
${\bf h}$, since globally rotating this state costs infinite energy
in the thermodynamic limit.
This is the so-called spontaneous symmetry breaking, see Fig.~\ref{fig:diagram1} (a).
Suppose that the ground state undergoes local rotation
and the resulting spin configuration varies over a spatial scale $\sim q^{-1}$.
The transformed state must have a larger energy. The energy increase
diminishes as the rotation becomes uniform, i.e., $q\rightarrow 0$.
That is, spatial fluctuations of spins generate soft modes in the sense that their energy
vanishes as $q^\nu$ at $q\rightarrow 0$ where the exponent $\nu>0$, i.e., these soft modes are
gapless. These gapless modes (sketched in Fig.~\ref{fig:diagram1} (b)) are the so-called Goldstone modes.
However, it is important to notice that not all the inhomogeneous rotation
costs energy. In fact, the ground state is invariant if the local transformation
is generated by the two-dimensional group of rotation, $H=O(2)$,
a subgroup of the full symmetry group $ G=O(3)$.
Therefore, the Goldstone modes are associated with the
breaking of the continuous symmetry belonging to the coset space
$G/H=O(3)/O(2)$ instead of the full symmetry group $G$ (cf. Fig.~\ref{fig:diagram1} (c)).

The supermatrix $\sigma$ model bears a formal analogy to
the Heisenberg ferromagnet. For the action (\ref{F}), in the absence of
the frequency term $\sim \omega \tilde \omega^+ {\rm str}\Lambda Q$ it is invariant
under the global gauge transformation:
$Q\rightarrow U^{-1}QU$ where $U$ is a constant matrix belonging to the symmetry group $G$,
which is $UOSP(2,2|4)$ for GOE systems and $U(1,1|2)$ for GUE systems.
Each `ground state', $Q_0$, solving the saddle point equation (\ref{eq:52}) breaks this
continuous symmetry: in general,
$U^{-1}Q_0U\neq Q_0$. Now, if we restore the frequency term, i.e.,
$\omega\tilde \omega^+\Lambda$ with $\tilde \omega\rightarrow 0$ in
the matrix Green function ${\cal G}_0$ (cf. Eq.~(\ref{eq:52})),
the system has a unique `ground state' $\propto\Lambda$.
In other words, the frequency (or the convergence-generating factor
$\delta$ if $\tilde \omega=0$) term plays the role of the `external field'
and causes the spontaneous symmetry breaking.

Similar to Heisenberg ferromagnet this `ground state'
is invariant under the local gauge transformation: $Q_0 \rightarrow U^{-1}({\bf r})
Q_0U({\bf r})$ where $U({\bf r})\in H$ with
$H=UOSP(2|2)\otimes UOSP(2|2)$ for GOE systems and
$H=U(1|1)\otimes U(1|1)$ for GUE systems. Then,
$T({\bf r}) \in G/H$ describes the Goldstone modes and
the action (\ref{action}) describes the energy required for exciting low-lying modes.
As expected $F[Q]|_{\tilde \omega^+=0}$ vanishes if $Q$ is homogeneous.


\section{Macroscopics of wave propagation in open media: general theory}
\label{sec:localdiffusion}

In Ref.~\cite{Tian07}, it was for the first time found at the full microscopic level
that the diffusion coefficient in open media is position-dependent.
However, very specific random media were studied in that work.
Later, the microscopic approach used there was generalized to realistic
random open media (see Sec.~\ref{sec:SUSY}) and was used to
systematically study transport of classical waves in these systems \cite{Tian08,Tian10}.
Armed with the first-principles theory developed, it was found that the wave intensity
correlation -- a key characteristic of wave propagation -- in open media
is very different from that in infinite media \cite{Woelfle80a,Woelfle80,Efetov83a}.
Specifically, such correlation function was found to obey Eq.~(\ref{eq:96}) in the
frequency ($\tilde \omega$) domain. Most strikingly, in
Ref.~\cite{Tian10} it was shown analytically and confirmed
numerically that in
the one-dimensional case, the static local (position-dependent)
diffusion coefficient exhibits a novel scaling behavior. In this section, we shall
establish the scaling theory in the more general case of slab open media in
high dimensions.

From the technical viewpoint, the source of the difficulty can be seen in that
even near the critical dimension
the ordinary renormalization group scheme does not apply:
one cannot replace the
Boltzmann diffusion coefficient by the renormalized one obtained for infinite
media because it does not account for effects arising from the openness of media
such as resonances \cite{Lifshits79,Azbel83,Azbel83a}.
We therefore resort to calculating the wave intensity correlation by using the
developed first-principles theory
and then proceed to establish the macroscopic (or `hydrodynamic',
the term commonly adopted by condensed matter physicists)
diffusion equation describing wave propagation.
Restoring the source term and taking the corresponding derivatives of the partition function
(\ref{average}),
we cast the correlation function (\ref{DCdefinition}) into a functional integral over the
supermatrix field $Q({\bf r})$, i.e.,
\begin{eqnarray}
{\cal Y} ({\bf r},{\bf r}';\tilde \omega) =\left(\frac{\pi \nu(\omega)}{8\omega}\right)^2
\int D[Q]{\rm str} (A_+ Q({\bf r})A_-Q({\bf r}')) e^{-F[Q]}
\label{DC}
\end{eqnarray}
for $\tilde \omega\ll \omega$, where $A_\pm =
\sigma_3^{\rm fb}\otimes ({\mathbbm{1}}\pm\sigma_3^{\rm ar})\otimes ({\mathbbm{1}}-\sigma_3^{\rm tr})$.
This expression is valid
for both infinite media and finite-sized open media.

\subsection{Wave propagation in infinite media}
\label{sec:correlator_infinite}

It is very illustrative to calculate Eq.~(\ref{DC}) first for infinite media.
In doing so, one may become familiar with perturbative
calculations of the supersymmetric field theory, and
appreciate -- at the perturbative level -- the main technical differences between
infinite and finite-sized open media. We shall explain in detail the calculations for
GOE systems and give the results for GUE systems directly.

\subsubsection{Parametrization of the supermatrix $Q$}
\label{sec:parametrization}

Notice
the following identity,
\begin{eqnarray}
\label{eq:133}
  T\equiv \left(
            \begin{array}{cc}
              T_{11} & T_{12} \\
              T_{21} & T_{22} \\
            \end{array}
          \right)^{\rm ar}=\left(
            \begin{array}{cc}
              T_{11} & 0 \\
              0 & T_{22} \\
            \end{array}
          \right)^{\rm ar}
          \left(
            \begin{array}{cc}
              1 & T_{11}^{-1}T_{12} \\
              T_{22}^{-1}T_{21} & 1 \\
            \end{array}
          \right)^{\rm ar}\equiv \left(
            \begin{array}{cc}
              T_{11} & 0 \\
              0 & T_{22} \\
            \end{array}
          \right)^{\rm ar}(1+iW).
\end{eqnarray}
Upon inserting it into $Q$, we find
\begin{eqnarray}
\label{eq:32}
  Q=(1+iW)^{-1}\Lambda (1+iW),\quad W=\left(
                   \begin{array}{cc}
                     0 & B \\
                     kB^\dagger & 0 \\
                   \end{array}
                 \right)^{\rm ar},
\end{eqnarray}
which is the so-called rational parametrization.
The structure imposed on the $W$ matrix
is fixed by two conditions. That is, $W$ anticommutes with $\Lambda$
and $Q^\dagger = KQK$
due to Eq.~(\ref{eq:124}). The $4\times 4$ matrix $B$ can be parameterized as
\begin{eqnarray}
B = \sum_{i=0}^3 \left(
\begin{array}{cc}
  a_i & i\sigma_i \\
 \eta_i & ib_i \\
\end{array}
\right)^{\rm fb}\otimes \sigma_i^{\rm tr},
\label{B}
\end{eqnarray}
where the entries of $a$'s and $b$'s ($\sigma$'s and $\eta$'s) are complex commuting
(anticommuting) variables. They are not independent due to the constraint,
\begin{equation}\label{eq:135}
    W^\dagger =-CW^{\rm T}C^{\rm T}
\end{equation}
introduced by the condition (\ref{eq:125}), and the number of independent
integral variables is reduced by one half. In Eq.~(\ref{B}),
the subscript $i=0,3$ ($1,2$) stands for the diffuson (cooperon).
An advantage of the rational
parametrization is that the corresponding Jacobian (the so-called `Berezinian') is unity.
For perturbative calculations below,
it is sufficient to take this as an exact mathematical theorem.
We refer the reader interested in its proof to Ref.~\cite{Efetov97}.

\begin{figure}
 \begin{center}
 \includegraphics[width=8.0cm]{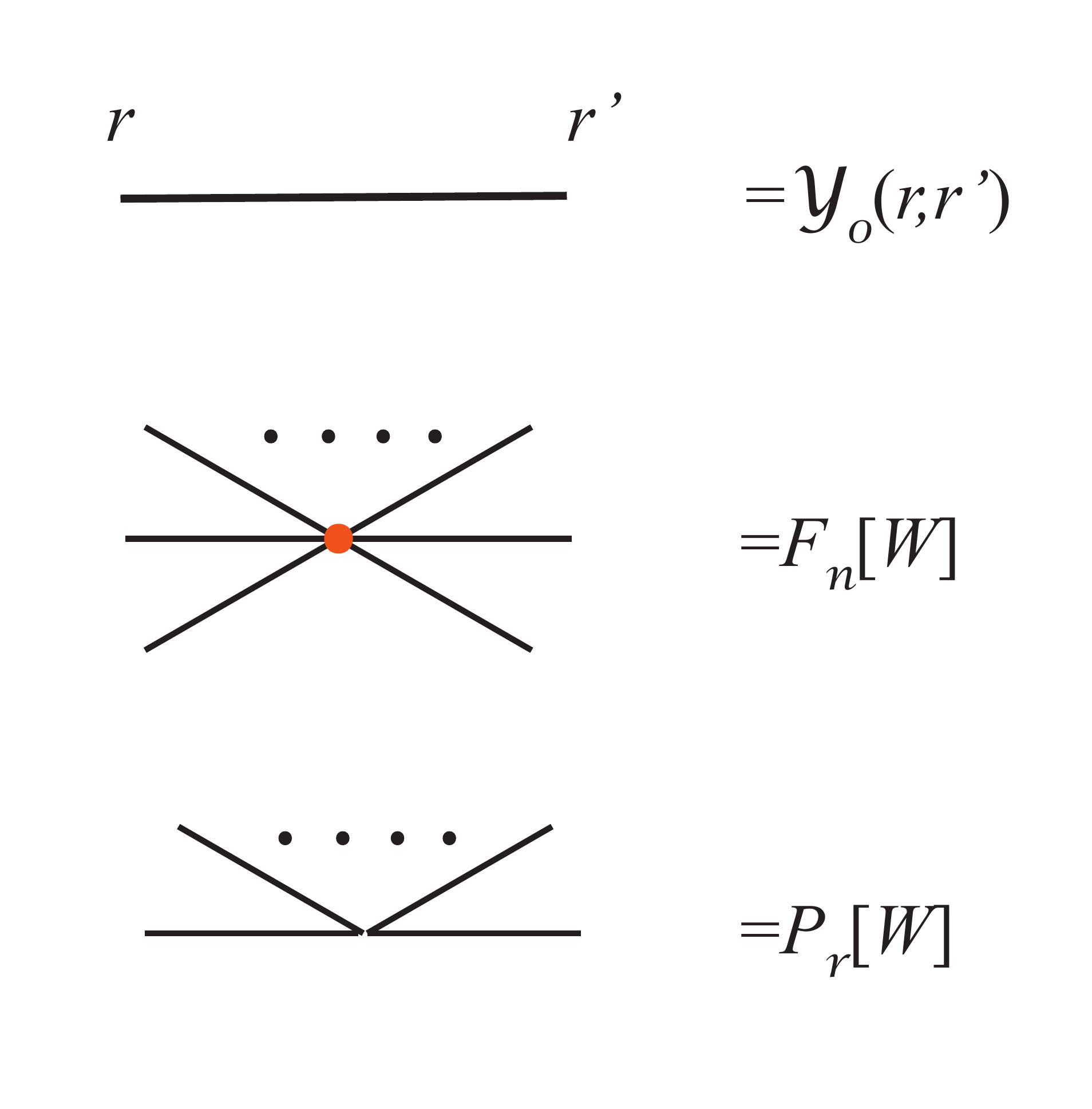}
\end{center}
 \caption{In the perturbation theory of the nonlinear supermatrix $\sigma$ model,
 each diagram is composed of the bare propagator (top) and the interaction vertexes
 (middle and bottom).}
 \label{fig:diagram3}
\end{figure}

\subsubsection{Wave interference corrections}
\label{sec:perturbation}

For large $\tilde \omega$ we may expand Eq.~(\ref{DC}) in terms of $W$.
The action $F[Q]$ is given by $\sum_{n=1}^\infty F_{n}[W]$ with $F_{n}[W]
\propto W^{2n}$. The pre-exponential factor
leads to another $W$-expansion, $\sum_{r,s=0}^\infty{\rm str}
(A_+P_{r}[W({\bf r})]A_-P_s[W({\bf r}')])$, where
$P_{r}[W] = {\cal O}(W^r)$. As a result,
\begin{eqnarray}
\label{eq:98}
  {\cal Y} ({\bf r},{\bf r}';\tilde \omega) =\left(\frac{\pi \nu(\omega)}{8\omega}\right)^2
  \left\langle \sum_{r,s=0}^\infty{\rm str}(A_+P_{r}[W({\bf r})]A_-P_s[W({\bf r}')])
  \sum_{m=0}^\infty\frac{1}{m!} \left(\sum_{n=2}^\infty F_{n}[W]\right)^m\right\rangle_0,
\end{eqnarray}
where the average is defined as
$\langle\cdot \rangle_0 \equiv \int D[W] (\cdot) \exp(-F_1[W])$, with
$F_1[W]=-\frac{\pi \nu}{2}
\int d{\bf r}{\rm str} (D_0(\nabla iW)^2 - i\tilde \omega (iW)^2)$.
This perturbative expansion describes the interaction between diffusons and cooperons,
and each term can be calculated by using Wick's theorem with the
help of the following contraction rules,
\begin{eqnarray}
\label{eq:84}
4\pi\nu \langle W({\bf r})M W({\bf r}')\rangle_0 &=& {\cal Y}_0({\bf r},{\bf r}';\tilde \omega)
\left[({\rm str} M + {\bar M}) - \Lambda ({\rm str} (M \Lambda) + {\bar M}\Lambda)\right], \nonumber\\
4\pi\nu \langle {\rm str}(W({\bf r})M) {\rm str}(W({\bf r}')M')\rangle_0 &=&
{\cal Y}_0({\bf r},{\bf r}';\tilde \omega) {\rm str}((M-{\bar M})(M'-\Lambda M'\Lambda)),
\end{eqnarray}
with $\bar M=KCM^{\rm T}C^{\rm T}K$. The
propagator, ${\cal Y}_0({\bf r},{\bf r}';\tilde \omega)$,
solves the normal diffusion equation,
\begin{eqnarray}\label{eq:29}
    (-i{\tilde \omega}-D_0\nabla^2){\cal Y}_0({\bf r},{\bf r}';{\tilde \omega})=\delta({\bf r}-{\bf r}'),
\end{eqnarray}
which, importantly, is translationally invariant, i.e.,
${\cal Y}_0({\bf r},{\bf r}';\tilde \omega)={\cal Y}_0({\bf r}-{\bf r}',0;\tilde \omega)$.
The contraction rules can be
proved straightforwardly by using the parametrization (\ref{eq:32}) and (\ref{B}).
With all the contractions made, each term
in the expansion (\ref{eq:98}) can be represented by a diagram composed of
the (bare) propagator ${\cal Y}_0({\bf r},{\bf r}';\tilde \omega)$
(representing the corresponding contraction),
and the `interaction' vertices $F_{n\geq 2}[W]$ and $P_r[W]$ (Fig.~\ref{fig:diagram3}).
The diagram is
either reducible or irreducible, depending on whether it may be
divided into two disconnected subdiagrams by cutting some bare propagator.

\begin{figure}
 \begin{center}
 \includegraphics[width=10.0cm]{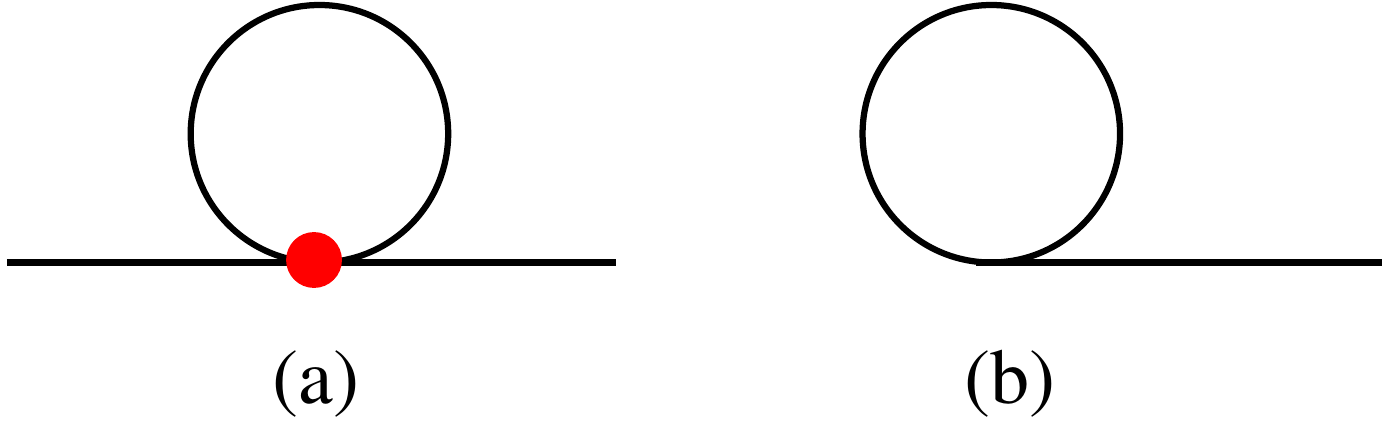}
\end{center}
 \caption{Diagrams give the leading order wave interference correction to the Boltzmann diffusion constant.}
 \label{fig:diagram}
\end{figure}

Let us start from the lowest order term (${\cal O}(W^2)$) in Eq.~(\ref{eq:98}), i.e.,
\begin{eqnarray}
\label{eq:108}
  -\left(\frac{\pi \nu(\omega)}{4\omega}\right)^2
  \int D[W]e^{-F_1[W]}
  {\rm str}(A_+ iW({\bf r}) A_- iW({\bf r}')).
\end{eqnarray}
With the help of the contraction rules
(\ref{eq:84}) we find that at this perturbation level, the correlator
${\cal Y}({\bf r},{\bf r}';\tilde \omega)$ is identical to
the bare correlator ${\cal Y}_0({\bf r},{\bf r}';\tilde \omega)$, which
excludes all wave interference effects.
The subleading term is ${\cal O}(W^4)$
with all the irreducible diagrams given in
Fig.~\ref{fig:diagram}. (Notice that $F_n[W]\propto W^{2n} ={\cal O}(W^{2(n-1)})$
for $F_1[W]={\cal O}(1)$.)
These terms contribute wave interference
corrections to the bare correlator. More precisely, at this perturbation level
the correlator is (see Appendix~\ref{sec:cancelation} for details)
\begin{eqnarray}
\label{eq:109}
  &&{\cal Y}_0 ({\bf r},{\bf r}';\tilde \omega) + \delta {\cal Y}_1 ({\bf r},{\bf r}';\tilde \omega) \nonumber\\
  &=& -\left(\frac{\pi \nu(\omega)}{4\omega}\right)^2
  \int D[W]e^{-F_1[W]}
  {\rm str}(A_+ iW({\bf r}) A_- iW({\bf r}'))
  \left(1+\frac{\pi\nu}{2}\int d{\bf r}_1 \delta D^{(1)}(\tilde \omega){\rm str}(\nabla iW({\bf r}_1))^2\right)
  ,
\end{eqnarray}
where
\begin{equation}\label{eq:100}
    \delta D^{(1)}({\tilde \omega})=-\frac{D_0}{\pi\nu}{\cal Y}_0(0,0;{\tilde \omega}).
\end{equation}
By re-exponentiating the last factor
(which is equivalent to summing up all the reducible diagrams
with the one-loop vertex) we rewrite Eq.~(\ref{eq:109}) as
\begin{eqnarray}
\label{eq:110}
  {\cal Y} ({\bf r},{\bf r}';\tilde \omega) \approx -\left(\frac{\pi \nu(\omega)}{4\omega}\right)^2
  \int D[W]e^{\frac{\pi\nu}{2}\int d{\bf r}_1 {\rm str} (D(\tilde \omega)
  (\nabla iW)^2 - i\tilde \omega (iW)^2)}
  {\rm str}(A_+ iW({\bf r}) A_- iW({\bf r}')).
\end{eqnarray}
Comparing the Gaussian wight with that of Eq.~(\ref{eq:108}),
we find that wave interference renormalizes the
bare diffusion coefficient, i.e., $D_0\rightarrow D(\tilde \omega)
=D_0+\delta D^{(1)}(\tilde \omega)$, and $\delta D^{(1)}(\tilde \omega)$ is
the well-known weak localization correction \cite{Woelfle80a,Woelfle80,Larkin79}
namely Eq.~(\ref{eq:123}).
The correlator (\ref{eq:110}) solves the macroscopic equation (\ref{eq:130}).

Importantly, the diagram (b) in Fig.~\ref{fig:diagram}
guarantees that the wave interference correction
to the bare propagator ${\cal Y}_0$ vanishes when ${\cal Y}_0$ is homogeneous in
space \cite{Tian08} (see also Appendix~\ref{sec:cancelation} for the proof).
In general, the roles of the expansion of the pre-exponential factor
are to conform with an exact Ward identity reflecting the energy conservation law,
\begin{eqnarray}\label{eq:101}
    {\cal Y}({\bf r},{\bf r}';{\tilde \omega})
    \stackrel{\nabla {\cal Y}=0}{=}\frac{1}{-i\tilde\omega}.
\end{eqnarray}
This must be respected at all levels of the perturbation.
In the diagrammatic perturbation theory,
conforming to this identity necessitates the
introduction of the well-known Hikami box \cite{Hikami81}.

For GUE systems the leading order weak localization correction (\ref{eq:100}) vanishes.
This certainly does not imply the absence of interference effects in
the weak (not strong) localization regime. The point is
that wave interference (see Fig.~\ref{fig:interference} for
example) can be much more complicated and quantitatively they
give higher order weak localization corrections.
It must be emphasized that these higher order corrections cannot be
obtained by calculating reducible diagrams with one-loop vertex.

\begin{figure}
 \begin{center}
 \includegraphics[width=10.0cm]{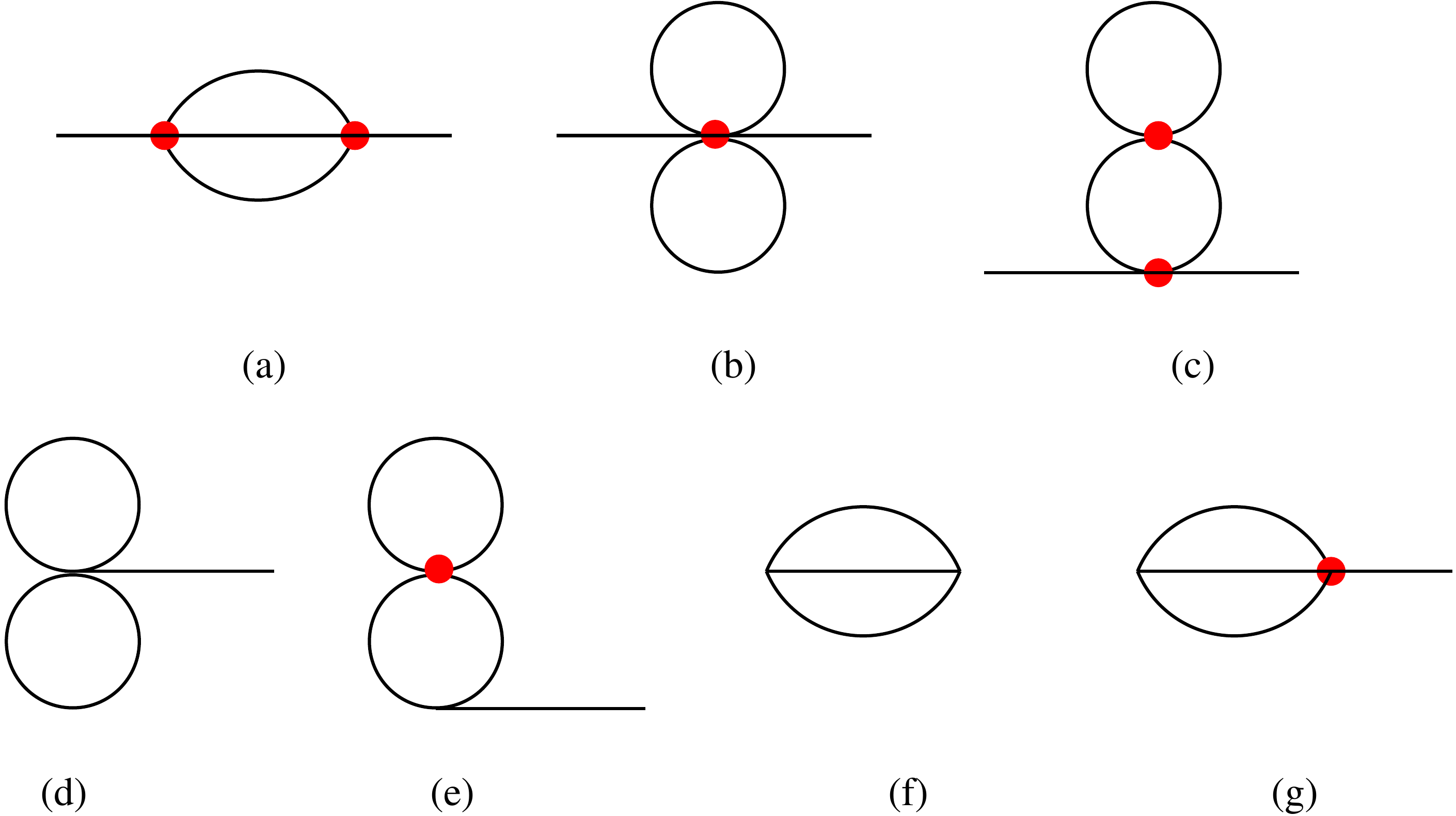}
\end{center}
 \caption{Diagrams give the subleading order wave interference correction to the Boltzmann diffusion constant.}
 \label{fig:diagram_a}
\end{figure}

Let us now consider the terms $\sim{\cal O}(W^6)$. The irreducible
diagrams of this order are given in
Fig.~\ref{fig:diagram_a}. (Recall that
$F_2[W] ={\cal O}(W^{2})$ and $F_3[W] ={\cal O}(W^{4})$.)
The derivation is similar to that
of deriving $\delta D^{(1)}(\tilde \omega)$ but is much more complicated. By using
the contraction rules (\ref{eq:84}) one may show that for GOE systems the
correction given by Fig.~\ref{fig:diagram_a}
(denoted as $\delta {\cal Y}_2({\bf r},{\bf r}';\tilde \omega)$) vanish, i.e.,
\begin{equation}\label{eq:198}
    \delta D^{(2)}(\tilde \omega)=0, \,
    {\rm for\, GOE}.
\end{equation}
For GUE systems here we will emphasize the key points and
only give the final result \cite{Tian12a}.
First of all, Fig.~\ref{fig:diagram_a} (d) - (g) again
conforms to the Ward identity (\ref{eq:101}). Secondly,
$\delta {\cal Y}_2({\bf r},{\bf r}';\tilde \omega)$ consists of
two contributions, i.e.,
\begin{eqnarray}
\label{eq:172}
  \delta {\cal Y}_2 ({\bf r},{\bf r}';\tilde \omega)
  &=& -\left(\frac{\pi \nu(\omega)}{4\omega}\right)^2\frac{\pi\nu D_0}{2}
  \int D[W]e^{-F_1[W]}
  {\rm str}(A_+ iW({\bf r}) A_- iW({\bf r}'))\int d{\bf r}_1\!\!\int d{\bf r}_2
  {\rm str}\left(\nabla^\alpha iW({\bf r}_1)\nabla^\beta iW({\bf r}_2)\right)\nonumber\\
  && \times
  \left[\frac{1}{2}\left(\frac{{\cal Y}_0({\bf r}_1,{\bf r}_2;{\tilde \omega})}{\pi\nu}\right)^2
  \delta({\bf r}_1-{\bf r}_2) \delta_{\alpha\beta} + D_0
  \nabla^\alpha \nabla^\beta
{\cal Y}_0({\bf r}_1,{\bf r}_2;{\tilde \omega})
 \left(\frac{{\cal Y}_0({\bf r}_1,{\bf r}_2;{\tilde \omega})}{\pi\nu}\right)^2\right].
  \end{eqnarray}
The first contribution leads to a wave interference
correction to $D_0$ which is local in space
thanks to the factor $\delta({\bf r}_1-{\bf r}_2)$,
while the second one to a correction non-local in space.
The non-locality arises from diagrams such as Fig.~\ref{fig:diagram_a} (a).
Because the system is isotropic,
the crossing terms, $\sim \nabla^{\alpha}\nabla^{\beta},\,
\alpha\neq \beta$, do not contribute upon integrating out the space coordinates.
As a result, Eq.~(\ref{eq:172}) is simplified to
\begin{eqnarray}
\label{eq:173}
  \delta {\cal Y}_2 ({\bf r},{\bf r}';\tilde \omega)
  &=& -\left(\frac{\pi \nu(\omega)}{4\omega}\right)^2\frac{\pi\nu D_0}{2}
  \int D[W]e^{-F_1[W]}
  {\rm str}(A_+ iW({\bf r}) A_- iW({\bf r}'))\int d{\bf r}_1\!\!\int d{\bf r}_2
  {\rm str} \nabla iW({\bf r}_1)\cdot \nabla iW({\bf r}_2)\nonumber\\
  && \times
  \left[\frac{1}{2}\left(\frac{{\cal Y}_0({\bf r}_1,{\bf r}_2;{\tilde \omega})}{\pi\nu}\right)^2
  \delta({\bf r}_1-{\bf r}_2) +
  \frac{D_0}{d}\nabla^2
{\cal Y}_0({\bf r}_1,{\bf r}_2;{\tilde \omega})
 \left(\frac{{\cal Y}_0({\bf r}_1,{\bf r}_2;{\tilde \omega})}{\pi\nu}\right)^2\right].
  \end{eqnarray}
In the hydrodynamic limit the spatial and time derivatives
in the macroscopic equation are decoupled.
Taking this into account, Eq.~(\ref{eq:173}) reduces to
\begin{eqnarray}
\label{eq:197}
  \delta {\cal Y}_2 ({\bf r},{\bf r}';\tilde \omega)
  &\rightarrow & -\left(\frac{\pi \nu(\omega)}{4\omega}\right)^2\frac{\pi\nu D_0}{2}
  \int D[W]e^{-F_1[W]}
  {\rm str}(A_+ iW({\bf r}) A_- iW({\bf r}'))\int d{\bf r}_1\!\!\int d{\bf r}_2
  {\rm str} \nabla iW({\bf r}_1)\cdot \nabla iW({\bf r}_2)\nonumber\\
  && \times
  \left[\frac{1}{2}\left(\frac{{\cal Y}_0({\bf r}_1,{\bf r}_2;{\tilde \omega})}{\pi\nu}\right)^2
  \delta({\bf r}_1-{\bf r}_2) +
  \frac{1}{d} (D_0 \nabla^2+i\tilde \omega)
{\cal Y}_0({\bf r}_1,{\bf r}_2;{\tilde \omega})
 \left(\frac{{\cal Y}_0({\bf r}_1,{\bf r}_2;{\tilde \omega})}{\pi\nu}\right)^2\right]\nonumber\\
  &=&-\left(\frac{\pi \nu(\omega)}{4\omega}\right)^2
  \int D[W]e^{-F_1[W]}
  {\rm str}(A_+ iW({\bf r}) A_- iW({\bf r}'))\frac{\pi\nu}{2} \int d{\bf r}_1\delta D^{(2)} (\tilde \omega)
  {\rm str} (\nabla iW({\bf r}_1))^2,
  \end{eqnarray}
where the wave interference correction to $D_0$ is local, read
\begin{equation}\label{eq:106}
    \delta D^{(2)}(\tilde \omega)=
    \left(\frac{1}{2}-\frac{1}{d}\right)D_0\left(\frac{{\cal Y}_0(0,0;{\tilde \omega})}{\pi\nu}\right)^2, \,
    {\rm for\, GUE}.
\end{equation}
At this perturbation level, the correlation function is
${\cal Y}={\cal Y}_0+\delta {\cal Y}_1+\delta {\cal Y}_2$
which solves the macroscopic diffusion equation (\ref{eq:130}),
with the dynamic diffusion coefficient replaced by
$D(\tilde \omega)=D_0 + \delta D^{(1)}(\tilde \omega) +
\delta D^{(2)}(\tilde \omega)$. At the
higher order perturbation level, Eq.~(\ref{eq:130}) remains valid with
$D(\tilde \omega)$ including corresponding higher order corrections.


Notice that there are two opposite contributions to $\delta D^{(2)}(\tilde \omega)$,
accounting for the factor $\frac{1}{2}$ and $-\frac{1}{d}$, respectively.
The wave interference picture
underlying the former \cite{Altshuler98,Tian04} (see also Fig.~\ref{fig:interference}
below) may be related to the interesting phenomenon of coherent forward scattering \cite{Mueller12}.
Importantly, the factor $(\frac{1}{2}-\frac{1}{d})$ ensures the renormalizability:
approaching two dimensions (from below), this infinitesimal factor cancels
one logarithmically diverging factor arising from ${\cal Y}_0(0,0;{\tilde \omega})$
and makes the two-loop
correction diverge only logarithmically.
Upon making the analytic continuation from dimensions below two to above, we find from
Eqs.~(\ref{eq:100}), (\ref{eq:198}) and (\ref{eq:106}) the
well-known scaling equation for the dimensionless Thouless conductance
$g(L_{\tilde \omega})
=\pi\nu D_0 L_{\tilde \omega}^{d-2}/\Omega_d$ with $L_{\tilde \omega}=\sqrt{
D_0/{\tilde \omega}}$ and $\Omega_d$ the volume of $d$-dimensional unit sphere, i.e.,
\begin{eqnarray}\label{eq:138}
    \frac{d\ln g}{d\ln L_{\tilde \omega}}=\bigg\{
\begin{array}{c}
  d-2 -\frac{1}{g}+{\cal O}(\frac{1}{g^3}),\quad {\rm for\, GOE}, \\
  d-2 -\frac{1}{2g^2}+{\cal O}(\frac{1}{g^3}),\quad {\rm for\, GUE}
\end{array}
\end{eqnarray}
near the critical dimension \cite{Efetov97,Anderson79}.


We wish to make some remarks here. (i) In the presence of magneto-optical effects
the one-loop
weak localization may be fully suppressed. In this case, the critical theory is
described by the second equation (\ref{eq:138}). This overcomes the notable difficulty
of the self-consistent diagrammatic approach
\cite{Woelfle80a,Woelfle80} which cannot be used to study localization of GUE systems.
This is a simple example showing the field theoretic approach
to be a powerful technique in studies of localization in
magneto-optical systems.
(ii) The criticality described by the scaling equation is quantitatively accurate
in $2+\epsilon$ ($\epsilon\equiv d-2$ which should not be confused with
the dielectric constant) dimensions since the critical point
$g^*={\cal O}(\epsilon^{-1})\gg 1$.
One may extrapolate the results into three dimensions by setting $\epsilon=1$. In particular,
we find that the Ioffe-Regel criterion is also valid for GUE systems. This
may guide experimental search of light localization in magneto-optical systems.
(iii) In combination with Eq.~(\ref{mfp}) this criterion predicts that there
is a mobility edge separating low-frequency extended states and
high-frequency localized states.
(iv) In low dimensions ($d<2$), the dimensionless conductance
decreases as the frequency $\tilde \omega$ is lowered. In this case,
all the higher order irreducible diagrams are important.
Similar to the above analysis, these give local corrections to the
Boltzmann constant. This implies the validity of the macroscopic equation (\ref{eq:130})
for low $\tilde \omega$. However, obtaining the explicit form of $D(\tilde \omega)$ is beyond the
perturbation scheme and one must resort to non-perturbative treatments
of the supersymmetric nonlinear $\sigma$ model. This indeed
has been done by Efetov and Larkin \cite{Efetov83a}
in the Q$1$D case. (v) According to the exact (non-perturbative) solution
given in Ref.~\cite{Efetov83a}, in Q$1$D the localization
length $\xi\sim \nu(\omega)D_0(\omega)$. This implies that
the localization length diverges as $\omega^{-2}$ in the limit $\omega\rightarrow 0$.
Notice that this scaling law is universal (but the coefficient not):
it is the same for both GOE and GUE systems.
(vi) In two dimension the localization length diverges exponentially in the low-frequency limit.
The behavior depends on system's symmetry: $\ln \xi \sim \omega^{-2}$ for GOE systems
and $\ln \xi \sim \omega^{-4}$ for GUE systems. The latter results from the
infinitesimal factor of $(\frac{1}{2}-\frac{1}{d})$.


\subsection{Wave propagation in open media}
\label{sec:open_media}

We now consider the case of open media and
investigate how the results above are modified.
Let us begin with a pictorial illustration in
terms of the supermatrix field of the
main difference between infinite and finite-sized open media. We then
substantiate the picture with explicit calculations of the wave intensity correlator.

\subsubsection{A pictorial illustration}
\label{sec:Heisenberg}

For simplicity let us consider the discrete space lattice (with
the lattice constant set to unity).
In this case, the nonlinear $\sigma$ model action
(\ref{action}) is written as
\begin{equation}\label{eq:26}
    F[Q]=-\nu D_0
    \sum_{\langle ij\rangle}{\rm str} Q_iQ_j+
    \frac{\pi\nu i\tilde \omega^+}{4}\sum_i {\rm str}\Lambda Q_i,\quad Q_i^2=1,
\end{equation}
where $i$ stands for the lattice site and the summation is over all
the nearest neighbors, $\langle ij\rangle$.
Comparing Eqs.~(\ref{eq:25}) and (\ref{eq:26}) we find that these two models bear
a firm analogy summarized in Table \ref{tab:1}.
Notice that the supermatrix structure
of $Q$ as well as the supertrace operation is irrelevant for
present discussions.

Identifying this analogy,
we may translate some well-known facts regarding
the ferromagnetic system \cite{Chaikin} to the present
supermatrix $\sigma$ model.
For the ferromagnetic system, if the temperature is sufficiently low,
spins tend to align in parallel as a result of the so-called spontaneous
symmetry breaking. In the presence of the external magnetic field ${\bf h}$
they are all parallel to the magnetic field. Such an ordered phase is the ferromagnetic
phase accompanied by
long-ranged order. Namely, the spatial correlation of two spins does not vanish
when their distance approaches infinity. If the temperature is
sufficiently high, spins strongly fluctuate in space and
the average magnetization is zero. Such a disordered
phase is the paramagnetic phase. In this phase, a nonvanishing average magnetization
can be triggered only by an external magnetic field. Unlike
the ferromagnetic phase, the spin-spin correlation decays exponentially
when the distance exceeds the correlation length.
At some critical temperature a transition from the ferromagnetic to paramagnetic phase occurs,
accompanied by large spin fluctuations.

This scenario has an analogy in the supersymmetric nonlinear $\sigma$ model.
As mentioned above, the supermatrix $Q$ field mimics the spin.
In high dimensions ($d>2$),
the metallic phase --
the `ferromagnetic phase' -- is formed,
if disorders are weak enough (large $\nu D_0$), in which
the `spins' $Q$ tend to align. In the presence of $\tilde \omega$ they all
point towards `north pole', $\Lambda$.
`Spin' fluctuations in space characterize
diffusive modes (diffuson and cooperon) and their interactions,
and long-range order is manifested in the spatial extension of
wave functions. In this ordered phase,
`spin' fluctuations are weak characterizing a metal suffering weak localization.
If disorder is strong enough (small coupling constant $g_0$) or in low dimensions ($d\leq 2$),
the system is in the insulator phase --
the `paramagnetic phase' -- where the `spin' $Q$ fluctuates strongly in space.
In this disordered phase,
the correlation of the `spin' $Q$ falls
exponentially at large distances with the `correlation length'
being the localization length $\xi$.
This is manifested in the exponential localization of wave functions.
As we have seen from the scaling theory (\ref{eq:138}), in dimension $d>2$,
a metal-insulator transition is triggered by lowering the coupling constant $g_0$:
this is the analogy of (thermodynamic) ferromagnetic-paramagnetic transition.
We wish to point out that the above supermatrix - spin analogy should not
be carried too far: notably, the supermatrix $Q$ field does not play the role of
the order parameter in the Anderson transition \cite{Efetov97}.
In fact, the average `magnetization', $k\Lambda Q$, does not vanish in both phases. Physically,
this is so because the density of states is not
critical in the Anderson transition.

\begin{figure}
 \begin{center}
\includegraphics[width=8.0cm]{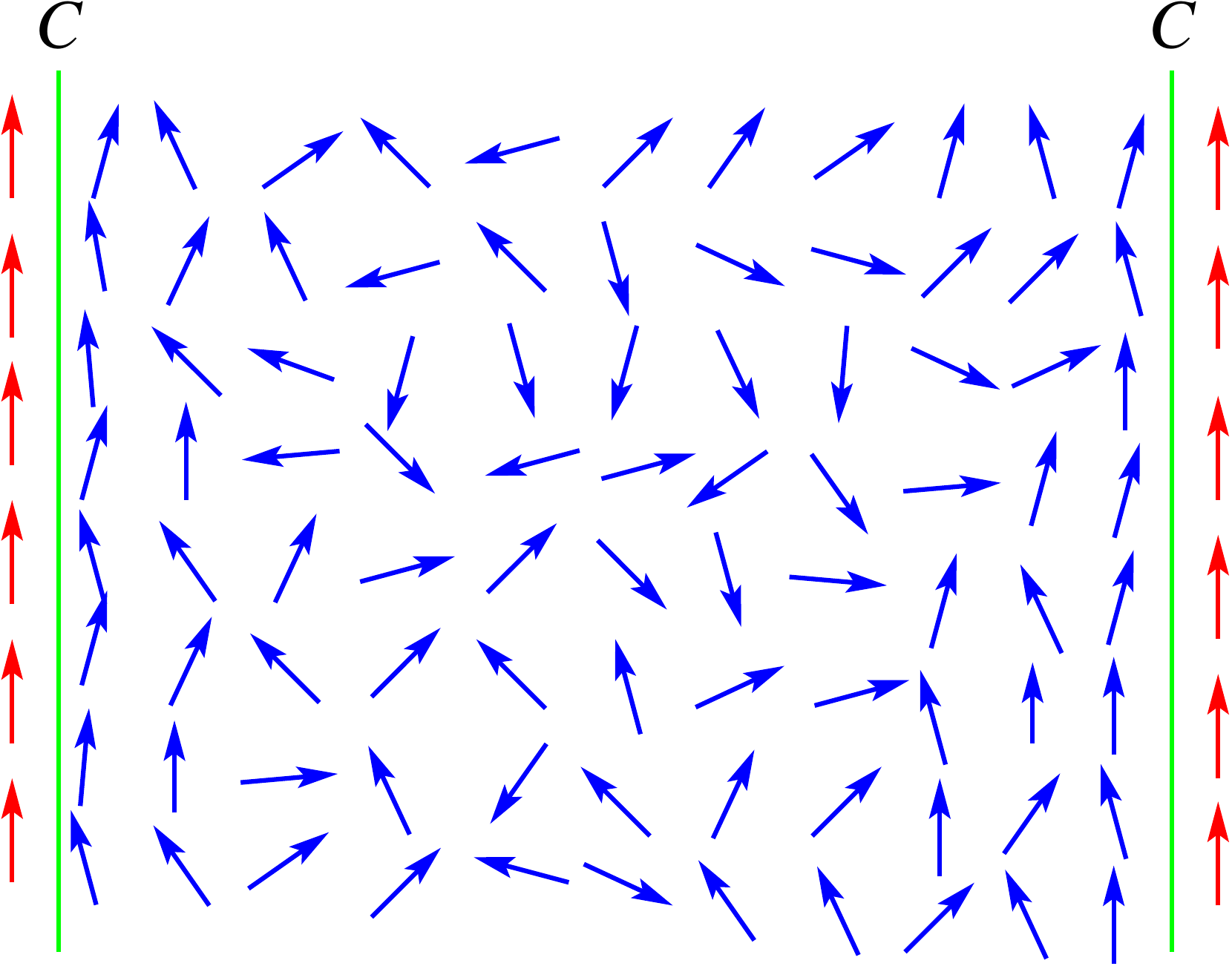}
\end{center}
 \caption{The supermatrix $Q$ and the
 supersymmetric nonlinear $\sigma$ model mimic the spin and the
 Heisenberg model of ferromagnetism, respectively. The
 `spin' $Q$ is frozen on the
 transparent interface (red), and fluctuations increase
 inside the medium (blue).}
 \label{fig:superspin}
\end{figure}

(For simplicity, below we focus on the static case, i.e., ${\tilde \omega}=0$.)
Open media (without internal reflection) differ from
infinite media in that outside the medium, the `spin' $Q$ is fixed, pointing towards `north pole' $\Lambda$
(the red arrows in Fig.~\ref{fig:superspin}).
Due to the `ferromagnetic' coupling,
near the boundary the `spin' $Q$ tends to align with
$\Lambda$. As the distance to the interface increases,
the $Q$-field fluctuations becomes stronger (Fig.~\ref{fig:superspin}).
Since such fluctuations characterize wave interference,
the interference strength increases as waves propagate from
the interface to the sample center.
Since homogeneous wave interference effects
lead to a spatially homogeneous diffusion coefficient, $D(\tilde \omega)$, in infinite media,
one may naturally expect that in open media a local
diffusion coefficient $D({\bf r};\tilde \omega)$ results.

\begin{table}
  \centering
  \caption{Ferromagnet-supermatrix $\sigma$ model analogy}
  \label{tab:1}
\begin{tabular}{|l|l|l|}
    \hline
    & Heisenberg ferromagnet & supermatrix $\sigma$ model  \\
\hline\hline system's energy & \qquad $\mathscr{H}[{\bf S}]$ & \qquad $F[Q]$ \\
\hline spin & \qquad $S_i$ & \qquad $Q_i$ \\
\hline constraint & \qquad $S_i^2=1$ & \qquad $Q_i^2=1$\\
\hline coupling constant & \qquad $J$ & \qquad $\nu D_0$ \\
\hline magnetic field & \qquad ${\bf h}$ & \qquad $-\frac{\pi\nu i\tilde \omega^+}{4}\Lambda$ \\
\hline ordered phase & \qquad ferromagnet & \qquad metal \\
\hline disordered phase & \qquad paramagnet& \qquad insulator\\
    \hline
  \end{tabular}
\end{table}

\subsubsection{Local diffusion equation and single parameter scaling of $D({\bf r};\tilde \omega)$}
\label{sec:scalinggeneral}

Let us first consider the boundary constraint. The role of
the second term of Eq.~(\ref{BC}) is to align $Q$ to $\Lambda$. Indeed,
if we ignore for the moment the first term and substitute into
it the parametrization (\ref{eq:32}),
we find $W({\bf r}\in C)=0$, i.e., $Q({\bf r}\in C)=\Lambda$.
For $\zeta$ much smaller than the inverse of the typical value
of the boundary derivative $Q\nabla_{{\bf n}({\bf r})}Q $ which is $\sim {\cal O}({\rm min}(L,\xi))$,
we may keep the leading order expansion in
$W$, obtaining
\begin{eqnarray}
(\zeta\nabla_{{\bf n}({\bf r})}-1)W=0.
\label{BCsimplify1}
\end{eqnarray}
This implies that $W
$ -- if extended to the air from the random medium -- would decay exponentially
for distances to the interface much larger than $\zeta$.
In other words, the Goldenstone modes tend to penetrate into the air
for a distance $\sim\zeta$. This translates the canonical
physical interpretation of the extrapolation length
into the field theoretical language. More precisely, for $\zeta \ll {\rm min}(\xi,L)$,
the boundary constraint (\ref{BC}) is simplified to
\begin{equation}
Q|_{{\bf r}\in C'}=\Lambda,
\label{BCsimplify2}
\end{equation}
where $C'$ is the trapping plane outside the medium and of a distance $\zeta$
to the interface $C$.
Then, effect of the simplified boundary condition (\ref{BCsimplify2})
is to enlarge the size of the medium by a thickness of $\zeta$,
which is well known for diffusive media \cite{Niuwenhuizen}.

It should be emphasized that the boundary condition (\ref{BCsimplify2})
also applies to the localized system, provided the condition $\zeta \ll \xi$ is met.
This implies that the common wisdom \cite{Niuwenhuizen}
for a diffusive (slab) sample -- that its effective thickness is larger than its
actual thickness by $2\zeta$ -- can be extended even to a localized sample. Most
importantly, such common wisdom should not be carried to the case
of strong internal reflection where $\zeta \gtrsim {\rm min}(\xi,L)$. In
this case, the perturbative expansion for Eq.~(\ref{BC})
ceases to work and Eq.~(\ref{BCsimplify2}) breaks down.

In this review, we are interested in the simplest case in which internal reflection
is absent. In this case, $\zeta \sim l$
is much smaller than any other macroscopic scale. For this reason, we may ignore
the slight difference of $2\zeta$ between the effective and actual
thickness of the sample, and from now on set $C'=C$.
We then repeat the procedures of Sec.~\ref{sec:correlator_infinite}
for deriving Eq.~(\ref{eq:130}).
A summary of the essential differences from Sec.~\ref{sec:correlator_infinite}
in the final results is as follows: (i) the propagator ${\cal Y}_0$ is no longer translationally invariant,
i.e., ${\cal Y}_0({\bf r},{\bf r}';\tilde \omega) \neq {\cal Y}_0({\bf r}-{\bf r}',0;\tilde \omega)$.
The translational symmetry is broken by the boundary condition:
${\cal Y}_0({\bf r}\in C,{\bf r}';\tilde \omega)=0$ which is a result of
Eq.~(\ref{BCsimplify2}).
(ii) Because of (i), the probability density
${\cal Y}_0({\bf r},{\bf r};\tilde \omega)$ is no longer homogeneous, but rather, depends on the
position ${\bf r}$ explicitly. (iii) Up to two loops the weak localization
correction is
\begin{equation}\label{eq:107}
    \delta D({\bf r}; \tilde \omega)=D_0\left[
    -\frac{{\cal Y}_0({\bf r},{\bf r};\tilde \omega)}{\pi\nu}+
    C \left(\frac{1}{2}-\frac{1}{d}\right)\left(\frac{{\cal Y}_0({\bf r},{\bf r};{\tilde \omega})}{\pi\nu}\right)^2\right].
\end{equation}
which as a result of (ii), is no longer homogeneous. Here, $C$ is zero (unity)
for GOE (GUE) systems. The first term was also obtained in Ref.~\cite{Skipetrov08}
by the refined diagrammatic perturbation theory.
(iv) Remarkably, the correlator ${\cal Y}$ solves Eq.~(\ref{eq:96})
instead of Eq.~(\ref{eq:130}), with the local
diffusion coefficient given by
\begin{eqnarray}\label{eq:30}
D({\bf r},{\tilde \omega})=D_0+\delta D({\bf r};{\tilde \omega}).
\end{eqnarray}
For ${\bf r}$ being at the sample center, sending $L$ to infinity (so that
effects arising from interfaces do not play any role), we recover ordinary single parameter scaling equation
(\ref{eq:138}) from Eq.~(\ref{eq:107}) for $d$ near the critical dimension.

Most strikingly, Eq.~(\ref{eq:107}) shows that the local diffusion coefficient
depends on ${\bf r}$ (as well as $\tilde \omega$)
via the scaling factor $\lambda({\bf r};{\tilde \omega})=
(\pi\nu)^{-1}{\cal Y}_0({\bf r},{\bf r};\tilde \omega)$. In fact,
such a dependence persists to higher orders. This
shows that the local diffusion coefficient exhibits novel single parameter scaling, i.e.,
\begin{equation}\label{eq:139}
    \frac{D({\bf r};{\tilde \omega})}{D_0}=D_\infty(\lambda({\bf r};{\tilde \omega})),
\end{equation}
where $D_\infty(\lambda)$ is some scaling function depending on
both symmetries and dimensions.
It is important that the novel scaling (\ref{eq:139}) is valid
also for large $\lambda$, the non-perturbative regime.
This scaling, as we will see in the next section, is the key component
in the description of macroscopic wave transport through one-dimensional open localized media.
It is missed by the SCLD model \cite{Lagendijk00}.
We emphasize that the inability of the SCLD model to capture this scaling
is intrinsic to the {\it phenomenological} assumption of
one-loop self-consistency in open media, but not to the
VW theory \cite{Woelfle80a,Woelfle80}
which is formulated for infinite systems. Indeed, if we set ${\bf r}$ to
the sample center and send $L$ to infinity,
the scaling factor diverges as
$ \ln {\tilde \omega}$ (${\tilde \omega}^{-1/2}$) in two (one) dimensions,
and the VW theory captures these infrared divergences and was designed
for the purpose of (partly) summing up these divergences.

The diffusion coefficient
(\ref{eq:30}) and the local diffusion equation (\ref{eq:96}) have
important implications. (For media without internal reflection,)
localization effects are absent on the interface,
i.e., $D({\bf r}\in C;\tilde \omega)=D_0$,
which is robust against wave interference. In particular,
there is no boundary
layer which scales in the
same way as the diffusion constant in an infinite medium, as was assumed by
the earlier
phenomenological model \cite{Berkovits87}.
As indicated by Eq.~(\ref{eq:107}), this arises from the inhomogeneity of wave interference effects
in open media. Near the interface,
wave interference characterized by
${\cal Y}_0({\bf r},{\bf r};\tilde \omega)$ is largely diminished due to
wave energy leakage through the interface.

\subsubsection{Origin of the novel scaling}
\label{sec:scaling}

\begin{figure}[h]
  \centering
\includegraphics[width=8.0cm]{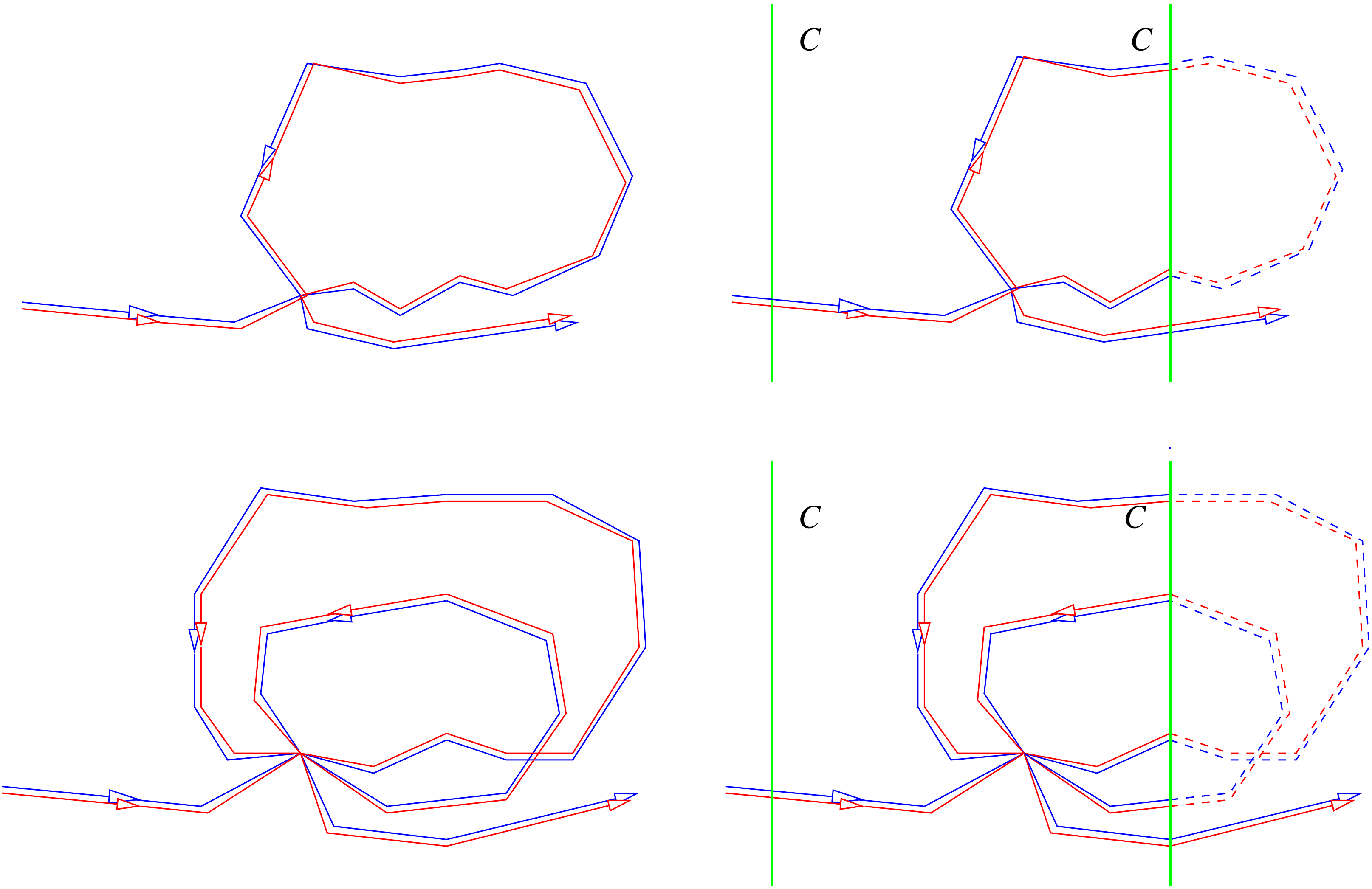}
 \caption{Left: examples of interference picture in
 infinite media.
Right: in open media, wave interference is suppressed because wave energies
leak out of the system (dashed line). Upper: interference
picture underlying one-loop weak localization.
Lower: examples of more complicated interference
picture. The green line, $C$, stands for the air-medium interface.}
  \label{fig:interference}
\end{figure}

It should be stressed that the novel scaling (\ref{eq:139})
applies to the local diffusion coefficient and therefore
describes the unconventional hydrodynamic behavior of wave propagation
in open media. It is fundamentally
different from the ordinary single parameter scaling theory of the
Thouless conductance \cite{Anderson79}, which
provides no information on how wave energies flow from one point to the other inside the medium.
As we will see in Sec.~\ref{sec:resonance},
the former leads to the latter and contains significantly more
information on wave transport. Now, let us discuss
the origin of the novel scaling. First of all,
optical paths may self-intersect, forming a loop.
It is well known that because of the time-reversal symmetry two optical paths may
counterpropagate along this loop and interfere constructively
with each other, leading to the (one-loop) weak localization
correction \cite{Altshuler83,Bergmann84} homogeneous in space
(Fig.~\ref{fig:interference}, upper left).
In the presence of open interfaces, the interference
picture is modified. That is,
wave energies leak out the system through the interfaces
(Fig.~\ref{fig:interference}, upper right) and as
such, at time $t$ the probability for a path to return to the
vicinity (of size $\sim \lbar^{d-1} dt$)
of its departure point explicitly depends on
the departure point ${\bf r}$: the returning probability is $\sim
{\cal Y}_0({\bf r},{\bf r}'={\bf r};t)\lbar^{d-1}dt
$ and increases as ${\bf r}$ moves into the medium. Here,
${\cal Y}_0({\bf r},{\bf r}';t)$
is the inverse Fourier transform of ${\cal Y}_0({\bf r},{\bf r}';{\tilde \omega})$.
Since wave interference enhances the backscattering
probability, the diffusion coefficient is suppressed by an amount
\begin{equation}\label{eq:5}
    \frac{\delta D^{(1)}({\bf r};{\tilde \omega})}{D_0} \sim - \lbar^{d-1} \int dt e^{i{\tilde \omega} t}
    {\cal Y}_0({\bf r},{\bf r};t) \sim -\lambda({\bf r};{\tilde \omega}).
\end{equation}

As waves penetrate deeper inside the medium, more complicated
wave interference effects take place. For example, optical paths
tend to
return to the vicinity of its departure point
more (say $n$) times
(e.g., Fig.~\ref{fig:interference}, lower panel) \cite{Altshuler98,Tian04},
with a probability
$\sim (\lbar^{d-1})^n\int_{t_1>t_2>\cdots >t_{n-1}} dt_1 dt_2 \cdots dt_{n-1} {\cal Y}_0
({\bf r},{\bf r};t-t_1){\cal Y}_0
({\bf r},{\bf r};t_1-t_2)\cdots {\cal Y}_0
({\bf r},{\bf r};t_{n-1})$. Two paths taking these loops as their routes
pass the $n$ loops in different orders. Having (almost) the same
phases, they also interfere constructively.
This results in a higher order wave interference correction,
\begin{eqnarray}\label{eq:6}
    \frac{\delta D^{(n)}({\bf r};{\tilde \omega})}{D_0} &\sim&
(\lbar^{d-1})^n\int dt e^{i{\tilde \omega} t}\int_{t_1>t_2>\cdots >t_{n-1}} dt_1 dt_2 \cdots dt_{n-1} {\cal Y}_0
({\bf r},{\bf r};t-t_1){\cal Y}_0({\bf r},{\bf r};t_1-t_2)\cdots {\cal Y}_0
({\bf r},{\bf r};t_{n-1})\nonumber\\
&\sim& \left(\lambda({\bf r};{\tilde \omega})\right)^n.
\end{eqnarray}
Thus, the position ${\bf r}$ enters into all the wave interference
corrections via the position-dependent return probability,
which justifies the novel scaling. It is important that for $n\geq 2$, the two optical paths,
though passing the loops in different orders,
may propagate along every loop in the same direction (clockwise or
anticlockwise). Such constructive
interference picture does not require
time-reversal symmetry, and is
key to establishing both the local diffusion and novel scaling in GUE systems.

\subsubsection{Unconventional Ohm's law}
\label{ohm's_law}

The existence of the local diffusion coefficient leads to
an interesting phenomenon -- the unconventional Ohm's law in the steady state
($\tilde \omega=0$).
Indeed, if the transverse plane of the slab is infinite, the local diffusion coefficient
is homogeneous in the transverse direction. As a result, $D({\bf r};\tilde \omega=0)
=D(x;\tilde \omega=0)\equiv D(x)$, where $x$ is the longitudinal coordinate $x$.
In the steady states, the system may sustain a uniform macroscopic current in
the longitudinal direction, $j$, going through the slab.
An intensity profile across the sample, $I(x)$, which is uniform in the transverse direction
is thereby built up. Eq.~(\ref{eq:96}) gives the local Fick's law,
\begin{equation}\label{eq:174}
    j=-D(x)\partial_x I(x),
\end{equation}
according to which the intensity bias across the sample is
$I(L)-I(0)=-j \int_0^L \frac{dx}{D(x)} $. Since
the ensemble-averaged transmission $\langle T(L)\rangle\propto j/(I(L)-I(0))$,
we find
\begin{equation}\label{eq:41}
    \langle T(L)\rangle^{-1} \sim
\int_0^L \frac{dx}{D(x)}.
\end{equation}
It suggests that the total resistance of the system, $\langle T(L)\rangle^{-1}$,
is the integration of the infinitesimal resistance, $\frac{dx}{D(x)}$. In this sense,
it is similar to the familiar Ohm's law.
However, it should be emphasized that in sharp contrast to the ordinary
Ohm's law, the resistivity $D^{-1}(x)$ is no longer inhomogeneous in space.
In particular, we will see in the next section that for localized samples
$D^{-1}(x)$ may increase by many orders of magnitude as the position
changes from the interface to the midpoint, and it is the
dramatic enhancement of $D^{-1}(x)$ near the sample center
that leads the system to exhibit a global localization
behavior namely the exponential decay of the average transmission in the sample length.
(For diffusive samples,
such position-dependence is weak and does not lead to any
interesting phenomena other than the scaling $\langle T(L)\rangle\sim 1/L$.)
For this reason, we term Eq.~(\ref{eq:41}) `unconventional Ohm's law'.
Such law was first conjectured in Ref.~\cite{Lagendijk00}.
The microscopic justification of Eq.~(\ref{eq:96}) puts this important conjecture
on a firm level.

\section{Local diffusion in one dimension}
\label{sec:localdiffusion1D}

In the one-dimensional case, the asymptotic behavior
of the scaling function $D_\infty(\lambda)$ at large $\lambda$ has been found
analytically \cite{Tian10}, which will be discussed in this section.
In this case, the macroscopic diffusion equation (\ref{eq:96}) is simplified to
\begin{eqnarray}
(-i\tilde \omega - \partial_x D(x;\tilde \omega) \partial_x) {\cal Y} (x,x';\tilde \omega) = \delta(x-x'),
\label{eq:2}
\end{eqnarray}
which is implemented with the boundary condition,
\begin{eqnarray}
{\cal Y} (x,x';\tilde \omega)|_{x=0\, {\rm or}\, L} = 0
\label{eq:8}
\end{eqnarray}
due to transparent interfaces.
Single parameter scaling (\ref{eq:139}) is simplified to
\begin{equation}\label{eq:28}
    \frac{D(x;{\tilde \omega})}{D_0}=D_\infty(\lambda(x;{\tilde \omega})),
\end{equation}
and the scaling factor to
\begin{eqnarray}\label{eq:34}
\lambda(x;\tilde \omega) \equiv \frac{{\cal Y}_0(x,x;\tilde \omega)}{\pi\nu}
= \frac{L}{\xi}\frac{\cosh\sqrt{\omega^*}
-\cosh(\sqrt{\omega^*}(1-2x/L))}{2\sqrt{\omega^*}\sinh\sqrt{\omega^*}},
\end{eqnarray}
with ${\omega^*}\equiv-i\tilde \omega L^2/D_0$.
Throughout this section the `effective' local density of states is the product
of the local density of states and the cross-sectional area of Q$1$D samples, $S$,
and we denote it as $\nu$ also in order to simplify notations.
The localization length $\xi\sim \nu D_0$ \cite{Efetov83a}.
We emphasize that Eqs.~(\ref{eq:2})-(\ref{eq:34}) are valid for both GOE and GUE systems.

Below, we will study in details the static case for which $\tilde \omega=0$.
We will present analytic results for the local diffusion coefficient,
$D(x)\equiv D(x,\tilde \omega=0)$ and then compare them with the results
obtained from numerical simulations.
As we will see, the analytic prediction of $D(x)$
is entirely confirmed by numerical simulations. Thus,
the novel scaling of the local diffusion coefficient
fully captures disorder-induced high transmission states (resonances) dominating
long-time wave transport. As such, the seemingly contradictory properties
of wave propagation, namely the local character of macroscopic diffusive behavior
and non-local character of resonances, are unified.
Here, by `local' it means that Fick's law is valid everywhere in space, while
by `non-local' that resonances result from
that waves propagate back and forth between two interfaces.

\subsection{Static local diffusion coefficient}
\label{sec:scaling1D}

For $\tilde \omega=0$,
Eqs.~(\ref{eq:28}) and (\ref{eq:34}) are further simplified to
\begin{eqnarray}
D(x)/D_0=D_\infty(\lambda(x)),\quad
\lambda(x;\tilde\omega=0)\equiv \lambda(x)
= x(L-x)/(L\xi).
\label{eq:4}
\end{eqnarray}
To proceed further, we introduce a new scaling function,
\begin{equation}\label{eq:132}
    {\cal G}(\lambda)\equiv \lambda^{-1}D_\infty(\lambda).
\end{equation}
The single parameter scaling (\ref{eq:4}) gives
the Gell-Mann--Low equation,
\begin{equation}\label{eq:40}
    \frac{d\ln {\cal G}}{d\ln \lambda}=\beta({\cal G}),
\end{equation}
with the $\beta$-function depending only on ${\cal G}$.

For ${\cal G} \gg 1$, according to Eq.~(\ref{eq:107})
the $\beta$-function is
expressed in terms of the $1/{\cal G}$-expansion,
\begin{equation}\label{eq:36}
    \beta({\cal G}) = -1+\frac{c_1}{\cal G} + \frac{c_2}{{\cal G}^2} +\cdots,\quad {\cal G} \gg 1.
\end{equation}
For GOE (GUE) systems, the nonvanishing subleading term is $\sim {\cal G}^{-1}$
(${\cal G}^{-2}$), and the corresponding coefficient $c_1\, (c_2)$ is negative
which arises from
the {\it inhomogeneous} leading weak localization correction.
As such, Eqs.~(\ref{eq:40}) and (\ref{eq:36}) bear an
analogy to the ordinary
single parameter scaling theory (\ref{eq:138}):
the scaling function ${\cal G}$ here plays the role of
the `Thouless conductance', $g$,
and $\lambda$ of the `size', $L_{\tilde \omega}$.
Identifying this analogy, one may
follow Ref.~\cite{Anderson80} to extrapolate the Gell-Mann--Low function
into the regime of small ${\cal G}$ and obtain
   \begin{eqnarray}
\label{eq:31}
  \beta({\cal G}) = \ln {\cal G},\quad {\cal G}\ll 1.
\end{eqnarray}
In fact, this non-perturbative $\beta$-function has been justified
by using the exact solution of the correlator (\ref{DC})
for special coset space leading to $Q\in {\bf GMat}(3,2|\Lambda)$ \cite{Zirnbauer91,Tian07}.
On the other hand, it has been well established \cite{Efetov97,Zirnbauer91,Zirnbauer92} that provided
strong Q$1$D localization exists,
its properties are insensitive to the symmetry of $Q$.
The latter only affects the numerical coefficient in
the analytic formula of the localization length \cite{Efetov83a,Zirnbauer91}.
Since introducing this exact solution requires very advanced mathematical knowledge
namely Fourier analysis on a hyperbolic supermanifold
with constant curvature
\cite{Zirnbauer91,Hueffmann90,Zirnbauer92} which is far beyond the reach of the present review,
we shall not discuss it further. Instead,
we will present the numerical confirmation of
the non-perturbative $\beta$-function (\ref{eq:31}) as well as a straightforward
physical interpretation below.

Notice that single parameter scaling (\ref{eq:4}) is valid for all the sample
lengths (larger than the transport mean free path).
Therefore, we may send $L\rightarrow \infty $ and the sample becomes semi-infinite.
In this case, the scaling factor $\lambda$ has a simple form of
$x/\xi$. On the other hand, the probability for waves to to penetrate into the sample
of a distance $x\gg \xi$
is exponentially small $\sim e^{-x/\xi}$ \cite{Lifshits79,Azbel83,Azbel83a,Freilikher03}.
Therefore, $D(x)$ decays exponentially for $x\gg \xi$, i.e.,
$D(x)/D_0 \sim e^{-x/\xi}$. This leads to
a large-$\lambda$ asymptotic scaling function, $D_\infty(\lambda) \sim e^{-\lambda}$.
Equation (\ref{eq:31}) is thereby reproduced and
thanks to the universality of Eq.~(\ref{eq:4}), is
independent of the sample length.

The novel single parameter scaling theory namely Eqs.~(\ref{eq:40})
and (\ref{eq:31}) -- for the local diffusion coefficient -- gives
a very remarkable prediction for
the local diffusion coefficient deep inside the localized sample, namely Eq.~(\ref{DSL}).
The result (solid lines in Fig.~\ref{fig:1Dlocal}) is
contrary to the result of SCLD \cite{Lagendijk00}: instead of simply decaying exponentially
from the interface (dashed lines in Fig.~\ref{fig:1Dlocal}), inside the sample $D(x)$
acquires an enhancement factor of $e^{x^2/(L\xi)}$, which drastically reduces the falloff of $D(x)$ into the sample.
Such a dramatic enhancement is also valid in one dimension where the
localization length is of the order of the transport mean free path.

\subsection{Numerical evidence of the novel scaling}
\label{sec:numerical}

Numerical experiments
on the spatially resolved wave intensity across a randomly
layered medium were performed in Ref.~\cite{Tian10} by adopting the method of
Ref.~\cite{Zhang09}. The random medium with layer thickness $a$ is embedded in an air
background. The relative permittivity
at each layer fluctuates independently, and is uniformly
distributed in the interval $[1-\sigma,1+\sigma]$ (the so-called `rectangular distribution'),
where $\sigma$ measures
the degree of randomness of the system and is
set to $0.7$ in numerical experiments. The air-medium interfaces are transparent.
Then, a plane wave of (angular) frequency $\omega$ is excited, and two frequencies, $\omega=1.65c/a$
and $0.72c/a$, are considered, where we have restored wave velocity in air.
For each $\omega$, two million dielectric
disordered configurations are realized for different sample lengths.
Then, for each disordered configuration,
the standard transfer matrix method is used to calculate the
transmission coefficient $T
$
and wave field $E(x)$, from which one obtains
the ensemble-averaged current $j \equiv \langle T
\rangle$
and wave intensity profile $I(x)\equiv \langle |E(x)|^2
\rangle$. Since the
current across the sample is uniform
in steady state, by further presuming the local Fick's law namely Eq.~(\ref{eq:174})
one may compute $D(x)$.

The numerical results of the local diffusion coefficient are shown in
Fig.~\ref{fig:1Dlocal} (squares and circles) for five different sample lengths.
In the limit $L\rightarrow \infty$, $D(x)$ decays
exponentially from the interface
(dotted line). This confirms the large ${\cal G}$-asymptotic of the $\beta$-function
namely Eq.~(\ref{eq:31}).
Most importantly, data points for
different frequencies, $1.65c/a$ (squares) and $0.72c/a$ (circles),
overlap which indicates that the scaling behavior is universal independent of the
microscopic parameters of media.
The universal curves are in excellent agreement with those
predicted by Eq.~(\ref{DSL}) (solid lines).
Obviously, except in the small regime
near the interfaces, the results from simulations are significantly
larger than those obtained by the prediction of the SCLD model (dashed lines)
and the deviation is more and more prominent as the ratio $L/\xi$ increases.

\begin{figure}
 \begin{center}
 \includegraphics[width=8.0cm]{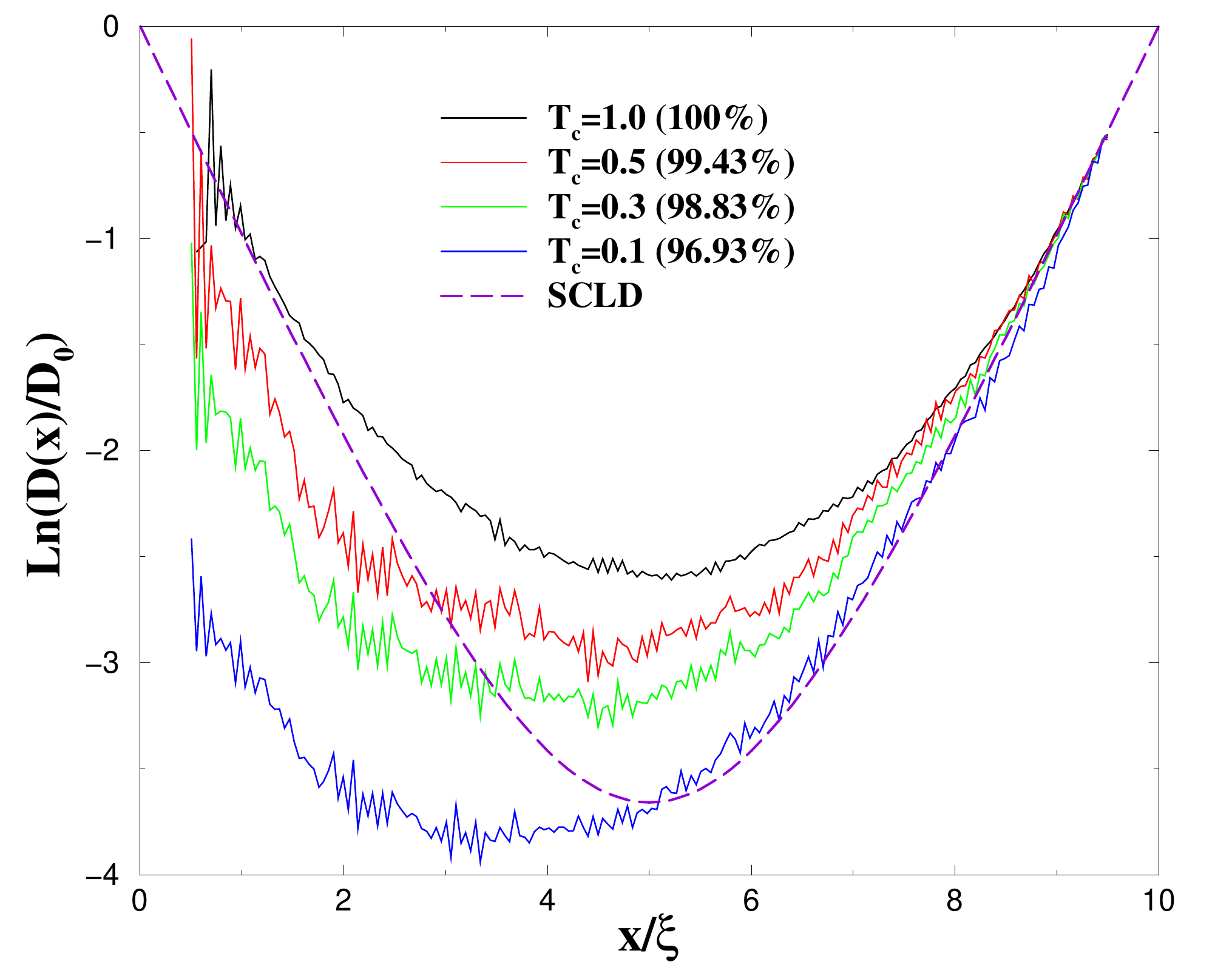}
\end{center}
 \caption{Local diffusion breaks down upon removing a small portion of
 high-transmission ($T>T_c$)
 states from the original ensemble. The source is placed at $x=0$.
(from Ref.~\cite{Tian10} with reproduction permission from C. S. Tian, S. K. Cheung, and Z. Q. Zhang
$\copyright$ The American Physical Society)}
 \label{fig:Tsuppression}
\end{figure}

\subsection{Roles of high transmission states}
\label{sec:resonance}

Dynamic studies of wave transport through localized samples \cite{Zhang09} have shown that
the long-time transport in these systems is dominated by rare long-lived modes
with high transmission coefficients (see Sec.~\ref{sec:DSPS}).
This indicates that in the strongly localized regime far-reaching consequences may arise from
the Lyapunov exponent fluctuations. To see this let us start from a semi-quantitative analysis.
A basic prediction of the SCLD model \cite{Lagendijk00} is as follows
that for localized samples $D(x)$ is
a sum of two symmetric exponential decays truncated at the sample mid-point
(see the dotted line in Fig.~\ref{fig:1Dlocal} which represents
the left truncated exponential decay).
This result may be written as
\begin{equation}\label{eq:39}
    D(x)\sim \int_0^\infty d\gamma P(\gamma)e^{-\gamma {\rm min}(x,\,L-x)}, \quad
    P(\gamma) = \delta (\gamma-\xi^{-1})
\end{equation}
for $x$ sufficiently away from the interfaces,
i.e., ${\rm min}(x,\,L-x) \gg \xi$.
Importantly, the above distribution $P(\gamma)$ reflects the absence of fluctuations in
the Lyapunov exponent. But this is correct only if
the sample length is infinite. (That is,
a semi-infinite sample is considered.)
For finite-sized localized samples, the distribution $P(\gamma)$
is given by Eq.~(\ref{eq:15}) instead.
Replacing the Dirac distribution in Eq.~(\ref{eq:39}) with the distribution
(\ref{eq:15}) we reproduce
the correct result for $D(x)$ namely Eq.~(\ref{DSL}), i.e.,
\begin{eqnarray}
D(x) &\sim& \int_0^\infty d\gamma\, e^{-\frac{\xi
L}{4}\left(\gamma-\xi^{-1}\right)^2} e^{-\gamma {\rm min}(x,\,L-x)} \nonumber\\
&\sim& e^{-L/(4\xi)} \int_0^\infty d\gamma\, e^{-\frac{\xi
L}{4}\gamma^2 + \gamma |L/2-x|} \sim e^{-\frac{x(L-x)}{L\xi}}.
\label{Lfluctuations}
\end{eqnarray}
This clearly shows that the SCLD model completely ignores the fluctuations of
the Lyapunov exponent. Interestingly,
the second line in Eq.~(\ref{Lfluctuations}) shows that the integral is
dominated by $\gamma\sim |1-2x/L|\xi^{-1}$. In particular, for $x$ closed to the sample center,
the integral is dominated by the small $\gamma$ tail
of the distribution $P(\gamma)$. This
suggests that the significant enhancement of the local diffusion coefficient
from the exponential decay near the sample center
is deeply rooted in the rare high transmission states.

In Ref.~\cite{Tian10}, the roles of
high transmission states in the static transport
are studied numerically, and data leading to results shown in Fig.~\ref{fig:1Dlocal}
are further analyzed. First of all, $\ln T$ follows the
normal distribution (\ref{eq:15})
\cite{Beenakker97,Anderson80}, with the average $\approx -L/\xi$ and
variance $\approx 2L/\xi$.
Then, to investigate the role of high transmission states
a small portion of them (with transmission $T>T_c$)
are removed from the original ensemble ($\omega=1.65c/a$ and $L/\xi=10$).
Most of removed states are singly localized states, and
the remainder (with a small portion) are necklace states \cite{Pendry94}.
$D(x)/D_0$ is re-computed for the new ensemble. As shown in
Fig.~\ref{fig:Tsuppression}, even when the fraction of states removed is as small as
$0.6\%$ (solid line, in red), the result deviates
drastically from the original one (solid line, in black) and is
asymmetric. This signals the breakdown of local diffusion and indicates
that rare high-transmission states are intrinsic to
local diffusion and novel scaling.

Substituting Eq.~(\ref{DSL}) into Eq.~(\ref{eq:41}), we find
\begin{equation}\label{eq:42}
    \langle T(L)\rangle\sim e^{-L/(4\xi)}.
\end{equation}
We see that the dramatic enhancement of
the resistivity near the sample center leads the system to exhibit a global
localization behavior namely the exponential decay of the average
transmission in the sample length, as we mentioned above.
On the other hand, from the normal distribution of $P(\gamma)$ it follows that
\begin{equation}\label{eq:143}
\frac{d\langle \ln T(L)\rangle}{dL} = -\xi^{-1}.
\end{equation}
Eqs.~(\ref{eq:42}) and
(\ref{eq:143}) show that (for both GOE and GUE systems,)
the localization lengths obtained by the arithmetic and geometrical means of $T(L)$
differ by a factor of $4$. This factor arises from
the rare disorder-induced resonant transmission, and previously
was discovered for the conductance distribution \cite{Beenakker97}.

\section{Conclusion}
\label{sec:conclusion}

How do classical electromagnetic waves propagate
through random open media? This is a long-standing fundamental issue
in Anderson localization. One might expect that like many complex systems
propagation of waves in random media bears
the microscopic and macroscopic description simultaneously. The
former is built upon the eigenmodes of
the Maxwell equation which provide complete information on
wave scattering (hence the term `microscopic'), while
the latter describes the system in terms of a single macroscopic variable --
the average energy density. It satisfies some macroscopic (generalized) diffusion equation
which is valid only on the scale much larger than
the mean free path (hence the term `macroscopic').
For open media, the microscopic description of
waves can be traded to the superposition of
excited quasi-normal modes.
In contrast, the rational of the macroscopic description especially for strongly localized
open media has been an open question explored by many researchers and had largely not been established as yet.
This review is devoted to substantial recent progress achieved in
the macroscopic (or `hydrodynamic' called by condensed matter physicists) description
of wave propagation in random open media.

We reviewed the recently developed first-principles theory
for classical wave localization in open media \cite{Tian08,Tian10} that justifies
unconventional macroscopic wave diffusion in these systems.
In essence, it describes localization physics in terms of an effective action
of the supermatrix $Q$ field.
The theory differs crucially from the supersymmetric field theory of (electron)
localization in infinite systems
in the air-medium coupling action
introduced by wave energy leakage through the interface.
The latter breaks the translational symmetry of the low-energy field theory and constrains
the supermatrix field by some boundary conditions.
We showed at a pedagogical level how to use this theory to explicitly calculate
the wave intensity profile and to establish a macroscopic description of wave propagation
at a firm level.

We reviewed highly unconventional macroscopic wave diffusion
in open media. This is described by a linear macroscopic equation --
the local diffusion equation. It
differs from the normal diffusion equation in that
the diffusion coefficient is inhomogeneous in space.
Most importantly, it was discovered in Ref.~\cite{Tian10}
that the local diffusion coefficient exhibits novel
single parameter scaling. That is, it depends on the distance to the interface
via a scaling factor proportional to the returning probability density,
and the universal scaling function
depends on both the system's symmetry and dimensionality.
In the static case (steady state), the scaling function in one dimensions
has been found analytically
giving the explicit analytic expression for the static local diffusion coefficient.
These results are fully confirmed by numerical simulations.
This suggests a profound new concept which is
contrary to previous expectations. That is, the
resonant transmission -- formed in the scale of system's size -- does not wash out
Fick's law (thereby macroscopic diffusion) which is valid locally in space; rather, it results in
novel scaling of the local diffusion coefficient which
plays a decisive role in the long-time transport of localized waves.

From the practical view point, the novel scaling of the diffusion coefficient (\ref{eq:139})
and the local diffusion equation (\ref{eq:96})
are sufficient for theoretical analysis of many experimental measurements involving
the averaged wave energy density.
Such approach based on the macroscopic equation has the great advantage of technical simplicity over
its sophisticated microscopic theory.
In this case, one has to find the scaling function of the
local diffusion coefficient by other methods
(e.g., numerical simulations or fitting experimental data), and
this is a relatively simple task.

We wish to mention some important directions in future studies.
While the novel scaling of local diffusion coefficient has been shown to
generally exist in the slab geometry in arbitrary dimensions, the
non-perturbative scaling function
remains to be worked out explicitly, and this could have far-reaching consequences.
In fact, the dynamic local diffusion coefficient
is related to a number of important issues.
These include time-resolved transmissions \cite{Maret06,Zhang09},
the statistics of quasi-normal modes \cite{Genack11},
dynamic single parameter scaling \cite{Zhang09}, anomalously localized states
and dynamic conductance \cite{Muzykantskii95}, etc..
Furthermore, the local diffusion coefficient
(both dynamic and static) in high dimensions may help to
reveal new physics due to the strong
interplay between the openness of the medium and criticality.
Indeed, we have seen that the novel scaling already
leads to highly non-trivial results in one dimensions. That is, the
localization length given by the mean transmission is four times
larger than the typical localization length.
We naturally expect that its effects in high dimensions
might be even more interesting. Whether and to what extent
the novel scaling of local diffusion coefficient
affects the size-dependence of the conductance near Anderson transition?
How does it lead to the pulsed wave decay on the output interface?
These questions are fully open.

In this review, we have focused on ideal optical systems ignoring
internal reflections at the boundary and absorption (gain) in the bulk.
In experiments, these can have significant effects.
Interestingly, there have been experimental and analytic evidence \cite{Tian12} showing that the common
wisdom in optics -- adding the extrapolation length to the sample length \cite{Niuwenhuizen} -- breaks down.
The interplay between internal reflection and wave interference may lead
to very rich localization behavior. The other issue,
effects of (linear) absorption and gain in random media
has received considerable attention \cite{John84,John85,Lai12,Kroha05,Kroha06},
but these works deal with infinite media. As far as realistic optical
devices or experimental environments are concerned, one often
deals with open media and this will lead to intriguing phenomena
(for examples, see Refs.~\cite{Genack06,Deysh05,Cao10,Zhang95,Beenakker96}).
A fundamental problem therefore is how these factors
interact with localization and unconventional macroscopic diffusion.

The supersymmetric field theory reviewed here is formulated for
scalar wave systems. As mentioned above, more complete studies of localization
of classical electromagnetic waves will require a formulation for vector wave
systems. What will be the symmetry of the supermatrix $Q$ and its low-energy action then?
To what extent will the vector nature modify the macroscopics of
localized waves in open media? These important questions
are central to studies of light localization and largely unexplored.
Throughout this review, we have focused on classical wave systems.
In principle, the macroscopics reviewed here
exists in electronic systems as well. As far as the latter is concerned, it is
well known that the Anderson transition is classified into ten symmetry classes
\cite{Altland96}.
Of particular interests is the sympletic class. For this symmetry class,
an intriguing phenomenon intrinsic to the openness of the system was
discovered some time ago \cite{Zirnbauer92}. That is, even though an infinite
Q$1$D disordered wire exhibits strong localization
\cite{Efetov83a}, the localization behavior is dramatically changed as long as
the system is open (namely coupled to ideal leads): the conductance
decreases to a constant instead of zero as the sample length goes to infinity.
It is therefore conceivable
that the symmetry may affect profoundly the behavior of the local diffusion coefficient.

Finally, we emphasize that throughout this review
we consider only ensemble averaged observables. The theory reviewed here provides no
information on individual disorder configurations.
The resonant properties of individual disorder configurations
in the localized regime can be studied by other approaches (see, e.g.,
Refs.~\cite{Freilikher03,Freilikher04,Freilikher12} and references therein).

\section*{Acknowledgements}

I am deeply grateful to Z. Q. Zhang and S. K. Cheung for
collaborations, with whom I have a joint work reviewed in this paper, and to
A. Z. Genack
for many important discussions and sharing his experimental data prior to publications.
I also would like to thank I. L. Aleiner, A. Altland, B. L. Altshuler,
K. Yu. Bliokh, S. Hikami, A. Kamenev, V. E. Kravtsov, A. A. Lisyansky, T. Nattermann, H. T. Nieh,
Y. Fyodorov, J. Zinn-Justin, and M. R. Zirnbauer for useful discussions and conversations, and
G. Maret for
his courtesy which makes the production of Figure 2 possible. Part of this work was done
during the 2012 trisemester `Disordered quantum systems'. I would like to thank the organizer
and Institut Henri Poincar$\acute{\rm e}$ for their hospitality.
Work supported by the NSFC (No. 11174174),
by the Tsinghua University Initiative Scientific Research Program (No. 2011Z02151) and
in part by the NSF (No. PHY05-25915).

\begin{appendix}

\section{The basics of Grassmann algebra and supermathematics}
\label{sec:supermathematics}

This Appendix includes some preliminary definitions and
basic theorems of Grassmann algebra and supermathematics which are required for following
(most) technical details of this review. We do not aim at a
complete introduction to mathematical foundation of the supersymmetric field theory,
for which we refer readers to
Refs.~\cite{Zirnbauer96,Efetov83,Efetov97,Berezin,Berezin61,Berezin87,Zirnbauer85,Zirnbauer97,Zinn-Justin}.
\\
\\
\noindent {\it Grassmann algebra.}
The Grassmann algebra ${\cal A}$ (over complex $\Bbb{C}$)
is an algebra constructed from $n$ generators
$\chi_i$.
These generators -- the so-called
Grassmannians -- satisfy the anticommuting relation, i.e.,
\begin{equation}\label{eq:85}
    \chi_i\chi_j+\chi_j\chi_i=0,\quad \forall{i,\,j}.
\end{equation}
From this definition it follows that $\chi_i^k=0$ if the integer $k\geq 2$.
This gives an important theorem: all elements in the Grassmann algebra ${\cal A}$ are first degree
polynomials in each generator $\chi_i$. In other words,
a generic function, $f(\chi_1,\chi_2,\cdots,\chi_n)$, must be
constructed from $2^n$ monomials, $\chi_1^{\alpha_1}\chi_2^{\alpha_2}\cdots\chi_n^{\alpha_n}$, where
the power $\alpha_k$ takes the value of either $0$ or $1$, i.e.,
\begin{equation}\label{eq:89}
    f(\chi_1,\chi_2,\cdots,\chi_n)=
    \sum_{\{\alpha_k\}} c_{\alpha_1 \alpha_2 \cdots  \alpha_n}
\chi_1^{\alpha_1}\chi_2^{\alpha_2}\cdots\chi_n^{\alpha_n},
\end{equation}
with the coefficient $c_{\alpha_1 \alpha_2 \cdots  \alpha_n} \in \Bbb{C}$.

Define the reflection:
\begin{equation}\label{eq:150}
    P(\chi_i)=-\chi_i.
\end{equation}
Then, a monomial of degree $k$ has a parity $(-1)^k$ for
$P(\chi_{i_1}\chi_{i_2}\cdots\chi_{i_k})=(-1)^k \chi_{i_1}\chi_{i_2}\cdots\chi_{i_k}$.
This divides the Grassmann algebra into two parts, ${\cal A}={\cal A}^+\cup {\cal A}^-, \,
{\cal A}^+ \cap {\cal A}^-=\emptyset$, where
${\cal A}^+$ (${\cal A}^-$) has even (odd) parity,
constructed from monomials of even (odd) number degree
and therefore composed of commuting (anticommuting) elements.
The former is a subalgebra in ${\cal A}$ whose elements are
similar to ordinary complex numbers.

The complex conjugate operation (adjoint) of the Grassmann algebra is defined by
\begin{equation}\label{eq:86}
    (\chi_i)^*=\chi_i^*,\quad (\chi_i^*)^*=-\chi_i,\quad
(\chi_i\chi_j)^*=\chi_i^*\chi_j^*.
\end{equation}
The `$-$' sign in the second relation makes $\chi_i^*\chi_i$ act like a `real number', i.e.,
$(\chi_i^*\chi_i)^*=\chi_i^*\chi_i$. Note that $\chi_i^*$ is independent of
$\chi_i$ and as such, the complex conjugate
of the Grassmann algebra is purely a formal definition.
\\
\\
\noindent {\it Calculus in the Grassmann algebra.} Eq.~(\ref{eq:89})
allows us to introduce the partial derivative with respect to
the generator $\chi_i$. Because any elements in the Grassmann algebra ${\cal A}$,
which are functions of the generators,
can be written as Eq.~(\ref{eq:89}), we
may move $\chi_i$ to the left for each term, obtaining
\begin{equation}\label{eq:151}
    f=f_0 +\chi_i f_1,
\end{equation}
where the coefficients $f_{0,1}$ are $\chi_i$-independent. Then, the left
partial derivative with respect to $\chi_i$ is defined as
\begin{eqnarray}
\label{eq:90}
\frac{\partial f}{\partial\chi_i} \equiv f_1.
\end{eqnarray}
(The right partial derivative can be introduced in the similar way.)
It is important that
in moving $\chi_i$ to the left, a fermion sign is left due to the
anticommuting relation (\ref{eq:85}). It is easy to check that the
partial derivative defined by Eq.~(\ref{eq:90}) is a nilpotent operator, i.e.,
$(\partial/\partial \chi_i)^2=0$. Furthermore, if we consider $\chi_i$ as an operator acting on ${\cal A}$
via left-multiplication, then the nilpotent operator $\partial/\partial \chi_i$ and
the operator $\chi_i$ constitute the Clifford algebra,
\begin{equation}\label{eq:152}
    \chi_i\chi_j+\chi_j\chi_i=0,\quad
    \frac{\partial }{\partial\chi_i}\frac{\partial }{\partial\chi_j}+
    \frac{\partial }{\partial\chi_j}\frac{\partial }{\partial\chi_i}=0,\quad
    \chi_i \frac{\partial }{\partial\chi_j}+\frac{\partial }{\partial\chi_j}\chi_i =\delta_{ij}.
\end{equation}

For Grassmann algebra, the integration operation is defined as the (left) derivative.
More precisely, the single variable integration is defined as
\begin{equation}\label{eq:88}
    \int fd\chi_i  \equiv \frac{\partial}{\partial \chi_i} f,
\end{equation}
while the multiple variable integration as
\begin{equation}\label{eq:9}
    \int f d\chi_{k}\cdots d\chi_{1} \equiv \frac{\partial}{\partial \chi_{k}
    }\cdots \frac{\partial}{\partial \chi_{1}} f.
\end{equation}
In fact, this formal definition satisfies the properties
of the ordinary definite integral. To see this let us denote the integration
defined in Eq.~(\ref{eq:88}) as $I$. First of all,
it is obviously a linear operator, $I[f_1+ f_2]=
I[f_1]+ I[f_2]$ for $f_{1,2}\in {\cal A}$. Secondly, because
the partial derivative is a nilpotent operator, we have
$I[\partial f/\partial \chi_i]=0$ implying that the integral of
a total derivative vanishes and $\partial I[f]/\partial \chi_i=0$
implying that a definite integral is a constant. Thirdly, if $f'$
is $\chi_i$-independent, we have $I[ff']=I[f]f'$: the overall factor can be pulled
out of the integral (from the right). Finally, if we make the replacement
$\chi_i\rightarrow \chi_i + \eta_i$, with $\eta_i$ a constant Grassmannian,
the integral remains the same implying $d\chi_i = d(\chi_i+\eta_i)$.
\\
\\
\noindent {\it Gaussian integrals with Grassmannians.} In this review we are
interested in the case where the generator
$\chi_i$ and its adjoint $\chi_i^*$ ($i=1,2,\cdots,n$) appear in pairs.
Consider an invertible $n\times n$ matrix, $M$, with the entries $M_{ij}\in \Bbb{C}$.
By Taylor expanding $e^{-\chi^*_i M_{ij} \chi_j}$ and using Eq.~(\ref{eq:9}), one finds an important identity,
\begin{equation}\label{eq:91}
\int e^{-\chi^*_i M_{ij} \chi_j}d\chi^*_1d\chi_1\cdots d\chi^*_nd\chi_n={\rm det} M.
\end{equation}
This is in sharp contrast to the Gaussian integral over ordinary complex variables,
$S_i^*,S_i,i=1,2,\cdots, n$, which is
\begin{equation}\label{eq:92}
\int e^{-S^*_i M_{ij} S_j}\frac{dS^*_1dS_1}{\pi}\cdots \frac{dS^*_ndS_n}{\pi}= ({\rm det} M)^{-1}.
\end{equation}
Importantly, the difference of the power of the determinant, $\pm 1$, on
the right-hand side of Eqs.~(\ref{eq:91}) and (\ref{eq:92}) reflect different rules of the change of variables
for Grassmiannians and complex variables. Indeed, the linear transformations:
$\chi'\equiv M\chi$ and $S' \equiv MS$
lead to $d\chi_n \cdots d\chi_1=\det M\, d\chi'_n \cdots d\chi'_1$
while $dS_n \cdots dS_1=(\det M)^{-1}\, dS'_n \cdots dS'_1$.
By using Eq.~(\ref{eq:91}), we
obtain another important identity:
\begin{eqnarray}
\label{eq:93}
  ({\rm det} M)^{-1} \int \chi_i\chi^*_j e^{-\chi_i^* M_{ij}\chi_j}  d\chi^*_1d\chi_1\cdots d\chi^*_nd\chi_n
  =(M^{-1})_{ij}.
\end{eqnarray}
\\
\\
\noindent {\it The supervector and the supermatrix.} A supervector $\phi$ is
composed of an $n$-anticommuting component vector $\chi$ and an
$m$-commuting component vector $S$, (We shall consider the case of $n=m$.)
\begin{eqnarray}
\phi \equiv \left(
       \begin{array}{c}
         \chi \\
         S \\
       \end{array}
     \right),\quad \chi=\left(
                          \begin{array}{c}
                            \chi_1 \\
                            \vdots \\
                            \chi_n \\
                          \end{array}
                        \right),\quad
                        S=\left(
                          \begin{array}{c}
                            S_1 \\
                            \vdots \\
                            S_n \\
                          \end{array}
                        \right),
\label{eq:153}
\end{eqnarray}
where $\chi_i$ ($S_i$) are anticommuting (commuting) variables.
All such supervectors constitute a linear space, $\Phi_{n,n}$.
The transpose of the supervector $\phi$ is defined by
\begin{equation}\label{eq:111}
    \phi^{\rm T}\equiv (\chi_1\cdots\chi_n S_1\cdots S_n),
\end{equation}
while its Hermitian conjugate by
\begin{equation}\label{eq:112}
    \phi^\dagger\equiv (\chi^*_1\cdots\chi^*_n S^*_1\cdots S^*_n).
\end{equation}
The inner product on the supervector space $\Phi_{n,n}$ is defined by
\begin{equation}\label{eq:113}
    (\phi,\phi')\equiv \phi^\dagger \phi'= \sum_{i=1}^n \chi^*_i\chi'_i + \sum_{i=1}^n S^*_i S'_i,\quad
    \forall \phi,\, \phi'\in \Phi_{n,n}.
\end{equation}
The norm $\sqrt{(\phi, \phi)}$ defines the `length' of a supervector $\phi$.

The linear transformation on the supervector space $\Phi_{n,n}$
is described by a supermatrix, $M$,
which transfers anticommuting (fermionic) into commuting (bosonic) components and
{\it vice versa}. It has the following structure,
\begin{eqnarray}
M\equiv \left(
  \begin{array}{cc}
    M_{\rm FF} & M_{\rm FB} \\
    M_{\rm BF} & M_{\rm BB}\\
  \end{array}
\right)^{\rm fb},
\label{eq:146}
\end{eqnarray}
with the superscript `fb' standing for the fermion-boson space
accounting for the supervector structure defined in the first relation
of Eq.~(\ref{eq:153}). Here,
$M_{\rm FF}$, $M_{\rm BB}$, $M_{\rm FB}$, and $M_{\rm BF}$ are $n\times n$
matrices. Furthermore,
the entries of diagonal (off-diagonal) blocks, $M_{\rm BB},\, M_{\rm FF}$ ($M_{\rm BF},\, M_{\rm FB}$), are commuting
(anticommuting) variables belonging to ${\cal A}^+$
(${\cal A}^-$). The supermatrix product is defined as usual.

The transpose of the supermatrix is defined by $({\rm M \phi})^{\rm T}\phi'\equiv \phi^{\rm T}
M^{\rm T} \phi',\, \forall \phi,\,\phi'\in \Phi_{n,n}$. This gives
\begin{eqnarray}
M^{\rm T} = \left(
  \begin{array}{cc}
    M_{\rm FF}^{\rm T} & -M_{\rm BF}^{\rm T} \\
    M_{\rm FB}^{\rm T} & M_{\rm BB}^{\rm T} \\
  \end{array}
\right)^{\rm fb},
\label{eq:70}
\end{eqnarray}
where the `$-$' sign in the upper right block
arises from the anticommuting relation, and for each block
$M_{\alpha\alpha'}$ the transpose is defined
as usual. By the definition
one may easily check that for two supermatrices, $M,\,M'$,
\begin{equation}\label{eq:69}
    (MM')^{\rm T}=M'^{\rm T}M^{\rm T},
\end{equation}
\begin{equation}\label{eq:122}
    (M\phi)^{\rm T}=\phi^{\rm T}M^{\rm T}.
\end{equation}

The Hermitian conjugate of the supermatrix is defined by
$({\rm M \phi},\phi')\equiv (\phi,
M^\dagger \phi'),\, \forall \phi,\,\phi'\in \Phi_{n,n}$.
This gives $M^\dagger = (M^{\rm T})^*$. The pseudounitary transformation
can be defined as such $U$ that conserves the metric tensor $K$, i.e.,
$(U\phi,KU\phi') \equiv (\phi,K\phi')$ or $U^\dagger KU=K$. If $K=\mathbbm{1}$, then
$U$ becomes unitary. Using the identity (\ref{eq:69})
and the third relation in Eq.~(\ref{eq:86}), we have
\begin{equation}\label{eq:115}
    (MM')^\dagger=M'^\dagger M^\dagger.
\end{equation}
By using Eq.~(\ref{eq:70})
and the second relation in Eq.~(\ref{eq:86}), we have
\begin{equation}\label{eq:116}
    (M^\dagger)^\dagger=M,
\end{equation}
which should be contrasted with $(M^{\rm T})^{\rm T} =\sigma_3^{\rm fb} M \sigma_3^{\rm fb}$.

For the supermatrix, we introduce the supertrace, denoted as `str',
\begin{equation}\label{eq:94}
    {\rm str} M\equiv {\rm tr}M_{\rm FF}-{\rm tr}M_{\rm BB},
\end{equation}
where `tr' stands for the normal trace. (This differs from that defined in some other references,
e.g., Refs.~\cite{Zirnbauer84,Zirnbauer05}, by a minus sign.)
By using the definition, it is easy to check the
following relations,
\begin{equation}\label{eq:114}
    {\rm str} (M M')={\rm str} (M'M)
\end{equation}
and
\begin{equation}\label{eq:162}
    (\phi, M\phi')=-{\rm str} (M\phi'\otimes \phi^\dagger).
\end{equation}

The superdeterminant is defined by
\begin{equation}\label{eq:117}
    \ln {\rm sdet}M \equiv {\rm str} \ln M,
\end{equation}
which is the direct generalization of the well-known identity
$\ln \det (\cdot) = {\rm tr} \ln (\cdot)$. This gives
\begin{equation}\label{eq:58}
    {\rm sdet}M\equiv {\rm det}(M_{\rm FF}-M_{\rm FB} M_{\rm BB}^{-1}M_{\rm BF})({\rm det}M_{\rm BB})^{-1}.
\end{equation}
By the definition we have
\begin{equation}\label{eq:118}
    {\rm sdet}(M M')= {\rm sdet}\, M{\rm sdet}M'.
\end{equation}
By using Eqs.~(\ref{eq:70}) and (\ref{eq:58}), we can show
\begin{equation}\label{eq:142}
    {\rm sdet}\,M^{\rm T}= {\rm sdet}\, M.
\end{equation}
\\
\\
\noindent {\it Gaussian integrals with supervectors.} Given the supermatrix $M$,
we have the following Gaussian integral over the supervector,
\begin{equation}\label{eq:119}
    \int e^{-\phi^\dagger M \phi}d\phi^\dagger d\phi={\rm sdet} M,
\end{equation}
where the measure $d\phi^\dagger d\phi\equiv \prod_{i=1}^n d\chi_i^* d\chi_i \prod_{i=1}^m
\frac{dS_i^*dS_i}{\pi}$, and
\begin{eqnarray}
\label{eq:120}
  ({\rm sdet} M)^{-1} \int \phi_i\phi^*_j e^{-\phi^\dagger M \phi}  d\phi^\dagger d\phi
  = (M^{-1})_{ij}.
\end{eqnarray}
In Eq.~(\ref{eq:120}), if the pre-exponential factor is a product
of $\phi$'s with the number $>2$, then the ordinary Wick theorem holds. The only difference is that
exchanging Grassmannians leads to a fermionic sign.
It is very important that here,
we have assumed the convergence of the integral, i.e.,
${\rm Re}\, S^\dagger M_{\rm BB} S>0$.

\begin{figure}[h]
  \centering
\includegraphics[width=10.0cm]{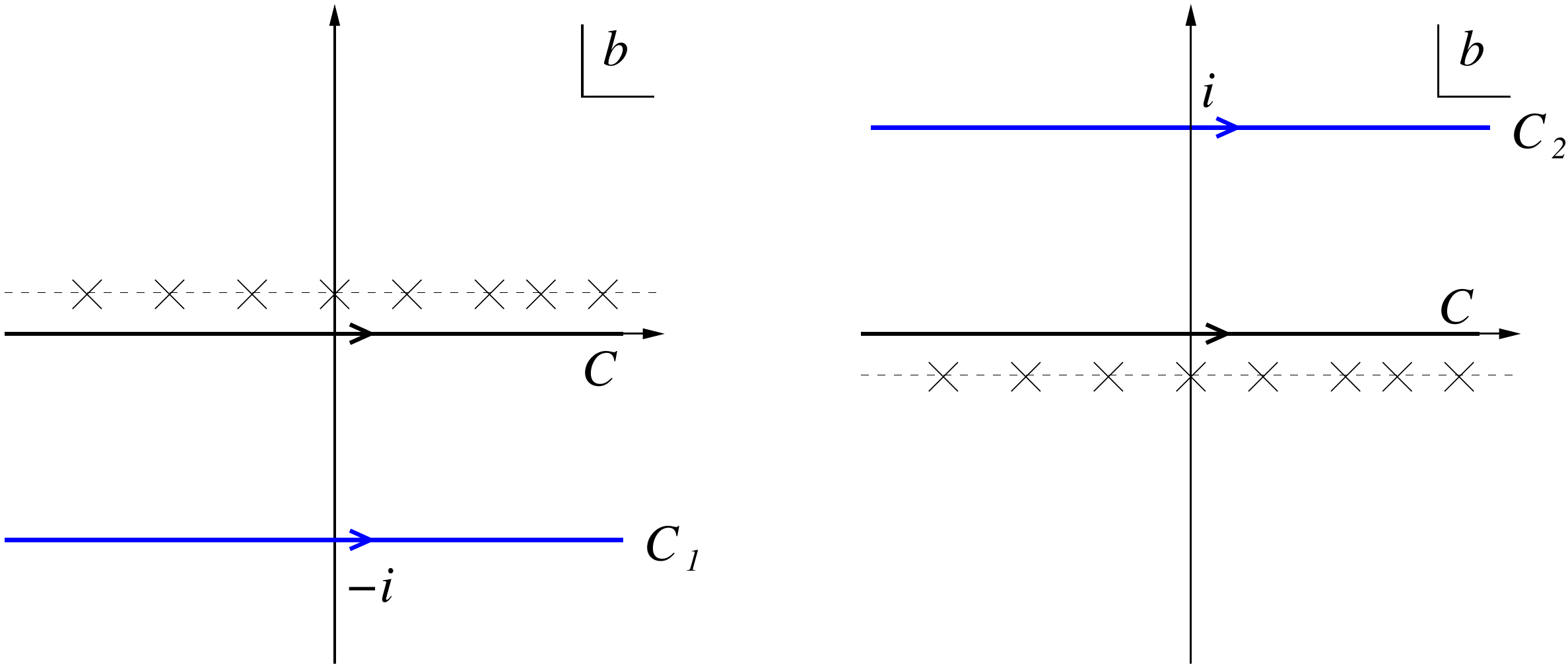}
 \caption{Contour deformation for the integral over the matrix element $b_{1t,1t}$ (left)
and $b_{2t,2t}$ (right).}
  \label{fig:analytic}
\end{figure}

\section{The diagonal mean-field saddle points}
\label{sec:saddlepoint}

In this Appendix we will find the diagonal solutions to
the saddle point equation (\ref{eq:52}),
i.e., $Q_0={\rm diag}\, \{q_i\}$. (Recall that the index $i=m\alpha t$.)
Substituting $Q_0$ into Eq.~(\ref{eq:52}) gives
\begin{eqnarray}\label{eq:56}
    q_i&=&
    \frac{1}{2}i\omega^2 \Delta \int\frac{d{\bf k}}{(2\pi)^d}\frac{1}{{\bf k}^2-\omega^2+
    2i\omega^2q_i}
=
\frac{1}{2}i\omega^2 \Delta\int_0^\infty dk \nu(k)\frac{1}{k^2-\omega^2+
2i\omega^2q_i}\nonumber\\
&=&\frac{1}{4}i\omega^2 \Delta\int_{-\infty}^\infty dk \nu(|k|)\frac{1}{k^2-\omega^2+
2i\omega^2q_i}.
\end{eqnarray}
Notice that depending on the dimension this integral may suffer ultraviolet divergence.
In this case, since the wavelength of photons has no lower bound,
one may cure this divergence by multiplying the integrand by a factor $e^{\pm i\delta \zeta}$, where
the `$+$'(`$-$') sign corresponds to ${\rm Re} q_i <0 \, (>0)$, and send
the positive infinitesimal $\delta$ to zero in the end.
(A more physical way of curing this divergence is to introduce
more realistic disorders as discussed in Sec.~\ref{sec:geomteric_optics}.
In any case, this divergence only enters into the renormalization of
the background value of the refractive index which is unimportant
for present discussions.)
In the limit $\Delta\omega \nu(\omega)\ll 1$ this transcendental equation is solvable, giving
\begin{eqnarray}\label{eq:181}
    q_i=\pm |{\rm Re}\, q_i| + i {\rm Im}\, q_i,\quad
    |{\rm Re}\, q_i| = \frac{\pi}{4}\Delta \omega \nu(\omega)
    + o(\Delta \omega \nu(\omega)),\,
    {\rm Im}\, q_i = o(\Delta \omega \nu(\omega)).
\end{eqnarray}
That is,
\begin{equation}\label{eq:57}
    q_i=\frac{\pi}{4}\Delta \omega \nu(\omega)\lambda_i=\tilde q\lambda_i,\quad
    \lambda_i=\pm 1.
\end{equation}

At the first glance, Eq.~(\ref{eq:57}) gives $2^8=256$ solutions. But
this is not true. First of all, due to the charge conjugation symmetry,
$Q=CKQ^{\rm T}KC^{\rm T}$ that inherits from the reality condition
(\ref{eq:23}),
\begin{equation}\label{eq:59}
\lambda_{m\alpha 1}=\lambda_{m\alpha 2}.
\end{equation}
Secondly, for the functional integral (\ref{average}) to converge,
it is necessary that $Q_{m{\rm B}t,m{\rm B}t}$ is purely imaginary.
Therefore, the saddle points (\ref{eq:57}) do not belong to
this manifold. To solve this problem,
in integrating out the matrix elements $Q_{m {\rm B}t,m{\rm B}t}$,
we deform the integral contour such that (i) no singularities
of the action $F[Q]$ are encountered, and (ii) the
new contour passes $q_{1{\rm B}t}=+\tilde q
$ ($\lambda_{1{\rm B}t}=+1$)
and $q_{2{\rm B}t}=-\tilde q
$ ($\lambda_{2{\rm B}t}=-1$), where the `$+$' (`$-$')
sign accounts for the analytic structure of the advanced (retarded) Green function.
To ensure the condition (i) we should first look for
the singularities of the integrand of
the partition function (\ref{average}). According to the definitions
(\ref{eq:117}) and (\ref{eq:58}), they
are the zeros of
${\rm det}(-\nabla^2-\omega^2+\omega\tilde\omega^+\sigma_3^{\rm ar}
\otimes {\mathbbm{1}}^{\rm tr} +2i\omega^2 Q_{\rm BB})$.
Defining $Q_{\rm BB}=ib$, we find that for the integral over the diagonal entries,
$b_{1t,1t}\in \Bbb{R}$, the singularities lie at the line ${\rm Im} b_{1t,1t}=\delta>0$. (Recall that
$\tilde \omega^+ =\tilde \omega +i\delta$.)
Taking this into account, we deform the contour, $C$, into $C_1$
that passes through $b_{1t,1t}=-i\tilde q
$,
corresponding to $\lambda_{1{\rm B}t}=1$
(Fig.~\ref{fig:analytic}, left).
Similarly, for the integral over the diagonal entries,
$b_{2t,2t}$, the singularities lie
at the line ${\rm Im} b_{2t,2t}=\delta<0$. Therefore,
we deform the contour into $C_2$ that passes
through $b_{2t,2t}=i\tilde q
$, corresponding to $\lambda_{2{\rm B}t}=-1$ (Fig.~\ref{fig:analytic}, right).
For the other cases, i.e., $\lambda_{1{\rm B}t}=-1$ and (or)
$\lambda_{2{\rm B}t}=1$, to perform the corresponding contour deformation
must cross the singularities. Such saddle points are
unphysical and must be discarded. Thus, we have
\begin{equation}\label{eq:60}
\lambda_{1{\rm B}t}=-\lambda_{2{\rm B}t}=1.
\end{equation}

Taking Eqs.~(\ref{eq:59}) and (\ref{eq:60})
into account, now we have only four possibilities
namely $\lambda_{1{\rm F}1}=\lambda_{1{\rm F}2}=\pm 1$
and $\lambda_{2{\rm F}1}=\lambda_{2{\rm F}2}=\pm 1$.
Integrating out fluctuations around the saddle points with $\lambda_{1{\rm F}t}=\lambda_{2{\rm F}t}$
leads to a negligible small factor because the
bosonic and fermionic degrees of freedom do not compensate.
Therefore, the two saddle points of $\lambda_{1{\rm F}t}=\lambda_{2{\rm F}t}$ do not play any roles.
Finally, only two saddle points out of $256$ possibilities are left:
$\Lambda$ and $-k\Lambda$.
Andreev and Altshuler
discovered \cite{Altshuler95,Altshuler96} that
both are the stationary points of
the action (\ref{action}). The former
(latter) determines the perturbative (non-perturbative) part
of the level correlator in finite closed systems.

\section{Derivation of the fluctuation action}
\label{sec:fluctuationaction}

The neighborhood of $Q_0=T_0^{-1}(\tilde q\Lambda) T_0$ in the
supermatrix space can be parametrized by $Q_0 + \delta Q
= (T_0(1+\delta T))^{-1} (\tilde q\Lambda +\delta \Lambda) (T_0(1+\delta T))$.
Here $\delta T$ and $\delta \Lambda$ parametrize fluctuations in
the coset space and eigenvalues, respectively.
Expanding $\delta T$ and $\delta \Lambda$, we obtain the fluctuation
$\delta Q\approx T_0^{-1} \delta \Lambda T_0+[Q_0,\delta T]$. It is easy to see that
the first (second) term (anti)commutes with $Q_0$. Therefore, the former (latter)
may be identified as $\delta Q^l$ ($\delta Q^t$). In other words,
the eigenvalue fluctuations generate the longitudinal components
while the fluctuations in the coset space generate the transverse components.
The former violates the nonlinear constraint (i.e., $Q^2$ is a
constant matrix)
and thereby brings $Q_0$ out of the saddle point manifold.

In the remainder of this Appendix we derive the fluctuation (\ref{eq:67}).
Upon the substitution of the decomposition $\delta Q_{\bf q}=\delta Q_{\bf q}^l+\delta Q_{\bf q}^t$,
we simplify Eq.~(\ref{eq:61}) to
\begin{eqnarray}
&&\int\frac{d{\bf q}}{(2\pi)^d} {\rm str}\left({\cal H}_{\bf q}^l \delta Q_{\bf q}^l
\delta Q_{-{\bf q}}^l+
{\cal H}_{\bf q}^t\delta Q_{\bf q}^t\delta Q_{-{\bf q}}^t\right), \label{eq:63}\\
&&{\cal H}_{\bf q}^l\equiv \Delta^{-1} + \omega^4\int\frac{d{\bf k}}{(2\pi)^d}
{\cal G}_0\left({\bf k}+\frac{{\bf q}}{2},Q_0\right)
{\cal G}_0\left({\bf k}-\frac{{\bf q}}{2},Q_0\right),\nonumber\\
&&{\cal H}_{\bf q}^t\equiv \Delta^{-1} + \omega^4\int\frac{d{\bf k}}{(2\pi)^d}
{\cal G}_0\left({\bf k}+\frac{{\bf q}}{2},Q_0\right)
{\cal G}_0\left({\bf k}-\frac{{\bf q}}{2},-Q_0\right).\nonumber
\label{eq:63}
\end{eqnarray}
To simplify the action of the longitudinal components, we notice
\begin{eqnarray}
&&\int\frac{d{\bf k}}{(2\pi)^d}{\cal G}_0\left({\bf k}+\frac{{\bf q}}{2},Q_0\right)
{\cal G}_0\left({\bf k}-\frac{{\bf q}}{2},Q_0\right)
=\int\frac{d{\bf k}}{(2\pi)^d}{\cal G}_0\left({\bf k}+\frac{{\bf q}}{2},T_0^{-1}(\tilde q\Lambda) T_0\right)
{\cal G}_0\left({\bf k}-\frac{{\bf q}}{2},T_0^{-1}(\tilde q\Lambda) T_0\right)\nonumber\\
&=& T_0^{-1}\left[\int\frac{d{\bf k}}{(2\pi)^d}{\cal G}_0\left({\bf k}+\frac{{\bf q}}{2},\tilde q\Lambda\right)
{\cal G}_0\left({\bf k}-\frac{{\bf q}}{2},\tilde q\Lambda\right)\right]T_0\nonumber\\
&=& T_0^{-1}{\rm diag}\left\{\int\frac{d{\bf k}}{(2\pi)^d}{\cal G}_0\left({\bf k}+\frac{{\bf q}}{2},\tilde q\lambda_i\right)
{\cal G}_0\left({\bf k}-\frac{{\bf q}}{2},\tilde q\lambda_i\right)\right\}T_0,
\label{eq:141}
\end{eqnarray}
where in deriving the last line, we have used the fact that
$\Lambda ={\rm diag}\, \{\lambda_i\}$ with $\lambda_{1\alpha t}=
-\lambda_{2\alpha t}=1$.
Because
\begin{eqnarray}
&&\int\frac{d{\bf k}}{(2\pi)^d}{\cal G}_0\left({\bf k}+\frac{{\bf q}}{2},\tilde q\lambda_i\right)
{\cal G}_0\left({\bf k}-\frac{{\bf q}}{2},\tilde q\lambda_i\right)
\approx \int\frac{d{\bf k}}{(2\pi)^d}{\cal G}_0\left({\bf k},\tilde q\lambda_i\right)
{\cal G}_0\left({\bf k},\tilde q\lambda_i\right)\nonumber\\
&=& \frac{1}{2}\int_{-\infty}^\infty dk \nu(|k|)
\frac{1}{(k^2-\omega^2+2i\omega^2 \tilde q\lambda_i)^2}=0,
\label{eq:64}
\end{eqnarray}
we have ${\cal H}_{\bf q}^l=\Delta^{-1}$. Here, the same ultraviolet
regularization as Eq.~(\ref{eq:56}) was used.

To simplify the action of the transverse components,
we notice
\begin{eqnarray}
&&\int\frac{d{\bf k}}{(2\pi)^d}{\cal G}_0\left({\bf k}+\frac{{\bf q}}{2},Q_0\right)
{\cal G}_0\left({\bf k}-\frac{{\bf q}}{2},-Q_0\right)\nonumber\\
&=& T_0^{-1}{\rm diag}\left\{\int\frac{d{\bf k}}{(2\pi)^d}
{\cal G}_0\left({\bf k}+\frac{{\bf q}}{2},\tilde q\lambda_i\right)
{\cal G}_0\left({\bf k}-\frac{{\bf q}}{2},-\tilde q\lambda_i\right)\right\}T_0\nonumber\\
&=& -T_0^{-1}{\rm diag}\left\{\int\frac{d{\bf k}}{(2\pi)^d}
\frac{1}{[({\bf k}+{\bf q}/2)^2-\omega^2 + 2i \omega^2 \tilde q\lambda_i
]
[({\bf k}-{\bf q}/2)^2-\omega^2 - 2i \omega^2 \tilde q\lambda_i
]}\right\}T_0.
\label{eq:65}
\end{eqnarray}
Since the integral over ${\bf k}$ is dominated by the shell of $|{\bf k}|\approx \omega$,
we have $|{\bf k}\cdot {\bf q}|\gg |{\bf q}|^2$. As a result, ${\cal H}_{\bf q}^t$ is simplified to
\begin{eqnarray}
{\cal H}_{\bf q}^t&=& \Delta^{-1} - \omega^4 T_0^{-1}{\rm diag}\left\{\int\frac{d{\bf k}}{(2\pi)^d}
\frac{1}{[{\bf k}^2 +{\bf k}\cdot {\bf q} -\omega^2 + 2i\omega^2 \tilde q \lambda_i
]
[{\bf k}^2 -{\bf k}\cdot {\bf q}-\omega^2 - 2i \omega^2 \tilde q \lambda_i
]}\right\}T_0\nonumber\\
&\approx& \Delta^{-1} - \Delta^{-1} T_0^{-1} {\rm diag}\left\{
\int\frac{d\Omega_d}{\Omega_d} \frac{1}{1-i\frac{2}{\pi \Delta \omega^2 \nu(\omega)}{\hat n}\cdot {\bf q}}\right\}T_0\nonumber\\
&=& \Delta^{-1} \left\{1-
\int\frac{d\Omega_d}{\Omega_d} \frac{1}{1-i\frac{2}{\pi \Delta \omega^2 \nu(\omega)}{\hat n}\cdot {\bf q}}\right\}
\nonumber\\
&\approx& \Delta^{-1} \left\{1- \left[1-\frac{1}{d}
\left(\frac{2}{\pi \Delta \omega^2 \nu(\omega)}\right)^2 {\bf q}^2\right]\right\}=
\frac{1}{d\Delta}\left(\frac{2}{\pi \Delta \omega^2 \nu(\omega)}\right)^2 {\bf q}^2,
\label{eq:129}
\end{eqnarray}
where ${\hat n}$ is the unit vector in the $d$-dimensional unit sphere of volume $\Omega_d$.

Substituting
Eq.~(\ref{eq:129}) into Eq.~(\ref{eq:63}) gives
\begin{equation}\label{eq:76}
\Delta^{-1}\left(
\int\frac{d{\bf q}}{(2\pi)^d}
{\rm str}\{\delta Q_{\bf q}^l\delta Q_{-{\bf q}}^l\}+
\frac{1}{d}
\left(\frac{2}{\pi \Delta \omega^2 \nu(\omega)}\right)^2\int\frac{d{\bf q}}{(2\pi)^d}|{\bf q}|^2
{\rm str}\{{
\delta Q}^t_{{\bf q}}{
\delta Q}^t_{-{\bf q}}\}\right).
\end{equation}
Upon inserting this action into the partition function
(\ref{average}), we may integrate out the longitudinal components, $\delta Q_{{\bf q}}^l$.
Such a Gaussian integral is unity because the
degrees of freedom of commuting and anticommuting
variables compensate each other. The remaining action
is a functional of the transverse components only which is
\begin{equation}\label{eq:66}
    \frac{1}{d\Delta}
\left(\frac{2}{\pi \Delta \omega^2 \nu(\omega)}\right)^2
\int d{\bf r}{\rm str}(\nabla {
\delta Q}^t({\bf r}))^2.
\end{equation}
Since
$(Q_0 + \delta Q^t({\bf r}))$
stays inside the manifold defined by Eq.~(\ref{eq:54}), we have
$\nabla {\delta Q}^t({\bf r}) = \tilde q \nabla (T^{-1}({\bf r}) \Lambda T({\bf r}))$.
Eq.~(\ref{eq:66}) then gives the fluctuation action (\ref{eq:67}).

\section{Derivation of the interface action}
\label{sec:interfaceaction}

In this Appendix, we will derive the interface action $F_{\rm int}[Q]$
with the help of the Zirnbauer-Efetov theorem.
Formally, Eq.~(\ref{theorem}) differs from Eq.~(\ref{Greenfunction}) only
in the non-Hermitian part of the effective Hamiltonian, i.e., $\pm i{\hat B}\delta_C$.
Therefore, we may repeat the procedures of Sec.~\ref{sec:HS_transformation}. As a result,
\begin{eqnarray}
F_{\rm int}[Q]
= -\frac{1}{2}\, {\rm str}_{{\bf r}}\ln\, \left(1-{\hat B} \Lambda {\cal
G}_0\right),
\label{interF}
\end{eqnarray}
where the supertrace `${\rm str}_{{\bf r}}$' includes the integral over
the spatial coordinates.
Near the interface we may choose locally the coordinate system as
${\bf r}\equiv ({\bf r}_\perp, z)$ where the first
$(d-1)$ coordinates give the projection to the interface $C$ and the last one
the distance to the interface. Note that the $z$-axis points towards the random medium.
As shown in Sec.~\ref{sec:fluctuation}, (inside the random
medium) the matrix Green function ${\cal G}_0$ exponentially
decays in space with a characteristic length of the order of
the mean free path. Moreover, both $Q$ and the
dielectric field vary over a scale much larger than the mean free path.
Taking these into account, we find
\begin{eqnarray}
{\cal G}_0({\bf r},{\bf r}';Q({\bf r}'))
\approx \frac{2
}{\pi i} \int
\frac{d^{d-1}{\bf k}_\perp}{(2\pi)^{d-1}}
\int_0^\infty dk \frac{e^{i{\bf k}_\perp\cdot ({\bf r}_\perp
-{\bf r}'_\perp
)}\cos (kz)\cos
(kz')}{(\omega({\bf r}'))^2-|{\bf k}_\perp|^2-k^2+i\frac{\omega({\bf r}')}{l({\bf r}')}
Q({\bf r}')}, \quad {\bf r},{\bf r}'\in {\cal V}_+.
\label{G0}
\end{eqnarray}
Here, $(\omega({\bf r}))^2= \omega^2(1+\overline{\epsilon({\bf r})})$, with
$(1+\overline{\epsilon({\bf r})})$ the local average refractive
index near the interface which generally causes the internal reflection.
Because $\omega({\bf r})$
locally depends on the average refractive index, from Eq.~(\ref{mfp}) we find that
near the interface the mean free path $l$ generally acquires a
${\bf r}$-dependence also.

On the other hand,
the Green function $g^R_{\omega^2}({\bf r},{\bf r}')$
is
\begin{eqnarray}
g_{\omega^2}^R({\bf r},{\bf r}')
=\frac{2}{\pi}\int \frac{d^{d-1}{\bf k}_\perp}{(2\pi)^{d-1}} \int_0^\infty dk
\frac{e^{i{\bf k}_\perp\cdot({\bf r}_\perp-{\bf r}_\perp')}\,\sin (kz)\sin
(kz')}{\omega^2-{\bf k}_\perp^2-k^2+i0^+}, \quad {\bf r},{\bf r}'\in {\cal V}_-.
\label{GFinterface}
\end{eqnarray}
Substituting it into Eq.~(\ref{effGF6}) gives
\begin{eqnarray}
({\hat B} f)\, ({\bf r}_\perp) =
\int
\frac{d^{d-1}{\bf k}_\perp}{(2\pi)^{d-1}}\int d^{d-1}{\bf r}_\perp '\,
\sqrt{\omega^2-|{\bf k}_\perp|^2}\, e^{i{\bf k}_\perp\cdot({\bf r}_\perp-{\bf r}_\perp')}\,
f({\bf r}_\perp ') \,.
\label{Bdefinition}
\end{eqnarray}

Let us insert Eqs.~(\ref{G0}) and (\ref{Bdefinition}) into Eq.~(\ref{interF}).
As a result,
\begin{eqnarray}
F_{\rm int}[Q] = -\frac{1}{2} \int d{\bf r}_{\perp}\int_{
|{\bf k}_\perp|\leq \omega}
\frac{d^{d-1}{\bf k}_\perp}{(2\pi)^{d-1}}\, {\rm str} \ln \left(1+
\alpha_{{\bf k}_\perp}({\bf r})\, \Lambda Q({\bf r})\right), \label{interF2}
\end{eqnarray}
where the coefficient
\begin{eqnarray}
\alpha_{{\bf k}_\perp}({\bf r})=\sqrt\frac{\omega^2-
|{\bf k}_\perp|^2}{\omega^2\left(1+\overline{\epsilon({\bf r})}\right)-|{\bf k}_\perp|^2}.
\label{alpha}
\end{eqnarray}
In deriving Eq.~(\ref{interF2}) we have adopted the trick used in
Appendix~\ref{sec:fluctuationaction}. That is, we
made the local gauge transformation
so that $Q({\bf r})$ becomes diagonal, and after integrating out $k$ we rotated the
diagonal matrix back to $Q({\bf r})$.
Eq.~(\ref{interF2}) can be rewritten as Eq.~(\ref{eq:131}) \cite{Tian08}, where
the internal reflection coefficient is
\begin{eqnarray}
R_{{\bf k}_\perp}({\bf r})=\left|\frac{1-\alpha_{{\bf k}_\perp}({\bf r})}{1+\alpha_{{\bf k}_\perp}({\bf r})}\right|^2,
\label{T}
\end{eqnarray}
in agreement with the well-known result \cite{Landau}.

\section{Derivation of Eq.~(\ref{eq:109})}
\label{sec:cancelation}

First of all, we have
\begin{eqnarray}
F_{2}[W] &=& \frac{\pi\nu}{2} \int d{\bf r}
{\rm str} \left(-2D_0(\nabla iW)^2(iW)^2 + i\tilde \omega (iW)^4\right)\nonumber\\
&=& \frac{\pi\nu}{2} \int d{\bf r}{\rm str}\left( 2 D_0\left(\nabla^2 iW
(iW)^3+\left(\nabla iW iW\right)^2+\left(\nabla iW\right)^2
(iW)^2\right) + i\tilde \omega (iW)^4 \right),
\label{F4transformation}
\end{eqnarray}
where in deriving the second line we have used the integral by parts.
Let us substitute it into the expansion (\ref{eq:98}).
This leads to the leading wave interference correction to the bare correlator
$\delta {\cal Y}_1 ({\bf r},{\bf r}';\tilde \omega)$. The latter has five nonvanishing
contributions, i.e., $\delta {\cal Y}_1 ({\bf r},{\bf r}';\tilde \omega)
=\delta {\cal Y}_{1a}+\delta {\cal Y}_{1b}+\delta {\cal Y}_{1c}+\delta {\cal Y}_{1d}+\delta {\cal Y}_{1e}$.
The first term is
\begin{eqnarray}
\label{eq:164}
  \delta {\cal Y}_{1a} \equiv -\left(\frac{\pi \nu(\omega)}{4\omega}\right)^2
  \left\langle{\rm str} A_+ iW({\bf r}) A_- (iW({\bf r}'))^3
+{\rm str} A_+ (iW({\bf r}))^3 A_- iW({\bf r}') \right\rangle_0
\end{eqnarray}
corresponding to the diagram given in Fig.~\ref{fig:diagram} (b).
The other terms are
\begin{eqnarray}
\delta {\cal Y}_{1b} &\equiv& \left(\frac{\pi \nu(\omega)}{4\omega}\right)^2 \pi\nu D_0
\int d{\bf r}_1\left\langle
{\rm str}A_+ iW({\bf r}) A_- iW({\bf r}'){\rm str} \nabla^2 iW({\bf r}_1)
(iW({\bf r}_1))^3 \right\rangle_0, \nonumber\\
\delta {\cal Y}_{1c}
&\equiv&  \left(\frac{\pi \nu(\omega)}{4\omega}\right)^2 \frac{i\pi\nu\tilde\omega }{2} \int d{\bf r}_1
\left\langle{\rm str}A_+ iW({\bf r}) A_- iW({\bf r}'){\rm str}(iW({\bf r}_1))^4 \right\rangle_0,\nonumber\\
\delta {\cal Y}_{1d} &\equiv& \left(\frac{\pi \nu(\omega)}{4\omega}\right)^2 \pi\nu D_0
\int d{\bf r}_1 \left\langle {\rm str}A_+ iW({\bf r}) A_- iW({\bf r}'){\rm str} (\nabla iW({\bf r}_1)
iW({\bf r}_1))^2 \right\rangle_0, \nonumber\\
\delta {\cal Y}_{1e} &\equiv& \left(\frac{\pi \nu(\omega)}{4\omega}\right)^2 \pi\nu D_0
\int d{\bf r}_1 \left\langle {\rm str}A_+ iW({\bf r}) A_- iW({\bf r}'){\rm str} (\nabla iW({\bf r}_1))^2
(iW({\bf r}_1))^2 \right\rangle_0
\label{eq:165}
\end{eqnarray}
given by Fig.~\ref{fig:diagram} (a).

For $\delta {\cal Y}_{1b}$ let us now make the first contraction by using the rules (\ref{eq:84}) and start from
the $W$-factor acted by the operator $\nabla^2$. There are three possibilities, leading to
\begin{eqnarray}
\delta {\cal Y}_{1b}
&  =  &  \left(\frac{\pi \nu(\omega)}{4\omega}\right)^2 \pi\nu D_0
\int d{\bf r}_1 \Bigg\{\left\langle{\rm str}A_+ iW({\bf r}) A_- \tkz{a}{iW}({\bf
r}') {\rm str}
\nabla^2 \tkz{b} {iW}({\bf r}_1) (iW({\bf r}_1))^3
\begin{tikzpicture}[remember picture,overlay]
    \draw (a) -- +(0,0.4) -| (b);
\end{tikzpicture}
\right\rangle_0
\nonumber\\
&  &  + \left\langle{\rm str}A_+ \tkz{a}
{iW}({\bf r}) A_- iW ({\bf r}') {\rm str}
\nabla^2 \tkz{b}{iW}({\bf r}_1) (iW({\bf r}_1))^3
\begin{tikzpicture}[remember picture,overlay]
    \draw (a) -- +(0,0.4) -| (b);
\end{tikzpicture}
\right\rangle_0 \nonumber\\
&  &  + \left\langle{\rm str}A_+ iW({\bf r}) A_- iW ({\bf r}') {\rm str}
\nabla^2 \tkz{a}{iW}({\bf r}_1) iW({\bf r}_1) \tkz{b}
{iW}({\bf r}_1) iW({\bf r}_1)
\begin{tikzpicture}[remember picture,overlay]
    \draw (a) -- +(0,0.4) -| (b);
\end{tikzpicture}
\right\rangle_0 \Bigg\}
= \delta {\cal Y}'_{1b} + \delta {\cal Y}''_{1b},
\label{Iaresult}
\end{eqnarray}
where
\begin{eqnarray}
\delta {\cal Y}'_{1b} &  =  &  \left(\frac{\pi \nu(\omega)}{4\omega}\right)^2 D_0\int d{\bf r}_1
\big\{\nabla^2
(-{\cal Y}_0({\bf
r}_1,{\bf r}';\tilde \omega)) \left\langle{\rm str} A_+ iW({\bf r}) A_- (iW({\bf r}_1))^3
\right\rangle_0\nonumber\\
&  & + \nabla^2
(-{\cal
Y}_0({\bf r}_1,{\bf r};\tilde\omega))\left\langle {\rm str}A_+ (iW({\bf r}_1))^3
A_-iW({\bf r}') \right\rangle_0 \big\},\nonumber\\
\delta {\cal Y}''_{1b} &  =  &  \left(\frac{\pi \nu(\omega)}{4\omega}\right)^2 D_0\int d{\bf r}_1
 \left\langle{\rm str}A_+ iW({\bf r}) A_- iW ({\bf r}') {\rm str}
\nabla^2 \tkz{a}{iW}({\bf r}_1) iW({\bf r}_1) \tkz{b}
{iW}({\bf r}_1) iW({\bf r}_1)
\begin{tikzpicture}[remember picture,overlay]
    \draw (a) -- +(0,0.4) -| (b);
\end{tikzpicture}
\right\rangle_0.
\label{eq:166}
\end{eqnarray}
Likewise, we have
\begin{eqnarray}
\delta {\cal Y}_{1c}
&=&  \left(\frac{\pi \nu(\omega)}{4\omega}\right)^2
i\pi\nu\tilde\omega \int d{\bf r}_1
\Bigg\{\left\langle{\rm str}A_+ iW({\bf r}) A_- \tkz{a}{iW}({\bf
r}') {\rm str} \tkz{b} {iW}({\bf r}_1) (iW({\bf r}_1))^3
\begin{tikzpicture}[remember picture,overlay]
    \draw (a) -- +(0,0.4) -| (b);
\end{tikzpicture}
\right\rangle_0
\nonumber\\
&  &  + \left\langle{\rm str}A_+ \tkz{a}
{iW}({\bf r}) A_- iW ({\bf r}') {\rm str}
\tkz{b}{iW}({\bf r}_1) (iW({\bf r}_1))^3
\begin{tikzpicture}[remember picture,overlay]
    \draw (a) -- +(0,0.4) -| (b);
\end{tikzpicture}
\right\rangle_0\Bigg\}\nonumber\\
&  =  &  \left(\frac{\pi \nu(\omega)}{4\omega}\right)^2 i\tilde \omega\int d{\bf r}_1
\big\{
(-{\cal Y}_0({\bf
r}_1,{\bf r}';\tilde \omega)) \left\langle{\rm str} A_+ iW({\bf r}) A_- (iW({\bf r}_1))^3
\right\rangle_0\nonumber\\
&  &  +
(-{\cal
Y}_0({\bf r}_1,{\bf r};\tilde\omega))\left\langle {\rm str}A_+ (iW({\bf r}_1))^3
A_-iW({\bf r}') \right\rangle_0\big\},
\label{Ib}
\end{eqnarray}
\begin{eqnarray}
\label{eq:168}
\delta {\cal Y}_{1d} &=& \delta {\cal Y}_{1d}'+\delta {\cal Y}_{1d}'', \nonumber\\
  \delta {\cal Y}'_{1d} &  =  &  \left(\frac{\pi \nu(\omega)}{4\omega}\right)^2 \pi\nu D_0\int d{\bf r}_1
 \left\langle{\rm str}A_+ iW({\bf r}) A_- iW ({\bf r}') {\rm str}
\nabla iW({\bf r}_1) \tkz{a}{iW}({\bf r}_1) \nabla iW({\bf r}_1) \tkz{b}
{iW}({\bf r}_1)
\begin{tikzpicture}[remember picture,overlay]
    \draw (a) -- +(0,0.4) -| (b);
\end{tikzpicture}
\right\rangle_0, \nonumber\\
\delta {\cal Y}''_{1d} &  =  &  \left(\frac{\pi \nu(\omega)}{4\omega}\right)^2 \pi\nu D_0\int d{\bf r}_1
 \left\langle{\rm str}A_+ iW({\bf r}) A_- iW ({\bf r}') {\rm str}
\nabla \tkz{a}{iW}({\bf r}_1) iW({\bf r}_1) \nabla \tkz{b}
{iW}({\bf r}_1) iW({\bf r}_1)
\begin{tikzpicture}[remember picture,overlay]
    \draw (a) -- +(0,0.4) -| (b);
\end{tikzpicture}
\right\rangle_0,
\end{eqnarray}
and
\begin{eqnarray}
\delta {\cal Y}_{1e} &=& \left(\frac{\pi \nu(\omega)}{4\omega}\right)^2 2 \pi\nu D_0
\int d{\bf r}_1 \left\langle {\rm str}A_+ iW({\bf r}) A_- iW({\bf r}'){\rm str}
\nabla iW({\bf r}_1)\nabla \tkz{a}{iW}({\bf r}_1) iW({\bf r}_1) \tkz{b}
{iW}({\bf r}_1)
\begin{tikzpicture}[remember picture,overlay]
    \draw (a) -- +(0,0.4) -| (b);
\end{tikzpicture} \right\rangle_0\nonumber\\
&=& -\left(\frac{\pi \nu(\omega)}{4\omega}\right)^2 \pi\nu D_0
\int d{\bf r}_1 \bigg\langle {\rm str}A_+ iW({\bf r}) A_- iW({\bf r}') \nonumber\\
&& \times {\rm str}\left(
iW({\bf r}_1) \nabla^2\tkz{a}{iW}({\bf r}_1) iW({\bf r}_1) \tkz{b}
{iW}({\bf r}_1)
\begin{tikzpicture}[remember picture,overlay]
    \draw (a) -- +(0,0.4) -| (b);
\end{tikzpicture}
+
iW({\bf r}_1)\nabla\tkz{a}{iW}({\bf r}_1) iW({\bf r}_1) \nabla\tkz{b}
{iW}({\bf r}_1)
\begin{tikzpicture}[remember picture,overlay]
    \draw (a) -- +(0,0.4) -| (b);
\end{tikzpicture} \right) \bigg\rangle_0.
\label{eq:169}
\end{eqnarray}
Here, we have used the integral by parts to obtain the second equality of Eq.~(\ref{eq:169}).

By using the definition of the bare propagator (\ref{eq:29}), we find
\begin{equation}\label{eq:171}
    \delta {\cal Y}_{1a}+\delta {\cal Y}_{1b}'+\delta {\cal Y}_{1c}=0
\end{equation}
which is a result of the energy conservation law (cf. Eq.~(\ref{eq:101})).
Furthermore,
$\delta {\cal Y}_{1b}''+\delta {\cal Y}_{1d}''+\delta {\cal Y}_{1e}=0$.
As a result, $\delta {\cal Y}_1 ({\bf r},{\bf r}';\tilde \omega)
=\delta {\cal Y}_{1d}'$. Applying the contraction rules (\ref{eq:84}) to Eq.~(\ref{eq:168})
and taking into account ${\bar W}=W$, we simplify $\delta {\cal Y}_{1d}'$ to
\begin{equation}\label{eq:170}
    \delta {\cal Y}'_{2d} =  \frac{1}{2}\left(\frac{\pi \nu(\omega)}{4\omega}\right)^2 D_0\int d{\bf r}_1
{\cal Y}_0({\bf r}_1,{\bf r}_1;\tilde \omega)
 \left\langle {\rm str}A_+ iW({\bf r}) A_- iW ({\bf r}') {\rm str}
(\nabla iW({\bf r}_1))^2 \right\rangle_0,
\end{equation}
which gives the second term of Eq.~(\ref{eq:109}).

\end{appendix}



\begin{thebibliography}{99}
\bibitem{note_Ward} This result corrects a mistake made in the original paper \cite{Tian08}.
However, it does not affect the main result of that paper.
\end{thebibliography}
\end{document}